\documentclass[10pt,prd,article,twocolumn,preprintnumbers,superscriptaddress,amsmath,amssymb,aps,longbibliography,nofootinbib]{revtex4-2}

\usepackage{dsfont}
\usepackage[utf8]{inputenc}
\usepackage{hyperref}
\usepackage{graphicx}
\usepackage[table,dvipsnames]{xcolor}
\usepackage{colortbl}
\usepackage{ragged2e}
\usepackage{amsthm}
\usepackage{amsmath}
\usepackage{url}
\usepackage[normalem]{ulem}
\usepackage[english]{babel}
\usepackage{slashed,centernot}
\usepackage{bbold}
\usepackage[capitalise]{cleveref}
\usepackage[noindentafter]{titlesec}
\usepackage{epigraph}
\usepackage{booktabs}
\usepackage{upgreek}
\usepackage{physics}
\usepackage{pifont}
\usepackage{booktabs}
\usepackage{braket}
\usepackage{float}
\usepackage{soul}
\usepackage{slashed}
\usepackage{mathtools}
\usepackage{bm}
\usepackage[compat=1.1.0]{tikz-feynman}

\usetikzlibrary{arrows.meta, fit}

\definecolor{red}{rgb}{0.6,.0706,.1373}
\definecolor{blue}{rgb}{0,0.396,0.741}
\definecolor{realblue}{rgb}{0,0,1}
\definecolor{green}{rgb}{0.25,0.6,0.2}
\definecolor{rossoc}{cmyk}{0,1,1,0.2}

\definecolor{teal}{rgb}{0, 0.5,0.5}
\colorlet{mylinkcolor}{blue}
\colorlet{mycitecolor}{blue}
\colorlet{myurlcolor}{blue}

\hypersetup{
  linkcolor  = mylinkcolor!,
  citecolor  = mycitecolor!,
  urlcolor   = myurlcolor!,
  colorlinks = true
}

\newcommand{\hc}{\; + \; \mathrm{h.c.} \;}

\newcommand{\leff}{\lambda_{\rm eff}}

\begin{document}

\title{Flavour Chains at the TeV Edge}

\author{Admir Greljo}
\email{admir.greljo@unibas.ch}
\affiliation{Department of Physics, University of Basel, Klingelbergstrasse 82,  CH-4056 Basel, Switzerland}

\author{Xavier Ponce D\'iaz}
\email{xavier.poncediaz@unibas.ch}
\affiliation{Department of Physics, University of Basel, Klingelbergstrasse 82,  CH-4056 Basel, Switzerland}

\author{Lia Schöneweiß}
\email{lia.schoeneweiss@unibas.ch}
\affiliation{Department of Physics, University of Basel, Klingelbergstrasse 82,  CH-4056 Basel, Switzerland}

\begin{abstract}
We study chains of vector-like fermions that generate the Yukawa hierarchies as products of small mixings, with the Higgs coupled at a single link. Minimal Flavour Protection allows these flavour messengers to live at the TeV scale, while Higgs naturalness offers a reason to place them there: in cosmological solutions based on Higgs criticality, their large multiplicity lowers the critical point of the Higgs potential to the TeV scale. Crucially, the same Higgs interactions induce tree-level shifts in the $Z$ and $W$ couplings of light fermions. Existing LEP and SLD measurements already bound $M_F/|y_F|\gtrsim 3$\,TeV, while a future Tera-$Z$ factory such as FCC-ee could probe up to $100$\,TeV. It would test the parameter space most relevant to Higgs naturalness, reaching scenarios with Higgs-mass tuning at the level of $10^{-4}$ or worse in minimal realisations. Longer chains partially alleviate this tuning but do not eliminate it, making a Tera-$Z$ factory a decisive test of the framework.
\end{abstract}
\maketitle

{\small \tableofcontents}
\pagebreak

\section{Introduction}
\label{sec:intro}

The Yukawa couplings of the Standard Model (SM) span almost six orders of magnitude, from the top quark to the electron, with the quark mixing angles small and hierarchical, while the lepton mixing angles are large. While the SM accommodates this pattern, it offers no explanation for its origin, leaving the hierarchy of fermion masses, the so-called \emph{flavour puzzle}, as one of the central open questions in particle physics~\cite{Feruglio:2015jfa, Altmannshofer:2024hmr}. Dynamical explanations of this puzzle have existed for a long time, the paradigmatic example being Froggatt--Nielsen (FN) models~\cite{Froggatt:1978nt, Leurer:1992wg}, in which the hierarchies arise as powers of a small symmetry-breaking parameter.\footnote{The horizontal symmetry can be global or gauged. In the former case the spectrum contains an axion or axion-like particle, which is strongly constrained~\cite{Ema:2016ops, Calibbi:2016hwq, Greljo:2024evt}; for gauged realisations see e.g.~Ref.~\cite{Smolkovic:2019jow}.} However, the same dynamics generically induces flavour-changing neutral currents (FCNC) that are severely constrained by kaon, $\mu$, charm, $\tau$ and $b$ observables, which typically push the scale of flavour dynamics far above the TeV scale~\cite{Cornella:2023zme, Bordone:2019uzc, Silvestrini:2018dos, Fedele:2020fvh}. This is referred to as the \emph{new-physics flavour puzzle}~\cite{Isidori:2010kg}.

This situation is in tension with the naturalness of the electroweak scale~\cite{Susskind:1978ms, tHooft:1979rat, Ellis:1986yg, Barbieri:1987fn, Anderson:1994dz, Halverson:2026cxe, Farina:2013mla, Giudice:2017pzm}, which favours new degrees of freedom coupled to the Higgs boson not far above the TeV scale. A TeV-scale flavour dynamics responsible for the observed hierarchies, linked to the microscopic theory of the Higgs boson, would therefore be doubly attractive~\cite{Davighi:2023iks}, if only it could be reconciled with the FCNC bounds. From a bottom-up perspective, the consistency of TeV-scale new physics with flavour data has been organised in terms of approximate flavour symmetries acting on the operators of the Standard Model Effective Field Theory (SMEFT): Minimal Flavour Violation (MFV), based on the $\mathrm{U}(3)^5$ global symmetry of the gauge sector~\cite{DAmbrosio:2002vsn, Cirigliano:2005ck}; its less restrictive $\mathrm{U}(2)^5$ version, which singles out the third family~\cite{Barbieri:2011ci, Barbieri:2012uh, Fuentes-Martin:2019mun}; and, more recently, Minimal Flavour Protection (MFP)~\cite{Greljo:2025mwj}, which identifies $\mathrm{SU}(2)_q\times\mathrm{U}(1)_X$ as the minimal continuous subgroup that must be approximately preserved by otherwise generic TeV-scale dynamics. Unlike MFV, MFP uses a few uniformly sized spurions of $\mathcal{O}(10^{-2})$ to reproduce all observed Yukawa hierarchies, as in the FN mechanism. Unlike FN, MFP provides the necessary protection against FCNCs for generic short-distance physics represented by SMEFT. (For related recent work on IR flavor symmetries, see~\cite{Greljo:2026pkz, Banks:2025pxa, Carrasco-Martinez:2026wzu, Glioti:2024hye, 
Antusch:2023shi,
Faroughy:2020ina, Greljo:2022cah}.)

In this paper, we study explicit ultraviolet examples of dynamics underlying flavour hierarchies entirely placed at the TeV scale, as recently proposed in Refs.~\cite{Greljo:2025mwj, Arkani-Hamed:2026wwy}. The basic building block is a \emph{chain} of vector-like fermions (VLFs) with mass mixings among themselves and with SM fermions (cf.~the universal seesaw~\cite{Berezhiani:1983hm, Davidson:1987mh,Kaplan:1991dc,Botella:2016ibj, FileviezPerez:2026dlz}), and a Higgs coupling $y_F$ at a single position along the chain. Integrating out the heavy states, the effective Yukawa coupling of a SM fermion is given by the product of the small mixing ratios $\mu/M$ encountered along the chain, so that the length of the chain controls the hierarchy, in close analogy with dimensional deconstruction and clockwork constructions~\cite{Arkani-Hamed:1999ylh, Arkani-Hamed:2001kyx, Choi:2015fiu, Kaplan:2015fuy, Giudice:2016yja, Alonso:2018bcg}. Locality along the chain, and carefully chosen links between chains, ensure that flavour- and CP-violating effects are suppressed even for VLF masses $M_F$ at the TeV scale, where direct searches at the LHC currently exclude coloured partners only up to about $1.5\,$TeV~\cite{Benbrik:2024fku}, and $\mathrm{SU}(2)_L$-singlet lepton partners only up to about $0.4$~TeV~\cite{ATLAS:2024mrr}. Finally, matching the theory to the SMEFT after integrating out the full chain yields an approximate flavour symmetry that structurally resembles MFP.

A complementary question is \emph{why the flavour messengers should lie near the electroweak scale in the first place}. The Yukawa hierarchies constrain ratios of chain mixings to masses, not the common mass scale, while flavour protection establishes that a low scale is allowed but not required. A possible motivation arises if these same Higgs-coupled fermions participate in a cosmological solution to the Higgs hierarchy problem. In particular, they can modify the Higgs metastability scale, relevant to a class of cosmological mechanisms that hinge on Higgs criticality~\cite{Bak:1987xua, Kartvelishvili:2020thd, Asadi:2022njl, Khoury:2021zao, Giudice:2021viw, Benevedes:2025qwt, Detering:2025lfn, Detering:2025nmh, Enguita:2025ybx, Benevedes:2024tdq}, such as the self-organised localisation~\cite{Giudice:2021viw}. We examine this connection in \cref{sec:sol_perspective}, keeping its model-dependent cosmological assumptions distinct from the electroweak predictions of the flavour sector.

The central observation of this work is that, in their minimal realisations, such models cannot escape electroweak precision observables (EWPO) at a future Tera-$Z$ factory such as the proposed FCC-ee~\cite{FCC:2025lpp, deBlas:2025gyz}. The same Higgs insertion in the chain that produces a SM fermion's Yukawa coupling also generates, at tree level, the dimension-six operators $(H^\dagger i\overleftrightarrow D_\mu H)(\bar f\gamma^\mu f)$ and their $\mathrm{SU}(2)_L$-triplet counterparts that modify SM fermion couplings to the $Z$ and $W$ bosons. For the fermion directly coupled to the Higgs, the corresponding Wilson coefficient scales as $|y_F|^2/M_F^2$ and carries no spurion suppression. These effects cannot be removed without disrupting the mechanism responsible for fermion-mass generation, and precision measurements at the $Z$ pole at LEP and SLD~\cite{ALEPH:2005ab, ALEPH:2013dgf} already probe VLF masses of several TeV for order-one Higgs couplings and exceed the direct-search bounds, but are still not decisive in testing the naturalness-preferred parameter space. However, FCC-ee is expected to collect of order $6\times 10^{12}$ $Z$ decays~\cite{FCC:2025lpp,deBlas:2025gyz}, compared with the $1.7\times10^{7}$ recorded at LEP~\cite{ALEPH:2005ab}, improving precision on electroweak observables by up to three orders of magnitude. The indirect reach of such a Tera-$Z$ programme for heavy new physics has been studied extensively within SMEFT~\cite{Fan:2014vta, deBlas:2022ofj, Allwicher:2023shc, Celada:2024mcf, Stefanek:2024kds, Allwicher:2024sso, Maura:2024zxz, Greljo:2024ytg, Gargalionis:2024jaw, Greljo:2025ggc, Born:2026tgm}; these studies find that FCC-ee could probe new physics to another order of magnitude in the effective scale.

We find that, for the models at hand, the implications go beyond a mere increase of the mass bounds: the Higgs couplings responsible for the electroweak effects also feed into the Higgs mass parameter, with a correction scaling as $|y_F|^2 M_F^2$. Since the EWPO constrain $M_F/|y_F|$, raising the bound by an order of magnitude increases the required cancellation in the Higgs sector by two orders of magnitude. A Tera-$Z$ factory would thus become a quantitative test of the naturalness of TeV-scale flavour models, in the same spirit as the analysis of supersymmetric scenarios in Ref.~\cite{Greljo:2025ggc}. In summary, the logic of this work can be condensed into:
\begin{center}
\begin{minipage}{\linewidth}\centering
\renewcommand{\arraystretch}{1.35}%
\small
$\begin{array}{r@{\;\Longrightarrow\;}l}
\text{SM flavour puzzle} & \text{Chains of VLFs,}\\
\text{Higgs naturalness (e.g.~criticality)} & \text{TeV scale \emph{required},}\\
\text{Minimal Flavour Protection} & \text{TeV scale \emph{allowed},}\\
\text{Tera-}Z & \text{the decisive test.}
\end{array}$\\[3pt]
\end{minipage}
\end{center}
The paper is organised as follows. In \cref{sec:model_and_flavour} we review the VLF-chain models of Refs.~\cite{Greljo:2025mwj,Arkani-Hamed:2026wwy}, present the tree-level matching onto the SMEFT, summarise the flavour and direct-search constraints, and introduce the measure of the electroweak tuning adopted throughout. In \cref{sec:EW_analysis} we derive the current bounds from a global fit to the EWPO, cross-check them with three independent public codes, and project them to the Tera-$Z$ stage of FCC-ee under several assumptions about theoretical uncertainties. In \cref{sec:general_lessons} we generalise the discussion to arbitrary chain lengths and positions of the Higgs coupling along the chain, showing how longer chains can reduce, but not remove, the tuning. In \cref{sec:sol_perspective} we critically assess a possible cosmological motivation for the TeV-scale flavour model based on the Higgs criticality approach to the hierarchy problem, exploring the role of heavy fermions and its implications for the interpretation of our bounds. We conclude in \cref{sec:conc}. \cref{app:VLF_chains} provides further details on the VLF chains and the flavour bounds, while \cref{app:EWPO_analysis} details the electroweak fits.

\section{Flavour hierarchies from TeV-scale dynamics}
\label{sec:model_and_flavour}

In this section, we review the features of the VLF chain models introduced in Refs.~\cite{Greljo:2025mwj, Arkani-Hamed:2026wwy} to explain the flavour hierarchies of the SM fermions. We assume the new fermions share a common mass scale $M_{F}$, while their mass mixings are either zero or of order $\mu$. The Yukawa hierarchies follow from consecutive mixings along the chain, $y^f_\textrm{eff} \sim (\mu/M_F)^{N_f}$, where $N_f$ is the length of the chain. The texture of the mass matrices, i.e.~the position of the zeros, is crucial not only for generating the Yukawa hierarchies but also for suppressing FCNCs. In this paper, the textures are controlled by the flavour symmetry of the MFP~\cite{Greljo:2025mwj}.

\begin{figure*}[t]
    \centering    \includegraphics[width=\linewidth]{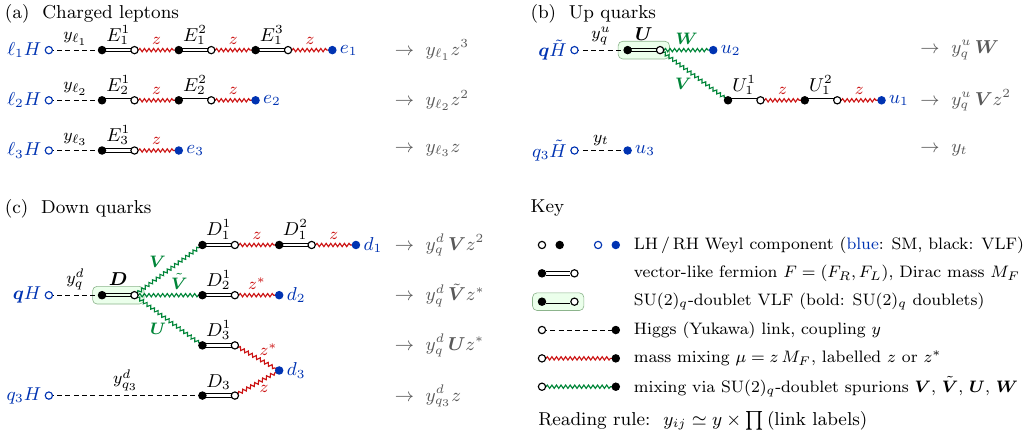}
    \caption{Chain diagrams of the benchmark model of Ref.~\cite{Greljo:2025mwj}, in the style of Ref.~\cite{Arkani-Hamed:2026wwy}: (a) charged leptons, (b) up quarks, (c) down quarks. Sites are the LH (open circles) and RH (filled circles) components of the SM fermions (blue) and of the VLFs (black). Each VLF is a pair of sites joined by its mass $M_F$ (double line); the $\mathrm{SU}(2)_q$-doublet VLFs are shaded. Dashed lines are Higgs couplings, wiggly lines are mass mixings $\mu_x\simeq x\,M_F$, labelled by their spurion $x$ (\cref{tab:field_content}). The effective Yukawa couplings on the right follow from \cref{eq:reading_rule}, up to signs and order-one factors.}
    \label{fig:benchmark_chains}
\end{figure*}

We describe these models with the \emph{chain diagrams} of Ref.~\cite{Arkani-Hamed:2026wwy}; see also~\cite{Arkani-Hamed:2026uos}. The sites of a chain are the chiral components of the SM fermions and of the VLFs, and its links are the couplings among them: Higgs couplings $y_F$, VLF masses $M_F$, and small mass mixings $\mu\ll M_F$. We will illustrate this on a single chain in \cref{sec:single_chain}, present the field content and the chains of the benchmark model in \cref{sec:benchmark_model} (see \cref{tab:field_content} and \cref{fig:benchmark_chains}), and finally in \cref{sec:moving_higgs}, discuss where the Higgs coupling can sit along a chain. The explicit mass matrices and the derivation of the CKM matrix are collected in \cref{app:VLF_chains}.

\subsection{The mechanism on a single chain}
\label{sec:single_chain}

The simplest example is the $\tau$ lepton. Its renormalisable Yukawa coupling is forbidden by the flavour symmetry of \cref{sec:benchmark_model}, and $\ell_3$ instead couples through the Higgs to a heavy VLF $E_3^1$ with the quantum numbers of $e_3$, which has a small mass mixing with $e_3$,
\begin{equation}
\label{eq:tau_chain}
    \mathcal{L} \supset - y_{\ell_3}\,\bar\ell_3 H E^1_{3R} - M_E\,\bar E^1_{3L} E^1_{3R} - \mu_z\,\bar E^1_{3L}\, e_3 \hc.
\end{equation}
For $\mu_z\ll M_E$, the light right-handed $\tau$ is mostly $e_3$, with an admixture $\mu_z/M_E$ of $E^1_{3R}$, so that integrating out $E_3^1$ gives $|y_\tau| \simeq |y_{\ell_3}|\,\mu_z/M_E\equiv |y_{\ell_3}|\,z$. Each further VLF inserted between the Higgs link and $e_3$ costs another factor of $z$. We now make this general.

\paragraph*{General setup.} These models follow the generic mass structure
\begin{equation}
    \mathcal{L}_\textrm{Yukawa} = -\bar{\Psi}^f_L \bm{m}_f \Psi^f_R \hc
\end{equation}
with
\begin{equation}
\label{eq:mf_generic}
    \bm{m}_f = \left(\begin{array}{c|c}
        \bm{0} & \bm{\mu}^{f_L} \\ \hline \\[-10pt]
        \bm{\mu}^{f_R} & \bm{M}^F
    \end{array}\right)\,,
\end{equation}
where $\bm\mu^{f_L}$ couples the SM left-handed (LH) fermions to the right-handed (RH) components of the VLFs, $\bm\mu^{f_R}$ couples the LH components of the VLFs to the SM RH fermions, and $\bm M^F$ is the VLF mass matrix.
For concreteness, we use partners of the RH fermions; that is, our VLFs have gauge representations corresponding to the SM RH fermions. Hence $\bm{\mu}^{f_L} \equiv \bm{Y}^{f_L} \varphi$, where $\varphi = H(\tilde{H})$ for $d, e \,(u)$ fermions, with $\tilde H = i\sigma_2 H^*$, is a Higgs coupling, while $\bm\mu^{f_R}$ is a mass mixing. The fermion multiplets for $f=u,\, d$ can be written as
\begin{align}
    \Psi_R^f &= \left(\begin{array}{c c c|c c c}
        f_{R,1} & f_{R,2 } & f_{R,3} & F_{R,1} & \cdots & F_{R, N_f}
    \end{array}\right)^T \, , \nonumber\\     \Psi_L^f &= \left(\begin{array}{c c c|c c c}
        f_{L,1} & f_{L,2 } & f_{L,3} & F_{L,1} & \cdots & F_{L, N_f}
    \end{array}\right)^T \, ,
    \label{eq:quark_multiplet}
    \end{align}
while there is a separate chain for each lepton $e_i = e,\mu,\tau$.
Here, $f_{R/L, i}$ correspond to the SM fermions, and $F_{R/L,i}$ are the heavy VLFs. A similar construction can be done by using the representations of the left-handed fermions of the SM as VLF partners, or a combination of both, see \cref{sec:moving_higgs}.

\paragraph*{Mass hierarchies mechanism.} Assuming a hierarchy in the scales $\langle H \rangle, \mu \ll M_F$, the resulting SM Yukawa interactions can easily be written as
\begin{align}
    \mathcal{L} &= -\bar{f}_L\, \bm{y}^f_\textrm{eff}\, f_R \varphi - \bar{q}_{L,3} y_t \tilde{H}\, u_{R,3} \hc \, , \nonumber \\
    \bm{y}^f_\textrm{eff} &= - \bm{Y}^{f_L} \left[\bm{M}^{F}\right]^{-1} \bm{\mu}^{f_R} \, .
    \label{eq:yeff_general}
\end{align}
 To generate the observed fermion mass hierarchies, we can simply take for $\bm M^F$ the form of a \emph{tridiagonal} matrix, that is, a matrix with a diagonal populated by mass entries $M_F$ and one super-diagonal made out of $\mu_F$ and sub-diagonal terms $\mu_{F}^{\ast}$. This texture can be enforced by a flavour symmetry~\cite{Greljo:2025mwj}, see \cref{sec:benchmark_model}. Then the start and the end of the chain are given by the first non-zero term of $\bm{Y}^{f_L}$ and the last non-zero term of $\bm{\mu}^{f_R}$, respectively. We choose the positions of the Yukawa and mixing vectors as $\bm{Y}^{f_L} = \left(\begin{array}{ccc}
    y_f & \cdots & 0
\end{array}\right)$ and $\bm{\mu}^{f_R} = \left(\begin{array}{ccc}
    0 & \cdots & \mu_f
\end{array}\right)^T$. With this form, the effective Yukawa Lagrangian is given by
\begin{align}
     \mathcal{L} &= \bar{f}_{L,i} \left[y_f \left(\bm{M}^{F}\right)^{-1}_{1N} \mu_f \right] f_{R,i} \varphi \nonumber \\  &= y_f (-1)^{N+1}\frac{\mu_f \prod_{k=1}^{N-1} \mu_{F,k}}{\det\bm{M}^F} \bar{f}_{L,i}  f_{R,i} \varphi  \label{eq:yeff_chain} \\ & \simeq y_f (-1)^{N+1}\frac{\mu_f}{M_F^{N-2}}\frac{\prod_{k=1}^{N-1} \mu_{F,k}}{(M_F^2 - \sum_{j=1}^{N-1} |\mu_{F,j}|^2)} \bar{f}_{L,i}  f_{R,i} \varphi \, . \nonumber
\end{align}
The Higgs coupling need not sit at the beginning of the chain, see \cref{sec:moving_higgs}.

\paragraph*{Reading rule.} Expanding $[\bm M^F]^{-1}$ in the off-diagonal mixings, each entry of $\bm y^f_\textrm{eff}$ in \cref{eq:yeff_general} becomes a sum over the paths that connect $f_{L,i}$ to $f_{R,j}$. At leading order only the shortest path contributes, and
\begin{equation}
\label{eq:reading_rule}
    \big[\bm y^f_\textrm{eff}\big]_{ij} \simeq (-1)^{N}\, y_F \prod_{k=1}^{N} \frac{\mu_k}{M_k}\,,
\end{equation}
where $N$ is the number of VLFs on the path, $M_k$ is the mass of the $k$-th one, and $\mu_k$ is the mixing that connects it to the next site away from the Higgs link. For a chain with universal entries this reproduces $|y_\textrm{eff}|\simeq|y_f|\,z^N$ of \cref{eq:yeff_chain}. The hierarchies thus follow from \emph{locality} along the chains~\cite{Arkani-Hamed:2026wwy}: a light fermion reaches the Higgs only through nearest-neighbour hops.

\subsection{The benchmark model}
\label{sec:benchmark_model}

In the benchmark model of Ref.~\cite{Greljo:2025mwj}, the allowed links are fixed by the MFP flavour group $\mathrm{SU}(2)_q\times\mathrm{U}(1)_X$, with the charges listed in \cref{tab:field_content}. The $\mathrm{SU}(2)_q$ factor acts on the two light LH quark doublets, $\bm q=(q_1,q_2)^T$, and on their VLF partners $\bm U$ and $\bm D$. The VLF masses and the Higgs couplings
\begin{equation}
\label{eq:higgs_links}
    \bar\ell_i H E^1_{iR}\,,\quad
    \bar{\bm q}\tilde H \bm U_R\,,\quad
    \bar{\bm q} H \bm D_R\,,\quad
    \bar q_3 H D_{3R}\,,\quad
    \bar q_3 \tilde H u_{3}\,,
\end{equation}
are invariant. As in Ref.~\cite{Greljo:2025mwj}, the symmetry is broken only softly by the mass mixings, each of which carries a single spurion: $z$ for the links along each chain, and the $\mathrm{SU}(2)_q$ doublets $\bm V$, $\bm W$, $\bm U$ (and $\tilde{\bm V}\equiv i\sigma_2\bm V^*$) for the links that attach $\bm U$ and $\bm D$ to the singlet chains.\footnote{Following Ref.~\cite{Greljo:2025mwj}, $\bm U$ denotes both the up-type VLF doublet and a spurion; the former is always a site and the latter a link in \cref{fig:benchmark_chains}.} This enforces the textures discussed below. The VLFs adjacent to a Higgs link, $E^1_i$, $\bm U$, $\bm D$ and $D_3$, define the six VLF classes $\{e,\mu,\tau,U,D,B\}$, with $B\equiv D_3$, of the electroweak analysis in \cref{sec:EW_analysis}. The resulting chains are shown in \cref{fig:benchmark_chains}.

\begin{table*}[t]
\centering
\small
\renewcommand{\arraystretch}{1.15}
\setlength{\tabcolsep}{4pt}
\begin{minipage}[t]{0.32\textwidth}\centering
\begin{tabular}[t]{@{}l c c@{}}
\toprule
SM field & $\mathrm{SU}(2)_q$ & $\mathrm{U}(1)_X$ \\
\midrule
$\bm{q}=(q_1,q_2)^T$ & $\bm{2}$ & $0$ \\
$q_3$ & $\bm{1}$ & $2X_z-X_U$ \\
$u_1$ & $\bm{1}$ & $-X_V-2X_z$ \\
$u_2$ & $\bm{1}$ & $-X_W$ \\
$u_3$ & $\bm{1}$ & $2X_z-X_U$ \\
$d_1$ & $\bm{1}$ & $-X_V-2X_z$ \\
$d_2$ & $\bm{1}$ & $X_V+X_z$ \\
$d_3$ & $\bm{1}$ & $X_z-X_U$ \\
$\ell_i$ & $\bm{1}$ & $X_i$ \\
$e_i$ & $\bm{1}$ & $X_i-(4-i)X_z$ \\
$H$ & $\bm{1}$ & $0$ \\
\bottomrule
\end{tabular}
\end{minipage}\hfill
\begin{minipage}[t]{0.64\textwidth}\centering
\begin{tabular}[t]{@{}l c c c@{}}
\toprule
VLF & $(\mathrm{SU}(3)_c,\mathrm{SU}(2)_L)_{Y}$ & $\mathrm{SU}(2)_q$ & $\mathrm{U}(1)_X$ \\
\midrule
$E_i^k$, $k\leq 4-i$ & $(\bm{1},\bm{1})_{-1}$ & $\bm{1}$ & $X_i-(k-1)X_z$ \\
$\bm{U}$ & $(\bm{3},\bm{1})_{2/3}$ & $\bm{2}$ & $0$ \\
$U_1^k$, $k=1,2$ & $(\bm{3},\bm{1})_{2/3}$ & $\bm{1}$ & $-X_V-(k-1)X_z$ \\
$\bm{D}$ & $(\bm{3},\bm{1})_{-1/3}$ & $\bm{2}$ & $0$ \\
$D_1^k$, $k=1,2$ & $(\bm{3},\bm{1})_{-1/3}$ & $\bm{1}$ & $-X_V-(k-1)X_z$ \\
$D_2^1$ & $(\bm{3},\bm{1})_{-1/3}$ & $\bm{1}$ & $X_V$ \\
$D_3^1$ & $(\bm{3},\bm{1})_{-1/3}$ & $\bm{1}$ & $-X_U$ \\
$D_3$ & $(\bm{3},\bm{1})_{-1/3}$ & $\bm{1}$ & $2X_z-X_U$ \\
\bottomrule
\end{tabular}

\vspace{8pt}
\begin{tabular}[t]{@{}c c c c@{}}
\toprule
Spurion & $\mathrm{SU}(2)_q$ & $\mathrm{U}(1)_X$ & Size \\
\midrule
$z$ & $\bm{1}$ & $X_z$ & $0.02$ \\
$\bm{V}$ & $\bm{2}$ & $X_V$ & $0.015$ \\
$\bm{W}$ & $\bm{2}$ & $X_W$ & $0.006$ \\
$\bm{U}$ & $\bm{2}$ & $X_U$ & $0.04$ \\
\bottomrule
\end{tabular}
\end{minipage}
\caption{Field content of the benchmark model of Ref.~\cite{Greljo:2025mwj} and charges under the MFP flavour group $\mathrm{SU}(2)_q\times\mathrm{U}(1)_X$: SM fields (left), VLFs (top right) and spurions (bottom right). The VLFs are Dirac fermions with the gauge quantum numbers of SM RH fermions; $k$ labels their position along a chain. The VLF masses and the Higgs couplings of \cref{eq:higgs_links} are invariant, and each mixing link of \cref{fig:benchmark_chains} carries one spurion: moving away from the Higgs link, a link with spurion $x$ lowers the charge by $X_x$. An example of a viable choice is $\{X_V,X_W,X_U,X_z\}=\{1,-2,-3,7\}$ and $\{X_1,X_2,X_3\}=\{-2,-3,-1\}$~\cite{Greljo:2025mwj}. The spurion sizes are the moduli of the benchmark values in \cref{eq:app_spurion_benchmark}.}
\label{tab:field_content}
\end{table*}

Three separate chains in the charged lepton sector forbid flavour violation by construction. However, the CKM mixing matrix requires intergenerational mixing in the quark sector; consequently, the generation-specific chains must be coupled. Neutrino masses and mixing are discussed in Ref.~\cite{Greljo:2025mwj} and are not relevant to this work.

\paragraph*{Charged leptons.} This structure suffices to explain the hierarchies of the charged leptons in the SM. Choosing all $\mu_F\sim \mu_f\sim\mu_z\ll M_F$, we can assume that the lepton hierarchies are generated by the minimal number of VLFs, that is, $N_\tau=1,\,N_\mu=2,\,N_e=3,$ see \cref{fig:benchmark_chains}(a),
\begin{equation}
    \mathcal{L}_\textrm{lept.} = -y_{\ell_i} z^{N_i} \bar{\ell}_{L,i} e_{R,i} H \hc \, , \, \textrm{with } z = \frac{\mu_z}{M_E},
\end{equation}
up to signs that can be absorbed in the order-one couplings $y_{\ell_i}$, with the hierarchies reproduced by $z\simeq 0.02$.

\paragraph*{Up quarks.} To generate the CKM mixing matrix along with the quark mass hierarchies, we must modify the previous basic structure. While the CKM matrix can be implemented in several ways, we follow the example of Ref.~\cite{Greljo:2025mwj}. This specific formulation guarantees flavour protection, even in the presence of generic TeV-scale new physics described by the SMEFT. In the up-quark sector, we accommodate the top Yukawa at the renormalisable level ($y_t \sim 1$) and introduce fermion chains for the two lighter generations. Generating Cabibbo mixing subsequently requires linking these chains. Constructing the chain with the fewest VLFs ($N_U = 3$, counting the doublet $\bm U$ once), we can choose $\bm{\mu}^{u}_{2} = \left(\begin{array}{ccc}\bm{\mu}_W & 0 & 0
\end{array} \right)^T$ and $\bm{\mu}^{u}_{1} = \left(\begin{array}{ccc} 0 & 0 & \mu_z
\end{array} \right)^T$ as the endpoints of the chain, in the basis $(\bm U, U_1^1, U_1^2)$, and the corresponding Yukawas $\bm{Y}^{q_u} = \left(\begin{array}{ccc} y_{q}^{u}\,\bm{1}_{2\times2} & 0 & 0 \end{array}\right)$.
In addition, there is also $\bm\mu_V$, which represents the mixing of the VLF doublet with the neighbouring VLF singlet in the chain. $\bm\mu_W$ ends the $u_{R,2}$ chain while $\bm\mu_V$ connects the chain ending on $u_{R,1}$. These chains are shown in \cref{fig:benchmark_chains}(b). The resulting Yukawa Lagrangian is then given as
\begin{equation}
\label{eq:yuk_u}
    \mathcal{L}_\textrm{up} =  -\bar{q}_{L,i} \bm{y}^u_{ij} u_{R,j} \tilde{H} \hc \, , \, \textrm{with } \bm{y}^u = -\begin{pmatrix}
        \hat{y}^u_1\, \bm{V} z^2 & \hat{y}^u_2\,\bm{W}
    \end{pmatrix} \, ,
\end{equation}
where $i,j=1,2$, $\bm{W} (\bm{V}) = \bm{\mu}_{\bm W (\bm V)}/M_{\bm U}$, $z = \mu_z/M_{\bm U}$ and the $\mathcal{O}(1)$ factors $\hat{y}^u_i$ denote the low-energy factors which are functions of the UV couplings. Both are related to the UV parameter $y^u_q$ but can include differences in the universality of the VLF masses and mixings.

The up and charm mass hierarchies are nicely explained by small $v, w \sim z$.  Without loss of generality, we choose the basis $\bm{V} = v\,\begin{pmatrix}
    1 & 0
\end{pmatrix}$ and $\bm{W} = w\,\begin{pmatrix}
     \sin\alpha   & \cos\alpha
\end{pmatrix}$, with a mixing $\alpha \approx -\lambda$ reproducing the Cabibbo angle~\cite{Greljo:2025mwj}.

\paragraph*{Down quarks.} The minimal Yukawa structure we aim for is
\begin{equation}
\label{eq:yuk_d}
    y^d_\textrm{eff} = \begin{pmatrix}
    - \hat{y}^d_1 \bm{V}z^2& \hat{y}_2^d z^*\tilde{\bm{V}} &  \hat{y}_q^d z^*\bm{U}
    \\
       0&0 & -z \hat{y}^d_3
    \end{pmatrix} \, ,
\end{equation}
where $\tilde{\bm V}\equiv i\sigma_2\bm V^*$ and we have kept only the leading order in the mixings. Note that this setup predicts alignment in the $1\!-\!2$ down-quark sector, motivated by strong bounds from kaons. As a result, the Cabibbo angle arises entirely in the up sector. Instead, the down sector generates mixing between the light generations and the third.

These entries follow from \cref{fig:benchmark_chains}(c). The light doublet $\bm q$ couples through the Higgs to the VLF doublet $\bm D$, which branches through the spurions $\bm V$, $\tilde{\bm V}$ and $\bm U$ into three short singlet chains ending on $d_1$, $d_2$ and $d_3$; by \cref{eq:reading_rule}, they give the first row of \cref{eq:yuk_d}, where each entry is a two-component vector in $\mathrm{SU}(2)_q$ space. The doublet $q_3$ has its own chain through $D_3$, which ends on $d_3$ as well, while no path connects $q_3$ to $d_{1,2}$. Hence, the mixing of the light families with the third one is controlled by $\bm U$ alone. Since $\bm D$ mixes with three VLFs, $\bm M_{\bm D}$ is no longer tridiagonal, but each mixing still carries a single spurion. The explicit matrices are given in \cref{eq:app_mass_matrices}.\footnote{One can impose exchange parity of the VLFs, demanding that the off-diagonal components be related by complex conjugation $\mu_i \to \mu_i^\ast$. Generally, these entries need not be the same. Such an assumption would impact CP violation at higher order in the spurion expansion. Phenomenologically, the diagonal Wilson coefficients playing into EWPOs and leading order Yukawa matrices are unaffected by the exchange parity assumption. For flavour observables it is relevant since the off-diagonal Wilson coefficients emerge at second spurion order and depend on the number of physical phases. In the following, we will assume exchange parity to hold.} The three branches share the Higgs coupling $y_q^d$, but their VLF masses and mixings may differ, which is needed to reproduce $m_d/m_s$, see \cref{eq:MFPcoeffs}. Analogous to the up sector, the $\hat{y}^d_{1,2,q}$ are proportional to the UV coupling $y^d_q$ and $\hat{y}^d_3$ is proportional to $y^d_{q_3}$.

In this way, we can associate $\bm{U} = u\begin{pmatrix}
    \sin\beta e^{i\phi} &\cos\beta
\end{pmatrix} $ with the mixing of the light families with the third one. Indeed, in the basis where $\bm V\propto(1,0)$ and $\tilde{\bm V}\propto(0,1)$, the first two families of \cref{eq:yuk_d} are diagonal and the third column is controlled by $\bm U$, which provides the CP-violating phase $\phi$ of the CKM matrix. The angle $\tan \beta \; e^{-i \phi} = (U_1/U_2)^* \sim V_{td}/V_{ts}$ is determined by the third row of the CKM matrix.
Together with the Cabibbo angle generated in the up sector, this reproduces the CKM matrix; the explicit computation using perturbative diagonalisation is presented in \cref{app:pert_diag}.
Counting phases in the UV model confirms that, with the exchange parity of the heavy mass matrices and the spurion identifications above, $\phi$ is the only physical phase. This number is independent of the universality of the VLF masses and of the $z$-couplings. Hence, all other unphysical signs in the low-energy Yukawa matrices can be rotated away by a field redefinition of the right-handed fields and the $q_3$, or equivalently by embedding minus signs into the UV mass matrices. Thus we can for example make all low-energy Yukawa parameters positive without changing the Wilson coefficients $C^{(1,3)}_{Hq}$ that are generated at tree level, see \cref{sec:matching_smeft}.

As mentioned before, these models are non-unique; for instance, we can add longer chains, making the suppression between families larger or smaller. For example, we could double the chain for each fermion generation by replacing $z^n \to z^{2n}$, allowing us to use larger values for the parameters $z, \, u$, etc.

\subsection{Moving the Higgs coupling along the chain}
\label{sec:moving_higgs}

\begin{figure*}[t]
    \centering    \includegraphics[width=\linewidth]{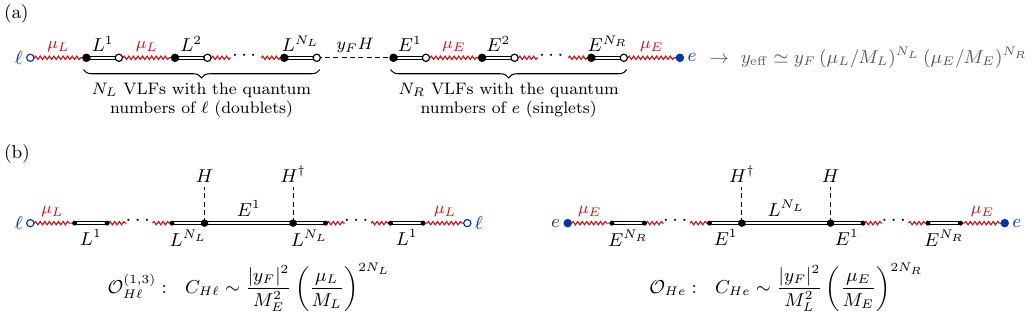}
    \caption{(a) Lepton chain with $N_L$ LH VLFs $L^k$ before the Higgs link and $N_R$ RH VLFs $E^k$ after it; the benchmark of \cref{fig:benchmark_chains} has $N_L=0$. The Yukawa coupling depends only on $N_L+N_R$, \cref{eq:yeff_moved}. (b) The same chain folded back onto a single SM fermion generates the operators that modify the $Z$ and $W$ couplings, \cref{eq:CHl_e,eq:CHe}: for $\ell$ the RH VLF $E^1$, and for $e$ the LH VLF $L^{N_L}$, is exchanged between the two Higgs insertions. For $N_L=0$, $\mathcal O_{He}$ is not generated at tree level. Conventions as in \cref{fig:benchmark_chains}.}
    \label{fig:chains_figure}
\end{figure*}

In the benchmark model presented above, all VLFs have the quantum numbers of SM RH fermions, and the Higgs link sits at the beginning of each chain. In general, we refer to VLFs with the quantum numbers of the SM LH (RH) fermions, i.e.~$\mathrm{SU}(2)_L$ doublets (singlets), as LH (RH) VLFs; each is a Dirac fermion with both chiralities. Consider a lepton chain with $N_L$ LH VLFs $L^k$ and $N_R$ RH VLFs $E^k$, as in \cref{fig:chains_figure}(a): $\ell$ enters the chain through a mass mixing with $L^1$, reaches the Higgs link $y_F\,\bar L^{N_L}_L H E^1_R$, and exits through the RH VLFs to $e$. By \cref{eq:reading_rule},
\begin{equation}
\label{eq:yeff_moved}
    y^e_\textrm{eff}\simeq y_F\left(\frac{\mu_L}{M_L}\right)^{N_L}\left(\frac{\mu_E}{M_E}\right)^{N_R}\sim y_F\, z^{N_L+N_R}\,,
\end{equation}
so that the hierarchy depends only on the total length of the chain, not on the position of the Higgs link. The benchmark corresponds to $N_L=0$; quark chains follow from $(\ell,L,E,e)\to(q,Q,D,d)$ or $(q,Q,U,u)$.

However, the position of the Higgs link does matter for the operators that modify the $Z$ and $W$ couplings. These are generated by chains that start and end on the \emph{same} SM fermion, \cref{fig:chains_figure}(b), and are suppressed only by the mixings on one side of the Higgs link: as $(\mu_L/M_L)^{2N_L}$ for the LH currents and $(\mu_E/M_E)^{2N_R}$ for the RH ones, see \cref{sec:matching_smeft}. For $N_L,N_R\geq1$, the product $C_{H\ell}\,C_{He}\sim|y_F|^2|y^e_\textrm{eff}|^2/M_F^4$ is therefore independent of the position of the Higgs link: the electroweak effects can be shifted between the two chiralities, but not removed. In the benchmark model the LH currents are thus unsuppressed, and they are the operators probed by the EWPO in \cref{sec:EW_analysis}. Longer LH chains suppress them, at the price of larger mixings, as explored in \cref{sec:general_lessons}.

\subsection{Matching onto the SMEFT}
\label{sec:matching_smeft}

At tree level, integrating out the VLFs generates dimension-six operators involving two SM fermions of the same chirality and two Higgs fields~\cite{deBlas:2017xtg}. Diagrammatically, they arise from chains that start and end on the same SM fermion: the chain is entered through the coupling of the SM fermion to the VLFs, reaches the site where the Higgs couples, and is closed back onto the initial fermion through a second Higgs insertion, as shown in \cref{fig:chains_figure}(b). The operators generated are those which modify the $Z$- and $W$-currents, namely
\begin{align}
[\mathcal O_{Hq}^{(1)}]_{ij}
&=
\left(H^\dagger i\overleftrightarrow D_\mu H\right)
\left(\bar q_i\gamma^\mu q_j\right),\nonumber\\
[\mathcal O_{Hq}^{(3)}]_{ij}
&=
\left(H^\dagger i\overleftrightarrow D_\mu^{\,I}H\right)
\left(\bar q_i\tau^I\gamma^\mu q_j\right),
\nonumber\\
[\mathcal O_{Hu}]_{ij}&=
\left(H^\dagger i\overleftrightarrow D_\mu H\right)
\left(\bar u_i\gamma^\mu u_j\right),\nonumber\\
[\mathcal O_{Hd}]_{ij}
&=
\left(H^\dagger i\overleftrightarrow D_\mu H\right)
\left(\bar d_i\gamma^\mu d_j\right),
\end{align}
with
\begin{equation}
    H^\dagger i\overleftrightarrow D_\mu^{\,(I)}H
\equiv
iH^\dagger(\tau^I) D_\mu H
-i(D_\mu H)^\dagger(\tau^I) H .
\end{equation}
For leptons the operators follow the analogous form,
\begin{align}
[\mathcal O_{H\ell}^{(1)}]_{ij}
&=
\left(H^\dagger i\overleftrightarrow D_\mu H\right)
\left(\bar \ell_i\gamma^\mu \ell_j\right),
\nonumber\\
[\mathcal O_{H\ell}^{(3)}]_{ij}
&=
\left(H^\dagger i\overleftrightarrow D_\mu^{\,I}H\right)
\left(\bar \ell_i\tau^I\gamma^\mu \ell_j\right),
\nonumber\\
[\mathcal O_{He}]_{ij}
&=
\left(H^\dagger i\overleftrightarrow D_\mu H\right)
\left(\bar e_i\gamma^\mu e_j\right).
\end{align}

Consider now the possibility of having LH and RH VLFs. Then, we have to split the different matrices into LH and RH components. Due to the gauge symmetry group we can formally rewrite the matrices living in different subspaces $\bm{\mu}^{f_R} \to \bm{\mu}^{f_R}\oplus \bm{Y}^{f_R}\varphi$ and $\bm{\mu}^{f_L} \to \bm{Y}^{f_L}\varphi\oplus \bm{\mu}^{f_L}$, where $\bm Y^{f_R}$ denotes the Higgs coupling of the SM RH fermions to the LH VLFs and $\bm\mu^{f_L}$ the mass mixing of the SM LH fermions with the LH VLFs, and the mass matrix
\begin{equation}
    \bm{M}_F=\begin{pmatrix}
        \bm{M}_{F_R} & \tilde{\bm{Y}}_F \varphi^\dagger\\ \bm{Y}_F \varphi & \bm{M}_{F_L}
    \end{pmatrix} \, ,
\end{equation}
where $\bm Y_F$ and $\tilde{\bm Y}_F$ are the Higgs couplings connecting the LH and RH VLF chains.

In this form, the Wilson coefficients of these operators are easily obtained by integrating out the chains as
\begin{widetext}
\begin{align}
\label{eq:CHq_d}
[C_{D}]_{ij}\equiv[C_{Hq}^{(1)}]_{ij}
+[C_{Hq}^{(3)}]_{ij}&=-\frac{1}{2}\left[
\bm{Y}^{q_d}
\bm{M}_{D}^{-1}
\bm{M}_{D}^{\dagger\,-1}
\bm{Y}^{q_d\,\dagger}
+
\bm{\mu}^{q_d}
\bm{M}_{Q}^{-1}
\bm{Y}_{D}
\bm{M}_{D}^{-1}
\bm{M}_{D}^{\dagger\,-1}
\bm{Y}_{D}^{\dagger}
\bm{M}_{Q}^{\dagger\,-1}
\bm{\mu}^{q_d\,\dagger}
\right]_{ij},
\\
\label{eq:CHq_u}
[C_{U}]_{ij}\equiv[C_{Hq}^{(1)}]_{ij}
-[C_{Hq}^{(3)}]_{ij}&=
\frac{1}{2}
\left[
\bm{Y}^{q_u}
\bm{M}_{U}^{-1}
\bm{M}_{U}^{\dagger\,-1}
\bm{Y}^{q_u\,\dagger}
+
\bm{\mu}^{q_u}
\bm{M}_{Q}^{-1}
\bm{Y}_{U}
\bm{M}_{U}^{-1}
\bm{M}_{U}^{\dagger\,-1}
\bm{Y}_{U}^{\dagger}
\bm{M}_{Q}^{\dagger\,-1}
\bm{\mu}^{q_u\,\dagger}
\right]_{ij},
\\
\label{eq:CHd}
[C_{Hd}]_{ij}
&=
\frac{1}{2}
\left[
\bm{Y}^{d\,\dagger}
\bm{M}_{Q}^{\dagger\,-1}
\bm{M}_{Q}^{-1}
\bm{Y}^{d}
+
\bm{\mu}^{d\,\dagger}
\bm{M}_{D}^{\dagger\,-1}
\bm{Y}_{D}^{\dagger}
\bm{M}_{Q}^{\dagger\,-1}
\bm{M}_{Q}^{-1}
\bm{Y}_{D}
\bm{M}_{D}^{-1}
\bm{\mu}^{d}
\right]_{ij},
\\
\label{eq:CHu}
[C_{Hu}]_{ij}
&=
-\frac{1}{2}
\left[
\bm{Y}^{u\,\dagger}
\bm{M}_{Q}^{\dagger\,-1}
\bm{M}_{Q}^{-1}
\bm{Y}^{u}
+
\bm{\mu}^{u\,\dagger}
\bm{M}_{U}^{\dagger\,-1}
\bm{Y}_{U}^{\dagger}
\bm{M}_{Q}^{\dagger\,-1}
\bm{M}_{Q}^{-1}
\bm{Y}_{U}
\bm{M}_{U}^{-1}
\bm{\mu}^{u}
\right]_{ij}, \\
\label{eq:CHl_e}
[C_{E}]_{ij}\equiv[C_{H\ell}^{(1)}]_{ij}
+
[C_{H\ell}^{(3)}]_{ij}
&=
-\frac{1}{2}
\left[
\bm{Y}^{\ell_e}
\bm{M}_{E}^{-1}
\bm{M}_{E}^{\dagger\,-1}
\bm{Y}^{\ell_e\,\dagger}
+
\bm{\mu}^{\ell_e}
\bm{M}_{L}^{-1}
\bm{Y}_{E}
\bm{M}_{E}^{-1}
\bm{M}_{E}^{\dagger\,-1}
\bm{Y}_{E}^{\dagger}
\bm{M}_{L}^{\dagger\,-1}
\bm{\mu}^{\ell_e\,\dagger}
\right]_{ij},
\\
\label{eq:CHe}
[C_{He}]_{ij}
&=
\frac{1}{2}
\left[
\bm{Y}^{e\,\dagger}
\bm{M}_{L}^{\dagger\,-1}
\bm{M}_{L}^{-1}
\bm{Y}^{e}
+
\bm{\mu}^{e\,\dagger}
\bm{M}_{E}^{\dagger\,-1}
\bm{Y}_{E}^{\dagger}
\bm{M}_{L}^{\dagger\,-1}
\bm{M}_{L}^{-1}
\bm{Y}_{E}
\bm{M}_{E}^{-1}
\bm{\mu}^{e}
\right]_{ij}.
\end{align}
\end{widetext}
If the SM fermion, either left- or right-handed, has a coupling to the Higgs, the operator is generated by just a single insertion of a VLF between the Higgs couplings (this corresponds to $N_L=0$ ($N_R=0$) for a LH (RH) SM fermion in \cref{fig:chains_figure}(b)). This is the first contribution of each of the previous expressions. However, when such a term is not present, the SM fermion first enters, through its mass mixing, the chain of VLFs carrying its own gauge quantum numbers, and propagates along it until it reaches the site with a non-vanishing Higgs coupling $\bm Y_F$ to the chain of opposite chirality. The two Higgs fields of the operator are attached at that site, connected through the propagator of the opposite-chirality VLF, and the same chain is then retraced back to the external fermion, so that each mixing insertion appears twice. This is the second contribution of each of the previous equations. If all the terms are of the same order, such an operator will contribute as
\begin{align}
    C_{H f_L} &\sim \frac{|y_{F}|^2}{M_{F_R}^2} \left(\frac{\mu_{f_L}}{M_{F_L}}\right)^{2 N_{F_L}} , \nonumber\\
    C_{H f_R} &\sim \frac{|y_{F}|^2}{M_{F_L}^2} \left(\frac{\mu_{f_R}}{M_{F_R}}\right)^{2N_{F_R}} \, .
    \label{eq:C_power_counting}
\end{align}
Here $N_{F_L}\equiv N_L$ and $N_{F_R}\equiv N_R$ of \cref{fig:chains_figure}.

In each of \cref{eq:CHq_d,eq:CHq_u,eq:CHd,eq:CHu,eq:CHl_e,eq:CHe}, the two contributions are never simultaneously relevant: whenever the SM fermion has a direct Higgs coupling to a VLF, the mixing-induced term is suppressed by additional powers of $\mu/M$ and can be neglected, while in the absence of a direct coupling only the mixing-induced term survives. In the benchmark model of \cref{fig:benchmark_chains}, which contains only RH VLFs, only the first (direct Higgs-coupling) terms of $C_D$, $C_U$ and $C_E$ are generated, while $C_{Hd}$, $C_{Hu}$ and $C_{He}$ vanish at tree level: the RH SM fermions have no heavy LH VLF to exchange between the Higgs insertions, see \cref{fig:chains_figure}(b) with $N_L=0$.

Finally, modified Yukawa operators $\mathcal{O}_{fH}= (H^\dagger H)\,\bar{f}_L f_R H$ are also generated. As shown in \cref{app:integrating_out}, these operators carry the same structure as the effective Yukawa, $C_{fH}\sim C_{Hf_L}\, y_\textrm{eff}^f + y_\textrm{eff}^f\, C_{Hf_R} $, and are inherently suppressed. Current LHC tests of the Higgs couplings, and even future high-luminosity runs, will not be capable of testing modifications of the Yukawas with these suppressions. Also at FCC-ee, EWPO precision will yield stronger constraints than modified Yukawas for these models.

\subsection{Flavour bounds and direct searches}
\label{sec:flavour_direct}

Regarding the flavour sector, the model is built to respect the MFP symmetry group at low energies, protecting it from generating large FCNCs. In this section, we review the leading flavour bounds that nevertheless constrain this particular implementation. For flavour observables, it is more transparent to work with the basis $\{C_D, C_U\}$ as defined in \cref{eq:CHq_d,eq:CHq_u} corresponding to the modifications of the $Z$ current of the down and up quarks, respectively. These bounds depend on how the CKM matrix is distributed between the up and down sectors, which MFP does not fix. For instance, generating the full CKM matrix in the up sector, would keep $C_D$ diagonal in the down mass basis and remove the tree-level $Z$-mediated $B_s\to\mu\mu$ and $K_L\to\mu\mu$ bounds discussed below, whereas the EWPO bounds of \cref{sec:EW_analysis} would be unchanged. The coefficients are diagonal at leading order in the spurion expansion with the $\mathrm{SU}(2)_q$-doublet block proportional to the identity. After matching, higher-order terms in the spurion expansion lead to off-diagonal coefficients and corrections to the diagonal terms,
\begin{equation}
\label{eq:WC_off-diagonals}
    \begin{aligned}
    [C_{U}]_{ab} &= \frac{c_{U}}{2}   \left[ \delta_{ab} +  \kappa_{1} V_{a}V^{\dagger}_{b} - \kappa_{2}W_{a}W^{\dagger}_{b} \right], \\
[C_{D}]_{ab} &= -\frac{c_{D}}{2}    \left[ \delta_{ab} +  \kappa_{3} V_{a}V^{\dagger}_{b} + \kappa_{4}\tilde{V}_{a}\tilde{V}^{\dagger}_{b} + \kappa_{5} U_{a} U^{\dagger}_{b} \right], \\
[C_{D}]_{33} &= -\frac{c_{B}}{2}    \left[ 1 - \kappa_{z} |z|^2 \right],
\end{aligned}
\end{equation}
for $a,b \in \{1,2\}$, with $c_U=|y_q^u|^2/M_U^2$, $c_D=|y_q^d|^2/M_D^2$ and $c_B=|y_{q_3}^d|^2/M_B^2$, where $M_B$ is the mass of the VLF $D_3$ coupled to $q_3$.
These terms arise from the same diagrams generating the leading-order coefficients, cf.~\cref{fig:chains_figure}(b), by inserting two additional spurions between the Higgs couplings. The $\kappa$ factors account for non-universality of the VLF masses and relations between the spurions in the up and down sector (see \cref{app:flavour_obs} for details). For instance, taking all VLF masses equal within the same chain and the $\bm{V}$ spurion universal across the up and down sector, we obtain $\kappa_{1,3,4,5} = 3$ and $\kappa_{2,z} = 2$.\footnote{If the masses in a chain are different, or the spurions between the up and down sectors differ by order-one factors, these constants are still close to $\kappa_{1,3,4,5} \sim 3$ and $\kappa_{2,z}\sim 2$ for comparable masses; e.g.~a heavy neighbour of mass $M_{F'}$ (spurion $\mu/M_{F'}$) gives $\kappa=1+2M_{F'}/M_F$, see \cref{app:flavour_obs}.} We will use these assumptions in the following analysis, noting that results do not qualitatively change.

However, these are not all the contributions: when going to low-energy couplings, we need to rotate the coefficients into the quark mass basis. This leads to corrections from the diagonal, which are rotated into the off-diagonal terms, with the same spurion counting as matching contributions. In these rotations, coefficients may pick up different complex phase contributions; however, the structure of the rotation matrices enforces each off-diagonal term after the rotation to have a single phase.
\begin{figure*}
    \centering
    \includegraphics[width=\linewidth]{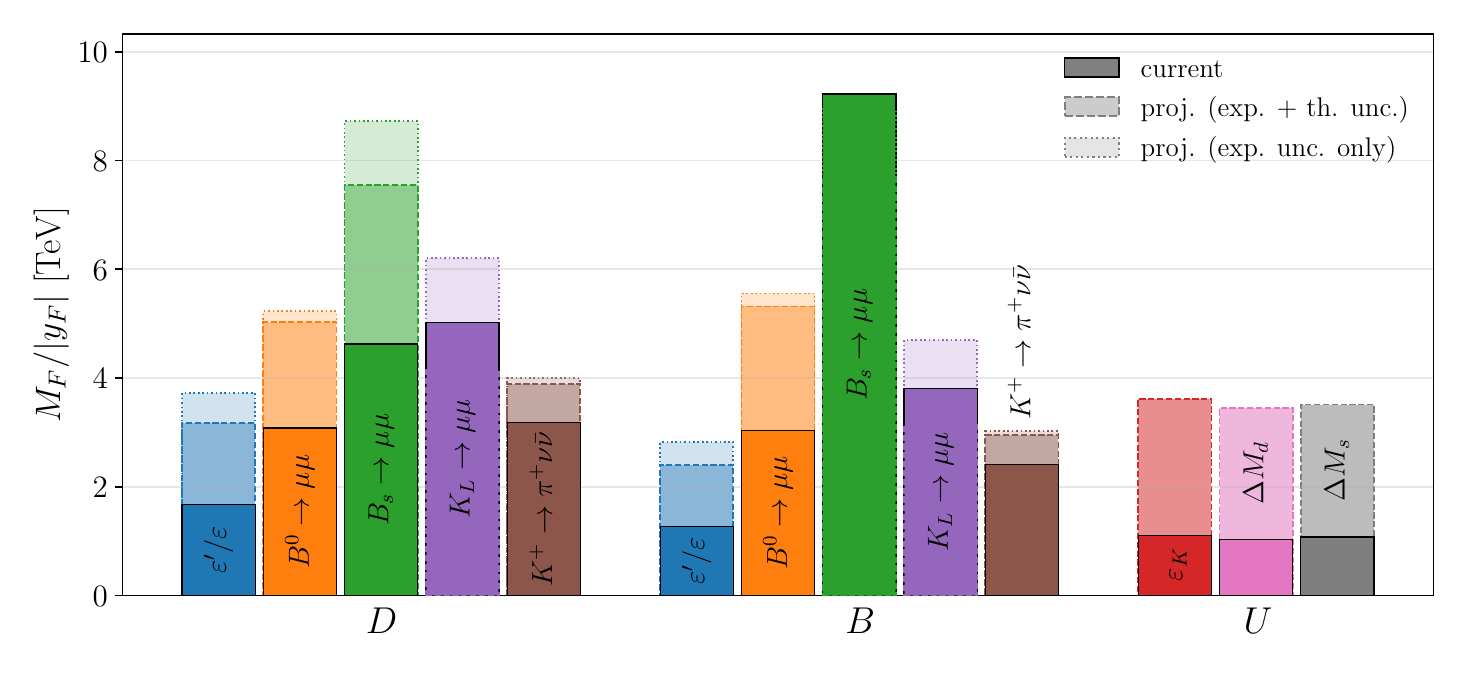}
    \caption{Current bounds and future projections on $M_F/|y_F|$ from selected observables at the scale $\mu_0 = 2\;$TeV assuming the spurion benchmark given in \cref{eq:app_spurion_benchmark}. Future projections are assumed to be SM-centred and their uncertainties reflect either improvements in experiment or theory.}
    \label{fig:flavour_bounds_projections}
\end{figure*}

Thus, the phases can be fixed up to a convenient rephasing choice, which we use to map the CKM matrix into the standard form. Hence, the phase of each Wilson coefficient is determined, and, for the observables shown in \cref{fig:flavour_bounds_projections}, we can determine the 95\% CL bounds from the one-sided chi-squared. The solid bars show the current bounds, while lighter shades show future projections based on different assumptions about theoretical and experimental improvements (see \cref{tab:flavour_refs_current,tab:flavour_refs_projections}). To obtain the bounds, we use the benchmark spurion values from Ref.~\cite{Greljo:2025mwj} in \cref{eq:app_spurion_benchmark}.

For the down-type VLFs, the leading constraints come from $\Delta F=1$ processes, $B_{s}\to\mu \mu$ and $K_{L}\to\mu \mu$, which arise from the modified $Z$ interaction in the low-energy EFT (LEFT)~\cite{Bobeth:2016llm}. The leading bound, from $\mathrm{BR}(B_{s}\to\mu \mu)$, is obtained by using the LHCb combined run 1+2+3 fit values~\cite{LHCb:2026oio}, leading to a constraint on $M_F/|y_F|\gtrsim 9\,$TeV for a single down-type class (the $Z$-mediated amplitude is proportional to $c_B-c_D$, see \cref{app:flavour_obs}), exceeding the current EWPO bounds from LEP/SLD derived in the next section.

On top of the $\Delta F = 1$ observables, there are also contributions to $\Delta F = 2$ meson mixing observables from RG running of the $C^{(1,3)}_{Hq}$ operators to the four-quark operators $C^{(1,3)}_{qq}$, see Refs.~\cite{Jenkins:2013zja,Jenkins:2013wua, Alonso:2013hga}. These effects
are shown for the $\epsilon_K$, $\Delta M_s$ and $\Delta M_d$ bars in \cref{fig:flavour_bounds_projections}, which, albeit generated at one loop, lead to the most constraining bounds for the $U$-type VLF, $M_U/|y_U|\gtrsim 1$--$2\,$TeV. In addition, these observables receive contributions from one-loop matching from the UV to SMEFT and from dimension-eight terms corresponding to double $Z\bar{d}d$ insertions, which are not shown in \cref{fig:flavour_bounds_projections}, but are discussed in \cref{app:flavour_obs}. One-loop contributions from SMEFT to LEFT matching are not considered in this analysis.
We also studied $D^0$-mixing observables, which, however, gave weaker bounds than the running effects mentioned above.

In this model, the lepton sector exhibits three disconnected chains for the three lepton generations so that lepton-flavour-violating (LFV) transitions are not generated. Additionally, the lepton chains contain no closed paths (given the exchange parity of \cref{app:integrating_out}), so we can redefine away all CP-violating phases; hence, electric dipole moments do not occur.

By the time of FCC-ee, the high-luminosity LHC and Belle~II will have collected most of their data, and future projections will yield the improvements shown in \cref{fig:flavour_bounds_projections}. However, most of these observables still have large uncertainties coming from hadronic uncertainties and errors in the determination of the CKM elements, see Ref.~\cite{deBlas:2025gyz} for related discussions. In particular, for the SM-centred projections we have used for $\textrm{BR}(B_s \to \mu\mu)$ and $\textrm{BR}(B^0 \to \mu\mu)$ the sensitivities of LHCb~\cite{ATLAS:2025lrr}\footnote{The new-physics contribution to $\mathrm{BR}(B_s\to\mu^+\mu^-)$ has a fixed sign, since $c_B\geq 0$, and the current measurement~\cite{LHCb:2026oio} lies on the opposite side of the SM prediction~\cite{Beneke:2019slt}. As a result, the present bound on $c_B$ is considerably stronger than it would be for an SM-centred measurement, and assuming an SM-centred future measurement leads to a weaker bound than today's, despite the improved precision.}, while for $\epsilon^\prime/\epsilon$ we keep today's experimental uncertainty and assume a relative theory uncertainty of 10\% in $\epsilon^\prime$ from lattice QCD~\cite{Blum:2022wsz}. The $\textrm{BR}(K_L\to\mu\mu)$ prediction takes into account only theory improvements in parametric uncertainties of the CKM parameters $|V_{cb}|$ and $\beta$ by LHCb and Belle II \cite{ATLAS:2025lrr} as well as the assumption that the hadronic uncertainties should be able to match the experimental ones in the future \cite{Hoferichter:2023wiy}.
For $\textrm{BR}(K^+\to\pi^+\nu\nu)$, we use the targeted experimental uncertainty of $15\%$ after the full data collection of the NA62 experiment \cite{Aebischer:2025mwl, Anzivino:2023bhp} rescaled to the newer measured central value~\cite{NA62:2026rwr}.
For the meson-mixing observables, we rescale the current bounds by assuming improved determination of the CKM parameters, as in Ref.~\cite{deBlas:2025gyz}.
The bounds show that, even with the future improvements, the main constraint for this model, protected by the MFP symmetry, will be the EWPO bounds from FCC-ee, as discussed in \cref{sec:tera-Z_projections}.

Finally, VLFs can also be constrained by direct collider searches. For VLFs in SM-like gauge representations, these searches target both pair and single production, with rates determined by their gauge quantum numbers and couplings to SM particles. Pair production is dominated by QCD interactions and therefore provides the most robust mass limits, largely independently of the mixing with SM fermions. Current ATLAS and CMS searches typically exclude singlet and doublet top and bottom partners with masses below $1.2$--$1.5\,$TeV, with limits reaching approximately $1.6\,$TeV in favourable scenarios involving degenerate multiplets~\cite{ATLAS:2022hnn,CMS:2022fck,CMS:2024xbc,ATLAS:2024zlo,ATLAS:2024gyc,ATLAS:2022tla}; see Ref.~\cite{Benbrik:2024fku} for a recent overview. Single-production searches are instead directly sensitive to the electroweak mixing and can extend the mass reach well beyond that of pair production when the couplings are sizeable. For vector-like top partners, single-production searches exclude masses up to around $2.3\,$TeV for sizeable couplings, while doublet scenarios are typically constrained up to $1.7\,$TeV~\cite{ATLAS:2023pja,ATLAS:2023bfh,ATLAS:2022ozf,CMS:2024qdd}. Comparable searches for bottom partners probe masses in the $1$--$2\,$TeV region~\cite{ATLAS:2023ixh}. If the VLF is far down the chain, its decays can become macroscopically displaced, weakening prompt-search limits but opening sensitivity to long-lived particle searches~\cite{Arkani-Hamed:2026wwy}.

Vector-like lepton doublets are also constrained by electroweak pair-production searches. ATLAS searches exclude mass-degenerate doublets coupling to electrons or muons below $1.22\,\mathrm{TeV}$ and $1.27\,\mathrm{TeV}$ respectively~\cite{ATLAS:2024mrr}. For doublets coupling to the third generation, ATLAS and CMS exclude vector-like $\tau$ leptons below roughly $1\,\mathrm{TeV}$~\cite{ATLAS:2023sbu,CMS:2022nty}. However, the singlet representations, due to fewer production channels and relatively larger decays into missing energy channels, have weaker bounds: $320 \,$GeV for partners coupling to the first and $400\,$GeV for those coupling to the second generation~\cite{ATLAS:2024mrr}. For the third family, the bounds are even weaker: singlet vector-like $\tau$ leptons are only excluded in the narrow window $125$--$150\,$GeV~\cite{CMS:2022nty}.
In the minimal model considered here, the VLFs carry the quantum numbers of the RH SM fermions, i.e.~they are $\mathrm{SU}(2)_L$ singlets, and the corresponding direct bounds are the ones quoted for the singlet representations shown as horizontal lines in \cref{fig:FCCee_fit}.

\subsection{Higgs tuning in VLF-chain models}
\label{sec:higgs_tuning}

An important aspect of these models is the tuning of the Higgs mass. Bringing the dynamics responsible for flavour close to the TeV scale is important not only for its experimental testability, but also to limit the radiative corrections to the Higgs mass parameter~\cite{Susskind:1978ms,tHooft:1979rat,Ellis:1986yg,Barbieri:1987fn,Anderson:1994dz,Halverson:2026cxe, Farina:2013mla}.\footnote{See also Refs.~\cite{Fichet:2012sn,Fowlie:2024nhs} for statistical interpretations of fine-tuning measures.} Current LEP constraints still allow regions of parameter space in which these corrections remain moderate (cf.~the ``LEP paradox''~\cite{Barbieri:2000gf}). As we will see in the following sections, future Tera-$Z$ factories will probe substantially larger VLF masses, putting these frameworks to the test. We therefore quantify, alongside the current and projected electroweak constraints, the tuning induced in the Higgs sector.

There is no unique and model-independent definition of fine-tuning~\cite{Ellis:1986yg,Barbieri:1987fn}. Its precise value depends on the parameters assumed to be independent, the renormalisation prescription and the way in which different sensitivities are combined. We adopt here a simple measure which directly quantifies the cancellation required between the Higgs mass parameter above the VLF threshold and the radiative contribution generated by the new fermions. For a class of VLFs $F$ containing $N_F$ degenerate states coupled to the Higgs with the same coupling $y_F$ (e.g.~$N_F=2$ for an $\mathrm{SU}(2)_q$ doublet; not to be confused with the chain length $N_f$), the
one-loop contribution to the Higgs mass parameter in the $\overline{\textrm{MS}}$ scheme can be written as
\begin{equation}
\label{eq:higgs_mass}
    \left.\delta m_H^2 \right|_F
    =
    -\frac{N_{c} N_F}{8\pi^2}
    \left|y_F \right|^2 M_{F}^2
    \left[
        1+\log\left(\frac{\mu^2}{M_{F}^2}\right)
    \right],
\end{equation}
where $y_F$ denotes the coupling of the corresponding VLF class to the Higgs and $N_c$ its number of colours. The logarithm depends on the renormalisation scale $\mu$, not to be confused with the mass mixings of the chains. In the matching convention used here, setting $\mu=M_F$ gives the finite contribution $\delta m_H^2|_F=-N_cN_F|y_F|^2M_F^2/(8\pi^2)$.
We use its mass dependence as a benchmark diagnostic of the cancellation required to keep the electroweak scale fixed. We quantify the Higgs-sector tuning using the Barbieri--Giudice (BG) measure~\cite{Barbieri:1987fn}, which measures the logarithmic sensitivity of an observable, here the Higgs mass, to a UV parameter, here the VLF mass scale. Applied to the threshold correction,
\begin{align}
\label{eq:BG_measure}
    \Delta_{H,F}
    &=
    \frac{2}{m_h^2}
    \left|
    \frac{\partial\,\big(\delta m_H^2|_{F}\big)_{\mu=M_F}}{\partial \log M_F}
    \right|
    = \frac{N_c N_F |y_F|^2 M_F^2}{2\pi^2 m_h^2}
\end{align}
where $m_h^2=2m_H^2$ is the physical Higgs mass squared at tree level and the factor of two in the numerator accounts for this relation.\footnote{We differentiate the finite threshold at $\mu=M_F$, giving twice the ratio $|\delta m_H^2|/(m_h^2/2)$. This captures the cutoff-independent threshold contribution in our convention; a full UV sensitivity calculation would also require the UV input parameters and their running.}
We also estimate the total tuning of the model by combining the sensitivities associated with the independent VLF classes in quadrature~\cite{Casas:2004gh},
\begin{equation}
    \Delta_{\textrm{total}}
    =
    \left[
        \sum_F \left(\Delta_{H,F}\right)^2
    \right]^{1/2}.
    \label{eq:tuning_total}
\end{equation}
This prescription treats the different VLF classes as independent sensitivity directions. In the text, we quote the tuning as $1/\Delta_{\textrm{total}}$.

The same Higgs couplings that generate this correction also enter the electroweak current operators. Parametrically, their Wilson coefficients and Higgs mass corrections scale in opposite ways with the VLF mass,
\begin{equation}
    C_{Hf}\sim\frac{|y_F|^2}{M_{F}^2},
    \qquad
    \Delta_{H,F}\sim |y_F|^2M_{F}^2.
\end{equation}
Consequently, at fixed Higgs coupling, increasing the lower bound on a VLF mass by one order of magnitude increases the corresponding Higgs tuning by two orders of magnitude. Equivalently, since the EWPO constrain the combination $M_F/|y_F|$, the tuning induced by a given bound scales as $|y_F|^4(M_F/|y_F|)^2$, and can be alleviated by lowering the Higgs coupling $y_F$ at the price of larger mixings $\mu/M$ to reproduce the same effective Yukawa couplings, against the logic of the mechanism. Future Tera-$Z$ factories can therefore turn apparently moderate present-day Higgs tuning into severe tuning requirements. This conclusion is based on the tuning measure adopted in this section. However, as we will discuss in \cref{sec:sol_perspective}, if cosmological dynamics selects that parameter, its naturalness must instead be assessed within the complete selection mechanism.

\section{Electroweak precision analysis}
\label{sec:EW_analysis}

Although these constructions are designed to suppress large flavour-changing neutral currents, and are only mildly constrained by direct searches (\cref{sec:flavour_direct}), they retain irreducible contributions to electroweak precision observables. As discussed in the previous section, the chains responsible for generating the low-energy Yukawa couplings necessarily contain the same structures that induce modifications of the $Z$- and $W$-boson currents. These effects therefore cannot be removed without simultaneously disrupting the mechanism responsible for fermion-mass generation. In this section, we review the current constraints from LEP and assess how the improved precision expected at future Tera-$Z$ factories could raise the scale of these models and, consequently, increase the associated Higgs-sector fine-tuning.
\subsection{Current constraints}
\label{sec:current_constraints}

We begin by considering the minimal realisation in which the chains contain only VLFs transforming in the same gauge representations as the SM right-handed fermions, i.e.~the benchmark model of \cref{sec:model_and_flavour}. Integrating out the heavy states generates the operators $\mathcal O_{H\ell}^{(1,3)}$ and $\mathcal O_{Hq}^{(1,3)}$, which modify the left-handed couplings of the SM fermions to the $Z$ and $W$ bosons. This setup captures the leading electroweak effects of the minimal chains in terms of six independent combinations of masses and Higgs couplings.

We assign an independent characteristic mass to each VLF class,
\begin{equation}
    \bm M_{\rm VLF}
    =\{M_e,\,M_\mu,\,M_\tau,\,M_U,\,M_D,\,M_B\}.
\end{equation}
Here, $M_e$, $M_\mu$ and $M_\tau$ denote the masses associated with the three charged-lepton chains. The up- and down-type partners of the first two quark families form doublets under the approximate $\mathrm{SU}(2)_q$ flavour symmetry and have common masses $M_U$ and $M_D$, respectively, while $M_B$ denotes the mass scale associated with the third-family down quark; these are the masses of the Higgs-coupled VLFs $E^1_{1,2,3}$, $\bm U$, $\bm D$ and $D_3$ of \cref{tab:field_content}. At leading order in the chain mixings, the electroweak currents depend on the VLF directly coupled to the Higgs, cf.~\cref{eq:C_power_counting}. Each class is therefore characterised by
\begin{equation}
\label{eq:ci_parameters}
    c_F\equiv\frac{|y_F|^2}{M_F^2}\geq0,
    \qquad F\in\{e,\mu,\tau,U,D,B\},
\end{equation}
where $y_F$ is the corresponding Higgs coupling. The relation to the Wilson coefficients is given in \cref{eq:app_EWPO_fit_params}. We present the constraints as lower limits on $M_F/|y_F|=c_F^{-1/2}$.

The modified currents enter the $Z$-pole widths and asymmetries, as well as the $W$-boson mass, width and leptonic branching fractions. The precision of these observables is largely driven by the LEP and SLD measurements~\cite{ALEPH:2005ab,ALEPH:2013dgf}. The observables particularly relevant for the present analysis and the projections below are
\begin{align}
    \mathcal O_{\rm EW}=\big\{&
    m_W,\,\Gamma_W,\,\Gamma_Z,\,\sigma_{\rm had}^{0},\,
    R_\ell^{0},\,R_b^{0},\,R_c^{0},\,
    \nonumber\\
    &\mathcal A_\ell,\,\mathcal A_b,\,\mathcal A_c,\,
    A_{\rm FB}^{0,\ell},\,A_{\rm FB}^{0,b},\,A_{\rm FB}^{0,c},\,
    \sin^2\theta^\ell_{\rm eff}\big\}.
    \label{eq:EWPOs}
\end{align}
The charged-lepton flavours are resolved separately to retain sensitivity to the flavour non-universality of the model. The inputs to the current fits are specified in \cref{app:ewpo_codes}, while \cref{eq:EWPOs} summarises the relevant observables. The leptonic widths and asymmetries primarily constrain the lepton chains, while the inclusive hadronic observables probe the light-quark chains. The third-family down-type chain is tested directly by $R_b^0$, $\mathcal A_b$ and $A_{\rm FB}^{0,b}$.

An additional correlation arises through the electroweak input parameters. In the $\{\alpha_{\rm em},m_Z,G_F\}$ scheme, the operators modifying the electron and muon charged currents contribute to muon decay through the combination~\cite{Berthier:2015gja,Berthier:2015oma,Brivio:2017vri}
\begin{equation}
    \delta G_F\propto
    [C_{H\ell}^{(3)}]_{ee}+[C_{H\ell}^{(3)}]_{\mu\mu}.
\end{equation}
The resulting shift in the Higgs vacuum expectation value inferred from $G_F$ propagates to the electroweak gauge couplings and the weak mixing angle. Consequently, the electron and muon chains also affect observables involving other fermions, allowing partial compensations among the contributions of different VLF classes.

We assess these correlations by comparing individual limits, obtained with a single $c_F$ different from zero, with limits from a simultaneous six-parameter fit. The likelihood is constructed from
\begin{align}
    \chi^2(\bm c)
    &=\bm r(\bm c)^T\bm V_{\rm EW}^{-1}\bm r(\bm c),\textrm{ with }
    \bm r(\bm c)
    =\bm{\mathcal O}_{\rm th}(\bm c)
      -\bm{\mathcal O}_{\rm exp},
\end{align}
where $\bm V_{\rm EW}$ contains the experimental and theoretical covariances. The theoretical predictions combine precise SM calculations with the leading SMEFT corrections induced by the matching of the VLF model. We evaluate these contributions using \texttt{flavio}/\texttt{smelli}~\cite{Straub:2018kue,Aebischer:2018iyb}, together with \texttt{wilson}~\cite{Aebischer:2018bkb} for the treatment of the Wilson coefficients, and obtain independent determinations with \texttt{SMEFiT}~\cite{Giani:2023gfq,Celada:2024mcf} and \texttt{HEPfit}~\cite{DeBlas:2019ehy}. This allows us to assess the dependence of the bounds on the implementation of the electroweak predictions and the statistical treatment. The datasets, statistical conventions and comparison of the implementations are collected in \cref{app:ewpo_codes}.

\begin{figure*}[tb]
    \centering
    \includegraphics[width=\linewidth]{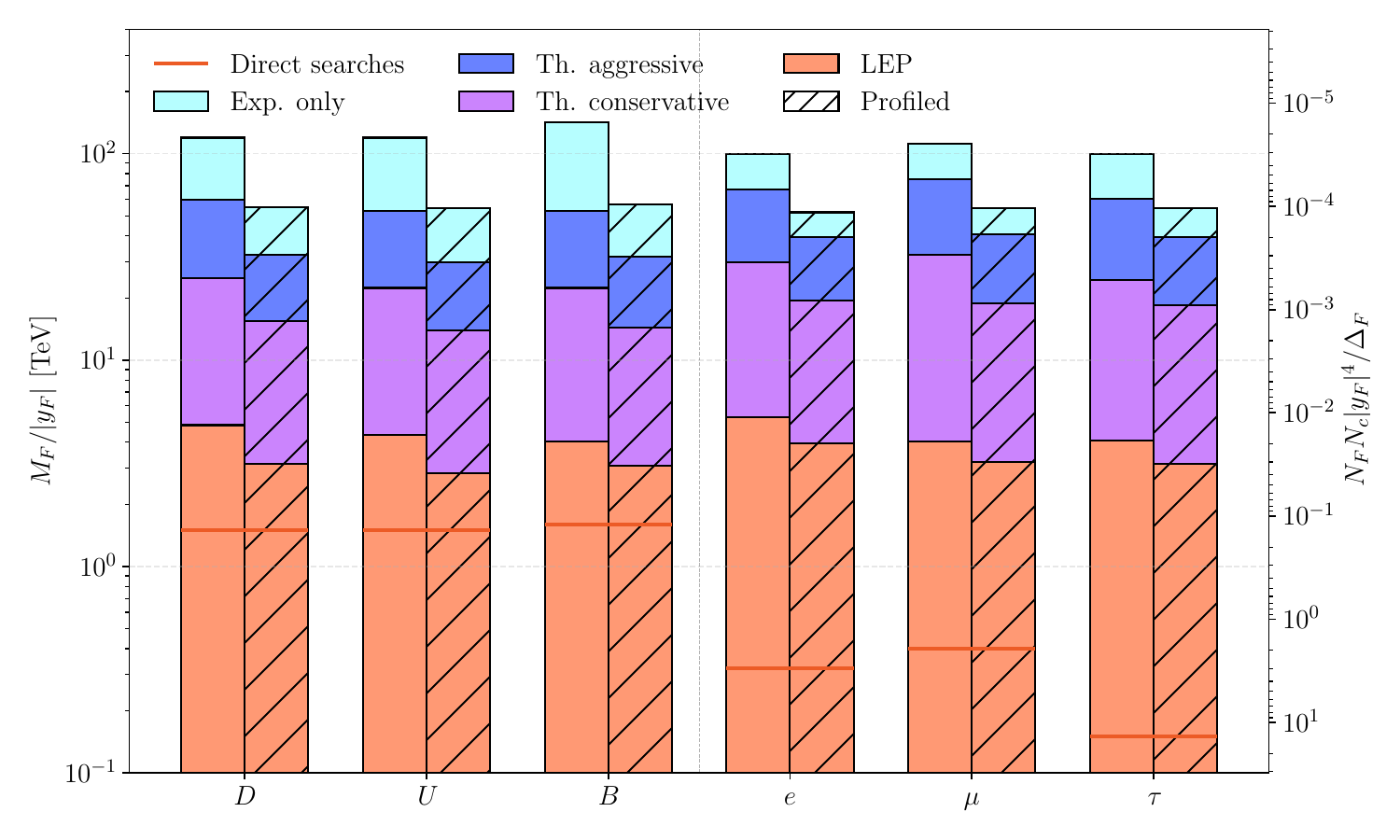}
    \caption{Projected $95\%$~CL lower bounds on $M_F/|y_F|$ at FCC-ee, assuming that no deviation from the SM is observed, for the three uncertainty scenarios described in the text (purple: S1, blue: S2, cyan: S3), compared with the bounds obtained with the current LEP/SLD uncertainties within the same framework (orange), which are SM-centred (expected) bounds. The left and right axes show $M_F/|y_F|$ and the associated Higgs tuning, $N_cN_F|y_F|^4/\Delta_{H,F}$, respectively. Unhatched columns correspond to the individual bounds, obtained by setting the remaining five parameters to zero, whereas the hatched columns correspond to the profiled bounds, in which the remaining parameters are allowed to float. Horizontal lines show the direct-search limits on $M_F$ of \cref{sec:flavour_direct} for $|y_F|=1$.}
    \label{fig:FCCee_fit}
\end{figure*}

The current constraints probe $M_F/|y_F|$ at scales of approximately $3$--$7\,\mathrm{TeV}$, with the detailed results collected in \cref{fig:LEP_fit}.
For order-one Higgs couplings, these indirect limits exceed the direct-search bounds discussed in \cref{sec:flavour_direct}. They already imply mild Higgs-sector tuning at the $\lesssim 0.1$ level. Combining the six contributions in quadrature according to \cref{eq:tuning_total}, each evaluated at its marginalised mass bound and with $|y_F|=1$, gives the indicative estimate $1/\Delta_{\rm total}\simeq0.3\%$. Note that, this is not a bound derived from the joint allowed region, since the separate marginal limits need not correspond to a simultaneously allowed point. However, note that smaller Higgs couplings alleviate the tuning: at fixed $M_F/|y_F|$, the sensitivity scales as $|y_F|^4$, at the price of larger chain mixings to reproduce the same fermion masses. For instance, $|y_F| \sim 0.2$ would completely remove the residual tuning.

\subsection{Future Tera-\texorpdfstring{$Z$}{Z} projections}
\label{sec:tera-Z_projections}

Here we want to investigate the improvements that could be made by future FCC-ee measurements.
The number of $Z$ bosons produced at a future circular $e^+e^-$ collider running at the $Z$ pole, such as the FCC-ee,
would provide a five orders of magnitude increase in statistics compared to LEP~\cite{ALEPH:2005ab,FCC:2025lpp,deBlas:2025gyz}, together with a much improved control of the systematic uncertainties, for instance, through the continuous calibration of the beam energy by resonant depolarisation.
As a result, the experimental uncertainties on the $Z$-pole observables are expected to improve by roughly two orders of magnitude with respect to LEP, e.g.~from about $2$~MeV to $\sim 10$~keV on $\Gamma_Z$ and from $1.6\times10^{-4}$ to a few $10^{-6}$ on $\sin^2\theta^\ell_{\rm eff}$~\cite{FCC:2025lpp}. Since the dimension-six contributions to the EWPO scale as $v^2/\Lambda^2$, such an improvement would bring the bounds from the few TeV of \cref{sec:current_constraints} to tens of TeV, as found in general SMEFT analyses of the Tera-$Z$ potential~\cite{Fan:2014vta,deBlas:2022ofj,Allwicher:2023shc,Celada:2024mcf,Stefanek:2024kds,Allwicher:2024sso,Maura:2024zxz,Greljo:2025ggc}.

Such a leap in precision, however, can only be exploited if the SM predictions are known with a comparable accuracy. Following Ref.~\cite{Greljo:2025ggc}, we therefore consider three scenarios for the total uncertainties on the observables, obtained by combining in quadrature the projected FCC-ee experimental uncertainties~\cite{FCC:2025lpp} with different assumptions on the theory uncertainties (see Refs.~\cite{Freitas:2019bre, deBlas:2025gyz,deBlas:2026vrz}):
a \emph{conservative} scenario (S1), in which the theory uncertainties are reduced only moderately with respect to the present ones; an \emph{aggressive} scenario (S2), in which the future theory calculations reach an accuracy close to the experimental one; and an \emph{experimental-only} scenario (S3), in which the theory uncertainties are neglected altogether and which defines the ultimate reach of the machine. The theoretical control of the $b$-quark observables deserves a special mention: since the third-family down chain is probed almost exclusively through $R_b^0$, its reach depends directly on the precision of the corresponding SM predictions and the $b$-tagging performance at FCC-ee~\cite{Greljo:2024ytg}.

For this analysis, we perform a frequentist fit to the EWPOs $\mathcal{O}_\mathrm{EW}^{\rm FCC}$ listed in \cref{eq:app_EWPO_list}, assuming that FCC-ee measurements agree with the SM predictions.
We therefore write the predicted observables as $\bar{X}(C)=\hat{X}+\delta X(C)$, where $\hat{X}=X_{\mathrm{SM}}=\bar{X}(C=0)$ and $C$ denotes the NP parameters. With this choice, the residuals entering the $\chi^2$ depend only on the NP contributions $\delta X(C)$. Furthermore, assuming Gaussian uncertainties and neglecting correlations between the observables, the global $\chi^2$ takes the form
\begin{equation}
    \chi^{2}_{\mathrm{global}}(C) = \sum_{X \in \mathcal{O}_\mathrm{EW}^{\rm FCC}} \left( \frac{\delta X(C)}{\sigma_{X}} \right)^{2},
\end{equation}
where $\sigma_X$ combines the experimental and theoretical uncertainties in quadrature. The NP contributions to the observables and the RG running are computed using the \texttt{flavio} and \texttt{wilson} packages~\cite{Straub:2018kue,Aebischer:2018bkb}, respectively.

Using the same six parameters as in \cref{sec:current_constraints}, we show our main results in \cref{fig:FCCee_fit}. The Wilson coefficients are defined at $\mu_0=50\,$TeV and run down to the $Z$ pole at one loop.
The bounds on the VLF masses $M_F/|y_F|$ are extracted by performing a scan over each of the six fit parameters to determine the 95\% confidence levels. These correspond to $\Delta \chi^2_{95\%}(1\,\rm{DoF}) \approx 2.71$ for a one-sided $\chi^2$ since the $c_F$ are bound to be positive by definition. At the same time, the remaining parameters are either set to zero (\textit{unprofiled}) or allowed to float (\textit{profiled}) while enforcing positivity for all parameters. In the profiled fit, we fix the parameter of interest at each point in the scan and vary the others to minimise the full $\chi^2$. This allows cancellations between their contributions to the observables, which can weaken the bounds compared to the individual fit.

The right $y$-axis allows one to read off the corresponding Higgs tuning contribution of the scanned VLF estimated by the inverse Barbieri--Giudice measure in \cref{eq:BG_measure} normalised by the representation-dependent factor $N_c N_F$, which is largest for the quark doublet VLFs $U$ and $D$, and by $|y_F|^4$, see \cref{eq:app_tuning_measure}.
We include all three scenarios of uncertainties (S1, S2, S3) as well as the LEP/SLD  and the direct-search bounds of \cref{sec:flavour_direct}.
Comparing the FCC-ee projections with the current bounds shows an increase in the required VLF masses up to approximately $100$\,TeV for the unprofiled and around $50\,$TeV for the profiled case. In S3 the profiled bounds become almost universal, $M_F/|y_F|\approx52$--$57\,$TeV for all six classes, about a factor of two below the individual ones, which points to an approximate flat direction of the likelihood within the physical region $c_F\geq0$ (\cref{app:tera-Z_details}). It originates from the ratios $R_f$ being approximately blind to the direction where all parameters are equal, which is explained in more detail in \cref{app:EWPO_analysis}. Since these are the only observables with an order of magnitude improvement between the aggressive and experimental-only scenario, their impact governs the behaviour of the global chi-squared in S3. Moving along the blind direction in the $R_f$ is therefore the only direction which avoids a very tight bound. This approximately flat direction could be lifted by comparable improvements in observables such as $\Gamma_Z$ which close the fit and do not exhibit the same degeneracy.

The increased mass bounds also lift the tuning to the $\mathcal{O}(10^{-3})$ level for S1 and $\mathcal{O}(10^{-4})$ for S2 in the profiled case, while S3 would push for an even higher tuning, of $\mathcal{O}(10^{-5})$ for the individual bounds.
We can conclude that the absence of NP at FCC-ee would, for order-one Higgs couplings, drive this model into a corner of significant fine-tuning. One remaining handle to alleviate this situation is the coupling strength $y_F$ between the VLF, the Higgs and the left-handed SM fields. Due to the quartic sensitivity of the tuning to this coupling at fixed $M_F/|y_F|$, a small reduction of $y_F$, e.g.~from $1$ to $0.6$, already reduces the tuning by almost an order of magnitude. The irreducible fact is that FCC-ee could probe up to three orders of magnitude further in terms of tuning measures compared to what we have today, covering what is arguably \emph{the most interesting} parameter space.

More precisely, at the S3 individual bound on the $B$ chain, $M_F/|y_F|\simeq140\,$TeV, $|y_F|=0.2$ gives $M_B\simeq28\,$TeV and $1/\Delta_{H,B}\simeq 3 \times 10^{-3}$. The Higgs-criticality motivation of \cref{sec:sol_perspective} requires $|y_F|\gtrsim 0.7$, which leads to~$1/\Delta_{H,B}\lesssim2\times10^{-5}$. Since the projected flavour bounds of \cref{sec:flavour_direct} reach at most $M_F/|y_F|\approx9\,$TeV (\cref{fig:flavour_bounds_projections}), the EWPO remain the strongest probe also in the future, consistent with the MFP protection of the model against large FCNCs.

\section{General lessons for TeV-scale flavour models}
\label{sec:general_lessons}

In the benchmark model the Higgs link sits at the start of every chain. The LH currents are then unsuppressed, and the EWPO bound the VLF mass directly, $M_F\geq|y_F|\Lambda_L$, where $\Lambda_L$ is the bound on $M_F/|y_F|$ shown in \cref{fig:FCCee_fit}. As discussed in \cref{sec:moving_higgs}, moving the link along the chain shifts the electroweak effects from one chirality to the other, but cannot remove them. Here we ask how far this can lower $M_F$, and with it the Higgs tuning.

Consider a chain with LH VLFs before the Higgs link and RH VLFs after it, as in \cref{fig:chains_figure}(a), and let $\epsilon_L$ ($\epsilon_R$) be the product of the mixing ratios $\mu/M$ on the LH (RH) side. The Yukawa coupling is $y\equiv|y^f_\textrm{eff}|\simeq|y_F|\,\epsilon_L\epsilon_R$, while for a common VLF mass the LH and RH currents, normalised as $c_F$ in \cref{eq:ci_parameters}, are
\begin{equation}
\label{eq:gl_chain}
 c_L\simeq\frac{|y_F|^2\epsilon_L^2}{M_F^2},\qquad
 c_R\simeq\frac{|y_F|^2\epsilon_R^2}{M_F^2},
\end{equation}
cf.~\cref{eq:C_power_counting}. Extending one side of the chain suppresses one current, but their product, $c_Lc_R\simeq|y_F|^2y^2/M_F^4$, is fixed by the Yukawa coupling. Imposing both bounds, $c_L\leq\Lambda_L^{-2}$ and $c_R\leq\Lambda_R^{-2}$, therefore gives
\begin{equation}
\label{eq:gl_floor}
 M_F\gtrsim\left(|y_F|\,y\,\Lambda_L\Lambda_R\right)^{1/2}.
\end{equation}
This floor is reached when the two currents are suppressed equally. For the $b$ quark and the $\tau$ lepton this requires only one additional LH partner; once the two currents are balanced, longer chains do not lower it further. The key point is that $M_F$ scales as $y^{1/2}$. Constraining only the LH currents would instead suggest a much lower floor, $M_F\sim y\Lambda_L$, but only at the price of RH currents far above their bound.

By unitarity, the Higgs couplings of all mass eigenstates of the chain add up to $|y_F|^2$, so that the chain contributes to the Higgs mass like a single VLF, whatever its length. \Cref{eq:BG_measure} then gives
\begin{equation}
\label{eq:gl_tuning}
 \Delta_{H,F}\gtrsim\frac{N_cN_F|y_F|^3\,y\,\Lambda_L\Lambda_R}{2\pi^2m_h^2},
\end{equation}
so that, compared with the minimal chain, rearranging the chain reduces the tuning at most by a factor $\sim1/y$.

We have not computed the bounds on the RH currents and assume $\Lambda_R\approx\Lambda_L$. This is a reasonable estimate, since $R_b$ and $R_\tau$, which drive the bounds, receive contributions of the same sign from both currents. For $|y_F|=1$ and the S3 bounds of \cref{fig:FCCee_fit}, the chains generating the $b$ and $\tau$ Yukawa couplings ($y\simeq0.012$ and $0.010$ at the TeV scale) must then lie above $15$ and $10\,$TeV, with $1/\Delta_{H,B}\approx4\times10^{-4}$ and $1/\Delta_{H,\tau}\approx3\times10^{-3}$. This is about a hundred times less tuned than the minimal chain, but still an order of magnitude above the reach of direct searches. With the present LEP/SLD bounds, instead, the floor lies below the direct-search limits.

The lesson is that the Tera-$Z$ test cannot be evaded by rearranging the chain: wherever the Higgs link sits, the EWPO probe the geometric mean of the LH and RH bounds. The only remaining handle is a smaller Higgs coupling, $\Delta_{H,F}\propto|y_F|^3$, which works against the Higgs-criticality motivation of \cref{sec:sol_perspective}.

\section{Flavour chains and Higgs criticality}
\label{sec:sol_perspective}

The flavour construction explains the hierarchy of the SM Yukawas through ratios of mixings and masses, but does not determine the overall VLF scale. MFP allows this scale to lie near the TeV without excessive flavour violation, but it does not necessarily explain why. A further motivation for such a low scale could arise if the VLFs participate in solving the electroweak hierarchy problem. In particular, their Higgs couplings modify the Higgs potential and can lower its instability scale. This is precisely the effect additional matter must provide in cosmological solutions to the electroweak hierarchy problem based on Higgs criticality~\cite{Bak:1987xua, Kartvelishvili:2020thd, Asadi:2022njl, Khoury:2021zao, Giudice:2021viw, Benevedes:2025qwt, Detering:2025lfn, Detering:2025nmh, Enguita:2025ybx, Benevedes:2024tdq}. For concreteness, we focus on the self-organised localisation (SOL) proposed in Ref.~\cite{Giudice:2021viw}. We therefore ask whether the fermions already present in the flavour model can supply this ingredient, and how this possibility relates to the electroweak bounds derived above.

In SOL, a light scalar $\phi$ undergoes quantum fluctuations during inflation, scanning a Lagrangian parameter.
If the theory admits different phases, this stochastic evolution, together with their different expansion rates, can favour values of this parameter near a phase boundary, or \textit{critical point}~\cite{Giudice:2021viw}.\footnote{More precisely, it is the volume-weighted probability distribution that can become localised near the critical point. This distribution weights regions by their physical volume, so faster-expanding regions contribute more as inflation proceeds.} For Higgs naturalness, the scanned parameter is the quadratic Higgs mass parameter. As it varies, the metastable minimum at small Higgs field values approaches the barrier separating it from a second vacuum at larger field values. The critical point is reached when this minimum and the barrier merge, so that the low-field vacuum ceases to exist. Considering only the SM matter content, the scale of this critical point is set by the running Higgs quartic when it crosses zero. Starting from a positive value at the weak scale, the quartic decreases through quantum corrections, particularly those involving the top quark, and crosses zero at a much higher scale $\sim10^{10}$--$10^{11}\,$GeV~\cite{Buttazzo:2013uya}.\footnote{The scale is particularly sensitive to the top Yukawa and strong coupling. Here we define $\Lambda_I$ by $\lambda_{\rm eff}(\Lambda_I)=0$, where the effective quartic includes both the running of the couplings and the explicit loop corrections to the Higgs potential.} As we show below, the minimum and barrier merge at a Higgs field value of the same order. Consequently, localisation near this endpoint would favour a Higgs expectation value far above the observed weak scale~\cite{Giudice:2021viw}.

The additional fermions present in our flavour model are useful because their Higgs couplings can make the quartic decrease more rapidly, bringing its zero and the associated critical endpoint to lower Higgs field values. They can therefore bring the scale favoured by SOL closer to the weak scale. Heavy fermions lowering the instability scale, and the additional bosonic states then needed to stabilise the potential, were studied in this context in Refs.~\cite{Detering:2025lfn,Detering:2025nmh}. The new element here is that these fermions are not introduced for this purpose, but are supplied by the flavour sector, with the multiplicity $\sum_F n_F=18$ discussed below.

To exhibit the scale entering the mechanism, write $H=(0,h/\sqrt2)^T$ and approximate the Higgs potential near the zero of its running quartic as
\begin{align}
 V_h&=-\frac{m^2(\phi)}{2}h^2+\frac{\leff(h)}4h^4,
 \nonumber\\
 \leff(h)&\simeq b\log\frac{h}{\Lambda_I},
 \qquad b\equiv\frac{d\leff}{d\log h}<0.
 \label{eq:sol_running_potential}
\end{align}
Here $m^2(\phi)$ is the scanned, renormalised mass parameter and $\Lambda_I$ is the scale at which the quartic vanishes. Treating $b$ as constant, the stationary condition and the curvature at a stationary point are
\begin{align}
 m^2&=h^2(\leff+b/4), \quad
 V_h''\big|_{V_h'=0}=h^2(2\leff+3b/2).
 \label{eq:sol_stationary_running}
\end{align}
The low-field minimum therefore merges with the barrier when both $V_h'(h_\textrm{crit})=0=V_h''(h_\textrm{crit})$, which in this case gives
\begin{align}
 h_{\rm crit}&=e^{-3/4}\Lambda_I,\quad
 m_{\rm crit}^2=\frac{|b|}{2}e^{-3/2}\Lambda_I^2.
 \label{eq:sol_endpoint}
\end{align}
The expression for $m_{\rm crit}^2$ coincides with the metastability bound derived in Ref.~\cite{Detering:2025nmh}. The relation for $h_{\rm crit}$ makes the connection to the flavour sector transparent: lowering the instability scale also lowers the Higgs field value associated with the critical configuration. Notice, however, that the curvature of the Higgs potential vanishes at the endpoint, whereas the observed Higgs mass requires a positive curvature at today's vacuum. The latter must therefore lie away from the exact critical configuration. Hence, reproducing the observed vacuum additionally requires a suitable displacement from the critical endpoint, controlled by inflationary fluctuations or further dynamics.

To show how our model may help in bringing the scale $\Lambda_I$ closer to the EW scale, we need to compute the corrections to the Higgs quartic. An easy way to see how this happens is to note that, above the relevant fermion thresholds, the direct one-loop contributions of the new Higgs Yukawas to the quartic running are
\begin{equation}
 \left.\Delta\beta_\lambda\right|_{\rm Yukawa}
 =\frac{1}{16\pi^2}
 \left(4\lambda\sum_F n_F|y_F|^2
             -2\sum_F n_F|y_F|^4\right),
 \label{eq:sol_beta_lambda}
\end{equation}
where $n_F=N_cN_F$ counts colour and independent flavour copies of a Higgs link. In the minimal model, $n_U=n_D=6$, $n_B=3$ and $n_e=n_\mu=n_\tau=1$.
Thus, taking $y_F=1$ for the Higgs couplings, the negative quartic term is $-0.228$, while the full contribution displayed in \cref{eq:sol_beta_lambda} is $-0.182$ at $\lambda=0.10$.
These fermions can therefore substantially accelerate the decrease of the Higgs quartic.

However, to perform the correct assessment of the criticality scale, and determine which VLF masses and couplings can bring the Higgs critical point closer to the EW scale, we need to compute their full one-loop contribution to the Higgs potential. Neglecting the mixings along the chain, the chiral SM fermion couples through the Higgs to the first massive partner. The resulting heavy eigenstate has a field-dependent squared mass $M^2_F+|y_F|^2h^2/2$, while the remaining partners are Higgs-independent spectators in this limit. For $n_F$ identical links, with colour multiplicities included in $n_F$, the Coleman--Weinberg contribution can be evaluated directly. At the renormalisation scale $\mu=M_F$, and after absorbing the field-independent and quadratic terms into the matched parameters, the one-loop Higgs potential takes the form
\begin{align}
 V_h&=-\frac{m^2}{2}h^2+\frac{\lambda_M}{4}h^4
             -\frac{n_F M^4_F}{16\pi^2}R(t),\nonumber\\
 R(t)&=(1+t)^2[\log(1+t)-3/2]+3/2+2t,\nonumber\\
 &\textrm{with}\quad t=\frac{|y_F|^2h^2}{2M^2_F},
 \label{eq:sol_cw_threshold}
\end{align}
where $\lambda_M$ is the matched quartic and the couplings are held fixed for this estimate. Since $R(t)=t^3/3+\mathcal O(t^4)$, the heavy contribution decouples at small field after matching.

\begin{figure*}
    \centering
    \includegraphics[width=\linewidth]{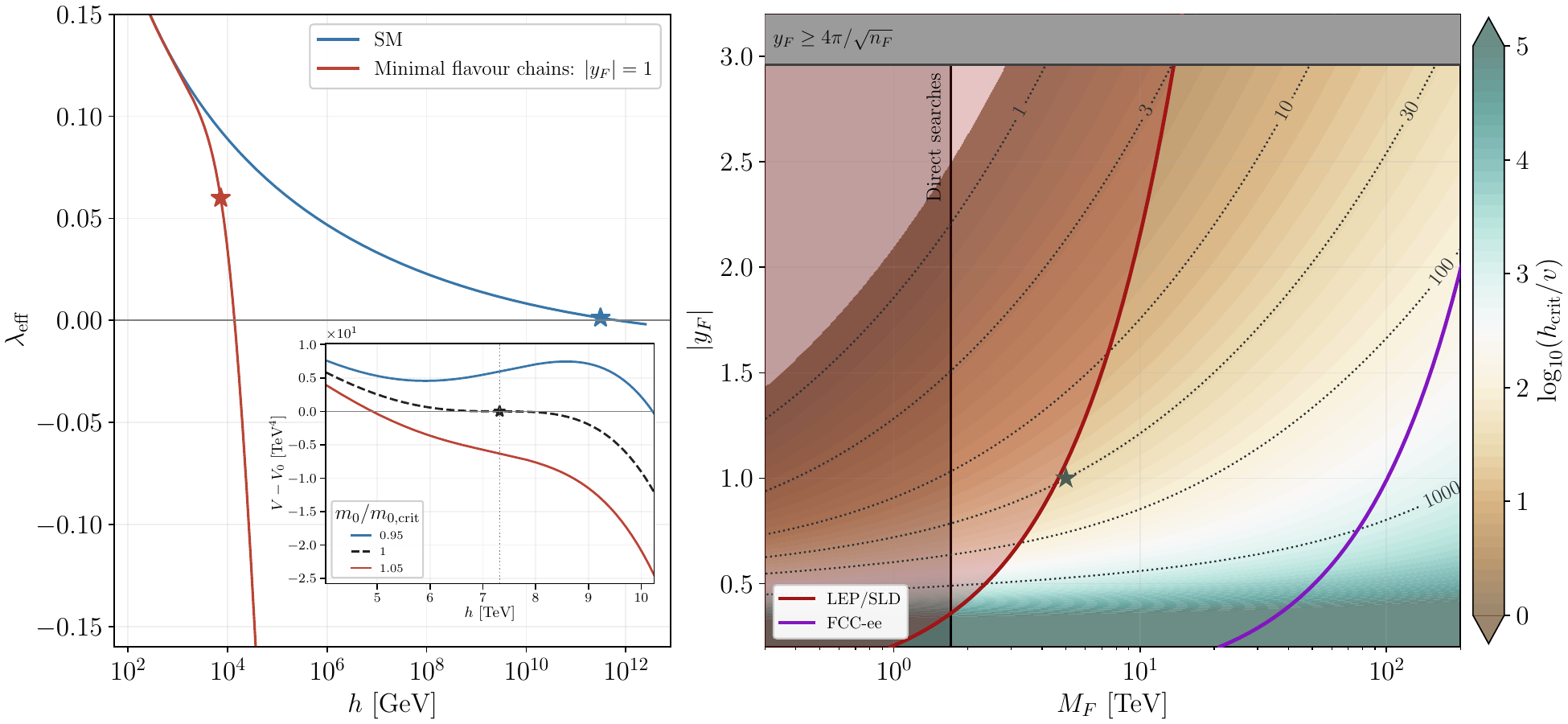}
    \caption{\textit{Left}. Higgs effective quartic as a function of the field value $h$ for the SM and our flavour chain model in blue and red respectively. The stars mark $h_{\rm crit}$ for the SM and for the benchmark $M_F=5\,$TeV, $|y_F|=1$. The inset shows the Higgs potential near the critical scale for different values of $m_0/m_{0,\rm crit}\equiv m/m_{\rm crit}$. \textit{Right.} We show the critical Higgs value as a function of the masses and couplings of the VLFs to the Higgs. We include the LEP/SLD bounds (red) and the future FCC-ee projections in the best-case scenario (purple) for the minimal chain model. The benchmark point of the left panel is shown as a star, while the dotted contours give $h_{\rm crit}/v$. The vertical line marks the direct-search limit and the grey band the non-perturbative region $y_F\geq4\pi/\sqrt{n_F}$.}
    \label{fig:higgs_criticality}
\end{figure*}

Solving $V_h'=V_h''=0$ gives a naive estimate of the critical configuration,
\begin{align}
 &t_{\rm crit}=\exp\!\left(\frac{16\pi^2\lambda_M}{n_F|y_{F}|^4}\right)-1,\quad
 h_{\rm crit}=\frac{\sqrt2 M_F}{|y_F|}\sqrt{t_{\rm crit}},
 \label{eq:sol_threshold_endpoint}\\
& m_{\rm crit}^2=\frac{n_F|y_F|^2 M_F^2}{8\pi^2}
       [t_{\rm crit}-\log(1+t_{\rm crit})].
 \label{eq:sol_threshold_mass}
\end{align}
For $n_F=18$, $y_F=1$ and $\lambda_M=0.10$, we obtain $h_{\rm crit}\simeq1.68M_F$ and $m_{\rm crit}^2\simeq0.120M_F^2$. The flavour fermions can therefore generate a critical configuration close to their mass scale. Together with the MFP protection of TeV-scale VLFs, this provides a possible connection between the flavour scale and Higgs criticality: for $M_F\sim1\,\mathrm{TeV}$, the critical value of the scanned Higgs mass parameter is $m_{\rm crit}\sim350\,\mathrm{GeV}$, bringing it towards the EW scale. A quantitative determination requires including the SM running and loop corrections, which modify the effective quartic and can appreciably shift the critical configuration.

In \cref{fig:higgs_criticality}, we repeat this computation including the SM running and loop contributions to the Higgs potential. We evolve the SM couplings and the quadratic Higgs mass parameter using the three-loop beta functions implemented in Ref.~\cite{Martin:2019lqd}. The boundary conditions at $\mu_0=173.1\,\mathrm{GeV}$ and the leading-order effective potential follow Ref.~\cite{Detering:2024vxs}. In particular, we include the Coleman--Weinberg contributions of the top quark and the $W$ and $Z$ bosons, evaluated with the running couplings at $\mu=h$. To this SM potential we add the finite VLF contribution derived above, neglecting the chain mixings and holding the VLF masses and Higgs couplings fixed; a complete evolution including the new fermions is beyond this estimate.

In \cref{fig:higgs_criticality}, the left panel compares the resulting effective quartic with its SM value, taking $n_F=18$, $M_F=5\,\mathrm{TeV}$ and $y_F=1$. The negative fermion contribution brings the decrease of the quartic to much lower Higgs field values. Solving $V_h'=V_h''=0$, we find that the critical field shifts from $h_{\rm crit}\simeq3.1\times10^{11}\,\mathrm{GeV}$ in the SM to $h_{\rm crit}\simeq7.3\,\mathrm{TeV}$ in this benchmark. This benchmark lies close to the boundary of the present experimental bounds. The inset shows the corresponding potential for different values of the Higgs mass-term, illustrating how the low-field minimum and the barrier approach, merge and disappear.

The right panel extends this calculation to different VLF masses and Higgs couplings, keeping the total number of degrees of freedom coupled to the Higgs fixed at $\sum_F n_F=18$. Larger Yukawa couplings or smaller masses make the quartic decrease faster and bring the critical point down towards the TeV scale. The current EWPO bounds (we take the LEP bound on the $U$ chain, $M_U/|y_U|\gtrsim 4.62\,$TeV, as a reference) already exclude the solutions that would bring it as low as $h_{\rm crit}\simeq 3\,v$.
On the other hand, FCC-ee, in the most optimistic scenario, will probe $M_F/|y_F|$ up to $100\,$TeV, i.e.~critical points as high as $h_{\rm crit}\sim 10^2\text{--}10^3\,v$ for $|y_F|\simeq2$--$0.75$ (about $460\,v\approx110\,$TeV for $|y_F|=1$).

Although this solution is appealing, and comes naturally from the explanation of the flavour puzzle, it still leaves a few unanswered questions in this setup. For instance, even if the critical field value is sufficiently low, localisation at the ideal endpoint would give zero Higgs curvature. Reproducing both the observed vacuum expectation value and $m_h\simeq125\,\mathrm{GeV}$ requires a specified displacement from criticality or subsequent evolution. This is the residual little hierarchy problem of this application of SOL~\cite{Giudice:2021viw}. For the benchmark of \cref{fig:higgs_criticality}, the full computation gives $m_{\rm crit}\simeq1.5\,$TeV (\cref{eq:sol_threshold_mass}: $1.7\,$TeV), while today's vacuum requires $m\simeq m_h/\sqrt2\simeq88\,$GeV, i.e.~$(m/m_{\rm crit})^2\simeq3\times10^{-3}$. This is of the same order as the conventional tuning estimates of \cref{sec:higgs_tuning}. A second requirement is vacuum survival. The negative fermion contribution that brings the instability down also makes decay towards larger Higgs fields more dangerous. The matched potential must therefore admit a present-day electroweak vacuum with an acceptable tunnelling rate and a controlled high-field phase. Survival over the age of the Universe requires stabilising bosonic states below $\sim 100\,$TeV~\cite{Blum:2015rpa}, with details left to the UV completion. Additional interactions that stabilise the high-field phase must be included when locating the critical configuration, since they can also shift or remove the barrier, as discussed in Ref.~\cite{Detering:2025nmh}.

In summary, the resulting connection is that the same Higgs interactions needed for flavour hierarchies can also supply the fermionic contribution needed to lower the critical Higgs scale. We have shown that MFP allows this matter to be light enough for that contribution to be relevant without conflicting with flavour bounds. In the minimal realisation of the flavour chains, the electroweak bounds constrain precisely the Higgs-coupled VLF masses, and raising these masses increases the critical scale. FCC-ee can therefore test the minimal low-scale realisation of this idea. The effect of these fermions on the Higgs potential provides a possible first step towards a cosmological solution to the Higgs hierarchy problem. However, whether this UV completion yields a completely successful explanation of the weak scale depends on vacuum survival, which requires new degrees of freedom to stabilise the potential at large field values.

\section{Conclusions}
\label{sec:conc}

The quest for a solution to the flavour puzzle at the TeV scale has a twofold motivation: present and future experiments can target such solutions, and the Higgs hierarchy problem is kept under control. Currently, models based on chains of heavy vector-like fermions~\cite{Arkani-Hamed:2026wwy}, together with a protecting flavour symmetry~\cite{Greljo:2025mwj}, can keep these solutions at the desired scales while remaining compatible with experimental searches, with a moderate tuning of the electroweak scale. However, the same Higgs couplings that generate the Yukawa hierarchies also shift the $Z$ and $W$ couplings of the light fermions at tree level, while the high-energy physics community envisions a future era of colliders aiming to push the precision of the EW sector of the SM well beyond the per-mille level~\cite{deBlas:2025gyz}.

In this paper, we derive current constraints on models based on chains of vector-like fermions and study how future precision machines operating at Tera-$Z$ can probe these constructions. Taking the \textit{minimal flavour protection} realisation as a benchmark~\cite{Greljo:2025mwj}, we find that current EWPO already provide the most comprehensive bounds, $M_F/|y_F|\gtrsim3$--$7\,$TeV, keeping the associated Higgs-sector tuning only at the $\lesssim 0.1$ level. In the future, precision measurements at the $Z$ pole will dramatically strengthen these bounds, reaching scales of order $100\,\mathrm{TeV}$ in the most optimistic scenario, and $\mathcal{O}(10$--$30)\,$TeV even with conservative theory uncertainties. Hence, if no signal is observed, future EWPO at Tera-$Z$ colliders would impose an unacceptable Higgs-sector tuning of order $10^{-4}$ or worse for order-one Higgs couplings. In other words, FCC-ee will be probing the most interesting region of parameter space indicated by naturalness. Smaller Higgs couplings soften this conclusion, since at fixed $M_F/|y_F|$ the tuning scales as $|y_F|^4$. However, order-one couplings are precisely those for which the flavour messengers can play a role in Higgs criticality, see below.

We further explored whether such prospects apply across the full class of models beyond the chosen benchmark. Extending the chains with LH partners before the Higgs insertion suppresses the corrections to electroweak currents, relaxing the corresponding mass bounds. Consequently, the strong tuning implied by the minimal realisation can be relaxed while keeping order-one Higgs couplings. However, this effect is limited and cannot fully decouple the tuning from the EWPO: as the chain is extended, the mixing required to reproduce the Yukawa hierarchies increases, and further extensions eventually provide little additional suppression of the EWPO corrections. Our estimate suggests that the tuning then remains at best at the $\mathcal{O}(10^{-3})$ level, dominated by the vector-like fermions coupled to the $b$ quark.

Finally, \textit{minimal flavour protection} allows the flavour sector to live at the TeV scale, where the Higgs tuning can remain modest, but leaves unanswered why nature should choose to place it there. In fact, these chains could be put at much larger scales, with a stabilising mechanism such as supersymmetry or compositeness screening the Higgs sector from their UV effects. (For recent selected works on the flavour problem within the SUSY or compositeness context, see~\cite{Davighi:2025cqx, Agashe:2026ggy}.) Cosmological solutions to naturalness problems, based on Higgs criticality~\cite{Giudice:2021viw}, offer a possible explanation for a lower scale. In these solutions, a light field scans the Higgs mass parameter during inflation. Its quantum fluctuations can favour parameter values on the verge of a phase transition, near a critical point. For the Higgs, this critical point marks the disappearance of its metastable vacuum. In the SM, the corresponding critical scale lies far above the EW scale, so additional ingredients are needed for this mechanism to explain the observed weak scale.

Our flavour model offers an intriguing coincidence: the multiplicity of Higgs-coupled fermions already required by the flavour construction, with order-one couplings and masses near the TeV scale, can bring the critical Higgs scale into the same neighbourhood. In fact, we have shown that for order-one couplings, the critical scale can be brought down to $\sim 7$\,TeV in the minimal chain with VLF masses at the present bounds, with the possibility of bringing it lower by extending the chain. FCC-ee will test these minimal scenarios by probing regions where the critical scale is pushed up to $10^3$ times above the EW scale.

A complete cosmological solution remains a task for future work. Beyond the residual little hierarchy problem comparable in size to the conventional tuning, the current setup implies that tunnelling into a larger-field true vacuum is unavoidable, signaling a need for additional bosons in the UV. Furthermore, the UV origin of the softly broken MFP flavour symmetry remains unspecified. The possibility that flavour and Higgs criticality share a common explanation at the TeV scale provides a compelling motivation for future model building.

In summary, TeV-scale solutions to the flavour puzzle that remain consistent with Higgs naturalness are still compatible with current data from flavour physics, direct searches, and EWPO. However, future high-precision colliders like the FCC-ee will severely challenge these constructions in their minimal formulations. In other words, the flavour puzzle motivates the chains, minimal flavour protection allows them at the TeV scale, Higgs naturalness orders them there, and a Tera-$Z$ factory will be the decisive test.

\section*{Acknowledgements}
We thank Alessandro Valenti and Tevong You for useful discussions. This work was supported by the programme ``Swiss
High Energy Physics for the FCC'' (CHEF).

\appendix
\crefalias{section}{appendix}
\crefalias{subsection}{appendix}
\onecolumngrid
\makeatletter
\let\set@footnotewidth\set@footnotewidth@one
\onecolumn@grid@setup
\makeatother

\section{Vector-like fermion chains}
\label{app:VLF_chains}

This appendix collects the technical details behind \cref{sec:model_and_flavour}: the explicit mass matrices corresponding to \cref{fig:benchmark_chains} and the tree-level matching (\cref{app:integrating_out}), the perturbative diagonalisation that fixes the benchmark spurions from the quark masses and the CKM matrix (\cref{app:pert_diag}), and the inputs to the flavour bounds of \cref{sec:flavour_direct} (\cref{app:flavour_obs}).

\subsection{Integrating out heavy fermions}
\label{app:integrating_out}

In this and the following subsection we present the explicit mass matrices of the benchmark model of \cref{sec:model_and_flavour}, describe the tree-level matching of the UV theory to the SMEFT, and show how the CKM matrix is reproduced. The vector-like fermions are denoted by $E_i^k$ (lepton partners of the $i$-th family), $U_1^k$, $D_i^k$ (up- and down-type partners), where the superscript labels the position along the chain, and $\bm U$, $\bm D$ for the $\mathrm{SU}(2)_q$ doublets coupled to the light quark doublet $\bm q$, see \cref{tab:field_content}.
The interactions of the VLFs with the SM particles are given by
\begin{equation}
    \mathcal{L}_{\mathrm{VLF, \alpha}} = -(\overline{\Psi}^{\alpha}_{L} \mathcal{M}_{\alpha} \Psi^{\alpha}_{R} \text{ + h.c.})
\end{equation}
where $\alpha$ differentiates between the different sectors $\alpha \in \{ e, \mu, \tau, u, d \}$. The corresponding spinors read
\begin{align}
\overline{\Psi}^{e}_{L}
&=
\begin{pmatrix}
\overline{l}_{1} &
\overline{E_{1}^{1}}_{L} &
\overline{E_{1}^{2}}_{L} &
\overline{E_{1}^{3}}_{L}
\end{pmatrix},
&
\Psi^{e}_{R}
&=
\begin{pmatrix}
e_{1} &
{E_{1}^{1}}_{R} &
{E_{1}^{2}}_{R} &
{E_{1}^{3}}_{R}
\end{pmatrix}^T,
\\
\overline{\Psi}^{\mu}_{L}
&=
\begin{pmatrix}
\overline{l}_{2} &
\overline{E_{2}^{1}}_{L} &
\overline{E_{2}^{2}}_{L}
\end{pmatrix},
&
\Psi^{\mu}_{R}
&=
\begin{pmatrix}
e_{2} &
{E_{2}^{1}}_{R} &
{E_{2}^{2}}_{R}
\end{pmatrix}^T,
\\
\overline{\Psi}^{\tau}_{L}
&=
\begin{pmatrix}
\overline{l}_{3} &
\overline{E_{3}^{1}}_{L}
\end{pmatrix},
&
\Psi^{\tau}_{R}
&=
\begin{pmatrix}
e_{3} &
{E_{3}^{1}}_{R}
\end{pmatrix}^T,
\\
\overline{\Psi}^{u}_{L}
&=
\begin{pmatrix}
\boldsymbol{\overline{q}} & \overline{q}_3 &
\boldsymbol{\overline{U}}_{L} &
\overline{U_{1}^{1}}_{L} &
\overline{U_{1}^{2}}_{L}
\end{pmatrix},
&
\Psi^{u}_{R}
&=
\begin{pmatrix}
u_1 & u_2 & u_3 &
\boldsymbol{U}_{R} &
{U_{1}^{1}}_{R} &
{U_{1}^{2}}_{R}
\end{pmatrix}^T,
\\
\overline{\Psi}^{d}_{L}
&=
\begin{pmatrix}
\boldsymbol{\overline{q}} & \overline{q}_3 &
\boldsymbol{\overline{D}}_{L} &
\overline{D_{3}}_{L} &
\overline{D_{3}^{1}}_{L} &
\overline{D_{2}^{1}}_{L} &
\overline{D_{1}^{1}}_{L} &
\overline{D_{1}^{2}}_{L}
\end{pmatrix},
&
\Psi^{d}_{R}
&=
\begin{pmatrix}
d_1 & d_2 & d_3 &
\boldsymbol{D}_{R} &
{D_{3}}_{R} &
{D_{3}^{1}}_{R} &
{D_{2}^{1}}_{R} &
{D_{1}^{1}}_{R} &
{D_{1}^{2}}_{R}
\end{pmatrix}^T
\end{align}
where we denote in bold the $\mathrm{SU}(2)_q$ flavour doublets $\boldsymbol{U}, \boldsymbol{D}, \boldsymbol{q}$. The mass matrices are
\begin{align}
&\mathcal{M}_{e} = \left(\begin{array}{c|ccc}
0 & y_{\ell_{1}} H & 0 & 0  \\ \hline
0 & M_{E} & \mu_{zE_{1}^{1}E_{1}^{2}} & 0 \\
0 & \mu_{zE_{1}^{1}E_{1}^{2}}^{*} & M_{E} & \mu_{zE_{1}^{2}E_{1}^{3}} \\
\mu_{zE_{1}^{3}e_{1}} & 0 & \mu_{zE_{1}^{2}E_{1}^{3}}^{*} & M_{E}
\end{array}  \right), \,
\mathcal{M}_{\mu} = \left(\begin{array}{c|cc}
0 & y_{\ell_{2}} H & 0 \\ \hline
0 & M_{E} & \mu_{zE_{2}^{1}E_{2}^{2}} \\
\mu_{zE_{2}^{2}e_{2}} & \mu_{zE_{2}^{1}E_{2}^{2}}^{*} & M_{E}
\end{array}  \right), \,
\mathcal{M}_{\tau} = \left(\begin{array}{c|c}
0 & y_{\ell_{3}} H \\ \hline
\mu_{zE_{3}^{1}e_{3}} & M_{E}
\end{array}  \right), \\[10pt]
&\mathcal{M}_{u} =\left(
\begin{array}{ccc|ccc}
 \bm0 & \bm0 & \bm0 & \bm{y}_{q}^{u} \tilde{H} & 0 & 0 \\
 0 & 0 & y_{q_{3}}^{u} \tilde{H}  & \bm0 & 0 & 0 \\
 \hline
 \bm0 & \bm{\mu_{W}}^u & \bm0 & \bm{M_U}  & \bm{\mu_{V}}^u & \bm0 \\
 0 & 0 & 0 & {\bm{\mu_{V}}^u}^{\dagger}  & M_{U} & \mu_{zU_{1}^{1}U_{1}^{2}} \\
 \mu_{zU_{1}^{2}u_{1}} & 0 & 0 & \bm0 & \mu_{zU_{1}^{1}U_{1}^{2}}^{*} & M_{U}
\end{array}
\right) \\[10pt]
&\mathcal{M}_{d} = \left(
\begin{array}{ccc|cccccc}
 \bm0 & \bm0 & \bm0 & \bm{y}^{d}_{q} H & \bm0 & \bm0 & \bm0 & \bm0 & \bm0 \\
 0 & 0 & 0 & \bm0 & y^{d}_{q_{3}} H & 0 & 0 & 0 & 0 \\
 \hline
 \bm0 & \bm0 & \bm0 & \bm{M_D} & \bm0 & \bm{\mu_{U}}^{d} & \bm{\mu_{\tilde{V}}}^{d} & \bm{\mu_{V}}^{d} & \bm0 \\
 0 & 0 & \mu_{zD_{3}d_{3}} & \bm0 & M_{B} & 0 & 0 & 0 & 0 \\
 0 & 0 & \mu_{z^{*}D_{3}^{1}d_{3}} & {\bm{\mu_{U}}^{d}}^\dagger & 0  & M_{D} & 0 & 0 & 0 \\
 0 & \mu_{z^* D_{2}^{1}d_{2}} & 0 & {\bm{\mu_{\tilde{V}}}^{d}}^\dagger & 0 & 0 & M_{D} & 0 & 0 \\
 0 & 0 & 0 & {\bm{\mu_{V}}^{d}}^\dagger & 0 & 0 & 0 & M_{D} & \mu_{zD_{1}^{1}D_{1}^{2}} \\
 \mu_{zD_{1}^{2}d_{1}} & 0 & 0 & \bm0 & 0 & 0 & 0 & \mu_{zD_{1}^{1}D_{1}^{2}}^{*} & M_{D}
\end{array}
\right),
\label{eq:app_mass_matrices}
\end{align}
where the entries in the rows and columns of the $\mathrm{SU}(2)_q$ doublets $\bm q$, $\bm U$ and $\bm D$ are to be understood as $2\times2$ blocks proportional to the identity ($\bm{y}_q^{u,d}$, $\bm{M_{U,D}}$) or as two-component flavour vectors ($\bm{\mu_{W,V}}^u$, $\bm{\mu_{U,\tilde V,V}}^d$). Each non-zero entry is one link of \cref{fig:benchmark_chains}: Higgs couplings in the rows of the SM doublets, VLF masses on the diagonal of the heavy blocks, and mass mixings off the diagonal. In the notation of \cref{sec:model_and_flavour}, $y^u_{q_3}\equiv y_t$.
The subscripts of the couplings $\mu$ already hint at the matching to the spurions of the low-energy theory performed in \cref{app:pert_diag}. Although in principle the lower off-diagonal couplings in the bottom-right block are independent of the upper off-diagonal ones, we relate them by complex conjugation by imposing the left--right exchange symmetry $F_L\leftrightarrow F_R$, under which the heavy mass matrix is Hermitian\footnote{Even when the lower off-diagonal elements are kept generic, they only contribute at subleading order to the Yukawa matrices and diagonals of the Wilson coefficients. However, they appear at higher spurion order, for example in the off-diagonal terms in $C^{(1,3)}_{Hq}$ or the corrections to the diagonals.}.
Furthermore, we set all VLF masses within a subblock to be equal for presentation purposes. Note that the $D_3$ is allocated its own mass $M_B$ since it only couples to the SM fields and therefore forms its own subblock.

The first step is integrating out the VLFs, which can be done sectorwise. For each sector, the Lagrangian for the VLFs can be written in terms of  $\Psi_{\alpha} = \begin{pmatrix}f_{\alpha} & F_{\alpha}\end{pmatrix}^T$ with $f_{\alpha}$ consisting of SM fields and $F_{\alpha}$ including the VLFs
\begin{equation}
\label{eq:VLF_Lagrangian}
\mathcal{L}_{\alpha} = \overline{\Psi_{\alpha}} i \slashed{D} \Psi_{\alpha} - \left( \begin{pmatrix}
\overline{f}_{L}^{\alpha}  &  \overline{F}^{\alpha}_{L}
\end{pmatrix} \begin{pmatrix}
0  & Y_{\alpha}^{L} \varphi \\
m_{\alpha}^{R} & M_{\alpha}
\end{pmatrix} \begin{pmatrix}
f^{\alpha}_{R} \\
{F}^{\alpha}_{R}
\end{pmatrix} + \text{h.c.}\right)
\end{equation}
where $Y^{L}_{\alpha}$, $M_{\alpha}$ and $m^{R}_{\alpha}$ are the Higgs couplings, the heavy mass matrix and the mass mixings of the VLFs with the RH SM fermions, respectively, and $\varphi \in \{H, \tilde{H}\}$. The SM-like Yukawa term for the top quark $y^u_{q_3} \bar{q}_3 \tilde{H} u_3$ is decoupled from the VLFs and hence does not contribute to the following discussion. Solving the equations of motion for the VLFs gives (for readability we omit the sector index $\alpha$ in the following)
\begin{align}
&F = - \mathcal{M}^{-1} \left( 1 + i \bar{\mathcal{M}}^{-1} \slashed{D} + \mathcal{O}(\mathcal{M}^{-2}) \right) (m^{R} P_{R} + (Y^{L} \varphi)^{\dagger} P_{L}) f
\end{align}
with $\mathcal{M} \coloneqq M P_{R} + M^{\dagger}P_{L}$, $\bar{\mathcal{M}} \coloneqq M^\dagger P_{R} + M P_{L}$ so that $\mathcal{M}^{-1} = M^{-1}P_{R} + (M^{\dagger})^{-1} P_{L}$ and $\bar{\mathcal{M}}^{-1} = (M^{\dagger})^{-1}P_{R} + M^{-1} P_{L}$.
Inserting this into \cref{eq:VLF_Lagrangian}, we obtain the dimension-four Yukawa contributions
\begin{equation}
\label{eq:Yukawa_terms}
\mathcal{L}_{\mathrm{Yukawa}} = (Y^{L}M^{-1}m^{R})_{ij} (\overline{f}_{L}^{i} H f_{R}^{j}) + \text{h.c.}\,,
\end{equation}
i.e.~$y_\textrm{eff}=-Y^LM^{-1}m^R$ in the convention $\mathcal L=-\bar f_L y_\textrm{eff}Hf_R$ of the main text, cf.~\cref{eq:yeff_general},
and a modification of the right-handed kinetic term of the same dimension
\begin{equation}
\mathcal{L}_{\mathrm{kin}} = \overline{f}_L i \slashed{D} f_L + \overline{f}_R i \slashed{D} f_R + \overline{f}_{R} {m^{R}}^{\dagger} (M^{\dagger})^{-1} i \slashed{D} M^{-1} m^{R} f_{R}.
\end{equation}
One can canonically normalise the kinetic term by redefining the SM fields
\begin{equation}
f_{R} \to A^{-1/2} f_R :=(\mathbb{1} + {m^{R}}^{\dagger} (M^{-1})^{\dagger} M^{-1} m^{R})^{-1/2} f_{R} \approx \left( \mathbb{1} - \frac{1}{2} {m^{R}}^{\dagger} (M^{-1})^{\dagger} M^{-1} m^{R} + \mathcal{O}(M^{-4}) \right) f_{R} \, .
\end{equation}
The redefinition of the RH SM fields modifies the Yukawas at subleading order in the spurion expansion
\begin{align}
    \mathcal{L}_{\mathrm{Yukawa}} &= (Y^{L}M^{-1}m^{R} A^{-1/2})_{ij} (\overline{f}_{L}^{i} H f_{R}^{j}) + \text{h.c.}\, \\
    &\approx \left[\left(Y^{L}M^{-1}m^{R}\right)_{ij}  - \frac{1}{2}\left(Y^{L}M^{-1}m^{R}(m^R)^\dagger (M^\dagger)^{-1}M^{-1}m^R\right)_{ij} \right](\overline{f}_{L}^{i} H f_{R}^{j}) + \text{h.c.}
\end{align}
which we will ignore for the proof of concept of reproducing the hierarchies in the CKM matrix.

For the dimension-six terms, we employ the equations of motion for the SM fields and the Fierz identity $T^{a}_{ij} T^{a}_{kl} = \frac{1}{2} \left( \delta_{il}\delta_{jk}-\frac{1}{N}\delta_{ij}\delta_{kl} \right)$ in order to match them to the Warsaw basis of Ref.~\cite{Grzadkowski:2010es}.
At tree level, this yields contributions to the operators $\mathcal O_{H\ell}^{(1)}$, $\mathcal O_{H\ell}^{(3)}$, $\mathcal O_{eH}$ for the leptons and $\mathcal O_{Hq}^{(1)}$, $\mathcal O_{Hq}^{(3)}$, $\mathcal O_{uH}$, $\mathcal O_{dH}$ for the quarks. The leading contributions of sector $\alpha$ to the singlet and triplet operators are
\begin{align}    [C_{Hq,\alpha}^{(1,3)}]_{ij} \propto \frac{1}{4} (Y_{\alpha}^{L} M_\alpha^{-1} (M_\alpha^{\dagger})^{-1} (Y_\alpha^{L})^{\dagger})_{ij}
\end{align}
so that the full Wilson coefficients read
\begin{align}    [C_{Hq}^{(1)}]_{ij} = -\frac{1}{4} \left[Y_{d}^{L} M_d^{-1} (M_d^{\dagger})^{-1} (Y_d^{L})^{\dagger} - Y_{u}^{L} M_u^{-1} (M_u^{\dagger})^{-1} (Y_u^{L})^{\dagger}\right]_{ij} \\
[C_{Hq}^{(3)}]_{ij} = -\frac{1}{4} \left[Y_{d}^{L} M_d^{-1} (M_d^{\dagger})^{-1} (Y_d^{L})^{\dagger} + Y_{u}^{L} M_u^{-1} (M_u^{\dagger})^{-1} (Y_u^{L})^{\dagger}\right]_{ij}.
\end{align}
See also Ref.~\cite{Palavric:2026vej} for a similar result. For the up-quark sector, where the Higgs enters as $\tilde H$, the coefficient of the singlet operator flips sign, $C^{(3)}_{Hq,u}=-C^{(1)}_{Hq,u}$, in agreement with \cref{eq:CHq_d,eq:CHq_u}. These expressions hold for a general heavy mass matrix $M$; for the Hermitian matrices adopted here, $M^\dagger=M$, and the combination $M^{-1}(M^\dagger)^{-1}$ reduces to $M^{-2}$.
The modified Yukawa operators $\mathcal O_{fH}$ arise at the same order from the terms with three Higgs fields, with coefficients
\begin{align}
C_{fH, \alpha} = \tfrac12 (Y_\alpha^L M_\alpha^{-1}(M_\alpha^{\dagger})^{-1}(Y_\alpha^{L})^{\dagger})\,y_{\rm{eff}}^\alpha A_\alpha^{-1/2}
\end{align}
i.e.~proportional to the effective Yukawa coupling itself, as anticipated in \cref{sec:matching_smeft}.

\subsection{Yukawa sector and perturbative diagonalisation}
\label{app:pert_diag}

As mentioned in \cref{app:integrating_out}, integrating out the heavy VLFs results in Yukawa terms which, at leading order, are given by \cref{eq:Yukawa_terms} for each sector. Writing the Yukawa matrices in terms of the couplings and VLF masses of \cref{eq:app_mass_matrices}, we can match this structure to the spurions of the low-energy MFP Lagrangian~\cite{Greljo:2025mwj}, i.e.~to the structures in \cref{eq:yuk_u,eq:yuk_d}, through
\begin{equation}
\begin{aligned}
z \sim \frac{\mu_{z}}{M},\quad z^* \sim \frac{\mu_{z^*}}{M} \sim \frac{\mu_z^*}{M}, \quad
\bm V \sim \frac{\bm{\mu_{V}}}{M}, \quad \bm W \sim \frac{\bm{\mu_W}}{M}, \quad \bm U \sim \frac{\bm{\mu_U}}{M}, \quad  \bm{\tilde{V}} \sim \frac{\bm{\mu_{\tilde{V}}}}{M} \sim i \sigma_2 \frac{\bm{\mu_V}^*}{M}.
\end{aligned}
\label{eq:app_spurion_dictionary}
\end{equation}
The exact relations are given only after fixing the relations between the spurions from the different sectors. For instance, the up and down sectors can in general have different $\bm{\mu_V}^u, \bm{\mu_V}^d$ which are related by a complex number $\bm{\mu_V}^u = c_1 \bm{\mu_V}^d$. The same can hold for $\bm\mu^d_V$ and $\bm{\mu_{\tilde V}}^d$ as well as for all $z$-couplings. This non-universality does not change the number of physical phases, just the number of physical moduli. The same holds for the VLF masses which for simplicity we took the same within each connected subblock of the mass matrices in the main text. However, in full generality these masses could be different, introducing a dependence of the MFP Yukawas on mass ratios and $z$-coupling ratios. This non-universality allows using the spurion benchmark in \cref{eq:app_spurion_benchmark}. Otherwise, in a fully universal model with equal $z$-couplings, VLF masses and $\bm V$ spurions, the relations between the UV parameters and the MFP Yukawas fix the $z$ spurion by the quark mass ratio $m_d/m_s$ and the benchmark in \cref{eq:app_spurion_benchmark} would have to be altered to fit this ratio.

In order to compute the CKM matrix and eigenvalues from the Yukawa matrices in \cref{eq:Yukawa_terms}, we use a perturbative singular value decomposition (SVD).
The perturbative SVD method applies consecutive rotations with small angles to the Yukawa matrices
\begin{align}
\label{eq:rotation_matrices}
& \hat{Y}^{d} = {L_{d}}^{\dagger} Y^{d} R_{d} = P^d_L R^{d} _{12}R^{d}_{23}R^{d}_{13}Y^{d}{\widetilde{R}^{d\, \dagger}_{13}}{\widetilde{R}^{d\, \dagger}_{23}}{\widetilde{R}^{d\,\dagger}_{12}} P^d_R \\  & \hat{Y}^{u} =  {L_{u}}^{\dagger} Y^{u} R_{u} = P^u_L R^{u}_{12} Y^{u}{\widetilde{R}^{u\, \dagger}_{12}} P^u_R
\end{align}
where $\hat{Y}^{u}$ and $\hat{Y}^{d}$ are diagonal.
We truncate the entries at linear order in the angles, each of which is chosen such that the off-diagonal entries $Y_{ij}$ and $Y_{ji}$ vanish after applying the rotation $R_{ij} \coloneqq R(\theta_{ij})$. For the up sector the angle $\alpha \approx -0.225$ is not small, so we do not expand in it and keep the full rotation $L_u$ given below. The matrices $P$ account for removing the phases so that $\hat{Y}^{u,d}$ are rendered positive and real.
Using the general form of the Yukawa matrices from \cref{eq:yuk_u,eq:yuk_d} with the unphysical signs rotated away
\begin{align}
    y^d_\textrm{eff} = \begin{pmatrix}
    \hat{y}^d_1 \bm{V}z^2& -\hat{y}_2^d z^*\tilde{\bm{V}} &  \hat{y}_q^d z^*\bm{U}
    \\
       0&0 & z \hat{y}^d_3
    \end{pmatrix} ,\quad y^u_{\rm{eff}} = \begin{pmatrix}
        \hat{y}^u_1\, \bm{V} z^2 & \hat{y}^u_2\,\bm{W} & \bm0 \\
        0 & 0 & \hat{y}^u_3
    \end{pmatrix} ,
\end{align}
where we kept the minus for $\hat{y}^d_2$ to compensate the minus from $\bm{\tilde{V}} = i \sigma_2 \bm V^*$.
The rotation matrices $L_u$ and $L_d$ can be expressed in terms of the spurions as

\begin{align}
\label{eq:app_rotation_matrices}
L_u &\approx
\begin{pmatrix}
\dfrac{W_2}{\sqrt{W_1^2+W_2^2}} & \dfrac{W_1}{\sqrt{W_1^2+W_2^2}} & 0 \\[2ex]
-\dfrac{W_1}{\sqrt{W_1^2+W_2^2}} & \dfrac{W_2}{\sqrt{W_1^2+W_2^2}} & 0 \\[2ex]
0 & 0 & 1
\end{pmatrix}
=
\begin{pmatrix}
\cos\alpha & \sin\alpha & 0 \\
-\sin\alpha & \cos\alpha & 0 \\
0 & 0 & 1
\end{pmatrix} \\
L_d &\approx
\begin{pmatrix}
1 & -\dfrac{(\hat{y}^d_q)^2}{(\hat{y}^d_3)^2}\, U_1 U_2^* & \dfrac{\hat{y}^d_q}{\hat{y}^d_3}\, U_1 \\[2ex]
0 & 1 & \dfrac{\hat{y}^d_q}{\hat{y}^d_3}\, U_2 \\[2ex]
-\dfrac{\hat{y}^d_q}{\hat{y}^d_3}\, U_1^* & -\dfrac{\hat{y}^d_q}{\hat{y}^d_3}\, U_2^* & 1
\end{pmatrix}
= \begin{pmatrix}
1 & -\dfrac{u^2 (\hat{y}^d_q)^2}{(\hat{y}^d_3)^2}\, s_\beta c_\beta\, e^{i\phi} & \dfrac{u\, \hat{y}^d_q}{\hat{y}^d_3}\, s_\beta\, e^{i\phi} \\[2ex]
0 & 1 & \dfrac{u\, \hat{y}^d_q}{\hat{y}^d_3}\, c_\beta \\[2ex]
-\dfrac{u\, \hat{y}^d_q}{\hat{y}^d_3}\, s_\beta\, e^{-i\phi} & -\dfrac{u\, \hat{y}^d_q}{\hat{y}^d_3}\, c_\beta & 1
\end{pmatrix}.
\end{align}

Note that the up rotation only includes the Cabibbo angle $\alpha \approx -\lambda$ and the down rotation is generated solely by the $\bm U$ spurion due to the alignment in the light block.
The CKM matrix is then given by
\begin{equation}
\begin{aligned}
V_{\mathrm{CKM}} = {L_{u}}^{\dagger} L_{d} \simeq{}&
\begin{pmatrix}
\dfrac{W_2}{|W|} & -\dfrac{1}{|W|}\left( W_1 + \dfrac{(\hat{y}^d_q)^2}{(\hat{y}^d_3)^2}\, U_1 U_2^*\, W_2 \right) & \dfrac{\hat{y}^d_q}{\hat{y}^d_3}\, \dfrac{U_1 W_2 - U_2 W_1}{|W|} \\[3ex]
\dfrac{W_1}{|W|} & \dfrac{1}{|W|}\left( W_2 - \dfrac{(\hat{y}^d_q)^2}{(\hat{y}^d_3)^2}\, U_1 U_2^*\, W_1 \right) & \dfrac{\hat{y}^d_q}{\hat{y}^d_3}\, \dfrac{U_1 W_1 + U_2 W_2}{|W|} \\[3ex]
-\dfrac{\hat{y}^d_q}{\hat{y}^d_3}\, U_1^* & -\dfrac{\hat{y}^d_q}{\hat{y}^d_3}\, U_2^* & 1
\end{pmatrix} \\[2ex]
\approx {}&
\begin{pmatrix}
\cos\alpha & -\sin\alpha & x\left(e^{i\phi}\sin\beta\cos\alpha-\cos\beta\sin\alpha\right)\\[1ex]
\sin\alpha & \cos\alpha & x\left(\cos\beta\cos\alpha+e^{i\phi}\sin\beta\sin\alpha\right)\\[1ex]
-x\,e^{-i\phi}\sin\beta & -x\cos\beta & 1
\end{pmatrix},
\end{aligned}
\end{equation}
with $|W| \equiv \sqrt{W_1^2+W_2^2}$ and $x \equiv \frac{u\, \hat{y}^d_{q}}{\hat{y}^d_{3}}$. In order to generate the CKM hierarchy and quark masses correctly, we use the approximate $\mathcal{O}(1)$ low-energy Yukawa expressions from \cite{Greljo:2025mwj}
\begin{align}
  \hat{y}^u_1 &\sim \frac{y_u}{v\,z^2} \simeq 1.03\,, &
  \hat{y}^u_2 &\sim \frac{y_c}{w} \simeq 0.51\,, &
  \hat{y}^u_3 &\sim y_t \simeq 0.84\,, \nonumber\\[4pt]
  \hat{y}^d_1 &\sim \frac{y_d}{v\,z^2} \simeq 2.19\,, &
  \hat{y}^d_2 &\sim \frac{y_s}{v\,z} \simeq 0.89\,, &
  \hat{y}^d_3 &\sim \frac{y_b}{z} \simeq 0.68\,, \nonumber\\[4pt]
  \hat{y}^e_1 &\sim \frac{y_e}{z^3} \simeq 0.37\,, &
  \hat{y}^e_2 &\sim \frac{y_\mu}{z^2} \simeq 1.55\,, &
  \hat{y}^e_3 &\sim \frac{y_\tau}{z} \simeq 0.52\,, \label{eq:MFPcoeffs}
\end{align}
changed to our example scale of $\mu = 2$ TeV with the quark masses at this scale obtained from \texttt{wilson} \cite{Aebischer:2018bkb}. The spurion benchmark values follow Ref.~\cite{Greljo:2025mwj}
\begin{equation}
\label{eq:app_spurion_benchmark}
  \bm V = v\begin{pmatrix}1\\0\end{pmatrix},\qquad
  \bm W = w\begin{pmatrix}\sin\alpha\\ \cos\alpha\end{pmatrix},\qquad
  \bm U = u\begin{pmatrix}\sin\beta\,e^{i\phi}\\ \cos\beta\end{pmatrix},\qquad
  \bm{\tilde{V}} = i\sigma_2 \bm V^*\,,
\end{equation}
with the values $z=0.02$, $v=0.015$, $w=0.006$, $u=0.04$. The light--heavy mixing coefficient and the spurion angles are fixed by the CKM matrix
\begin{equation}
  \hat{y}^d_q \simeq \frac{A\lambda^2\,\hat{y}^d_3}{u} \sim \frac{A\lambda^2\,y_b}{u\,z} \simeq 0.71\,,\qquad
  \alpha \simeq -\lambda\,,\qquad
  \tan\phi \simeq \frac{\eta}{1-\rho}\,,\qquad
  \beta \simeq -\lambda\,\frac{1-\rho}{\cos\phi}.
  \label{eq:MFPangles}
\end{equation}
For the Wolfenstein parameters we use the same values as in the reference, given by $\lambda = 0.225$,  $A= 0.826$, $\rho= 0.163$ and $\eta = 0.357$ from \cite{Workman:2022ynf}.

\subsection{Flavour-violating effects}
\label{app:flavour_obs}

In this subsection we provide more details on the analysis of the flavour observables presented in \cref{sec:flavour_direct}.

At tree level and leading order in the spurions, the matching of \cref{app:integrating_out} gives, in the interaction basis of \cref{eq:app_mass_matrices},
\begin{equation}
C_{H\ell}^{(1)}=C_{H\ell}^{(3)}=-\tfrac14\,\mathrm{diag}(c_e,c_\mu,c_\tau)\,,\qquad
C_D=-\tfrac12\,\mathrm{diag}(c_D,c_D,c_B)\,,\qquad
C_U=+\tfrac12\,\mathrm{diag}(c_U,c_U,0)\,,
\label{eq:app_C_interaction_basis}
\end{equation}
with $c_F=|y_F|^2/M_F^2$ and the basis $C_D = C^{(1)}_{Hq} + C^{(3)}_{Hq}$ and $C_U = C^{(1)}_{Hq} - C^{(3)}_{Hq}$, while $C_{Hd}$, $C_{Hu}$ and $C_{He}$ vanish. On top of that, at order $\mathcal{O}(z^2)$ in the spurion expansion, we obtain corrections to the diagonals and off-diagonal terms, as given in \cref{eq:WC_off-diagonals}. They originate from mixings between the doublet VLFs $\bm U, \bm D$ and a singlet VLF  or between a singlet VLF and a light right-handed SM field in the chain, which allows for a flavour change.
The general form of the Wilson coefficients in the interaction basis, without assuming universal masses or mixings, is then given by
\begin{align}
    [C_U]_{ab} &= \frac{c_U}{2}\left[\delta_{ab}
 - 2\,\frac{(\mu^u_{\bm{W}})_a(\mu^u_{\bm{W}})_b^*}{M_{\bm U}^2}
 + \frac{(\mu^u_{\bm{V}})_a(\mu^u_{\bm{V}})_b^*}{M_{U_{11}}^2}\left(\frac{2 M_{U_{11}}}{M_{\bm U} } + 1 \right)\right],
\qquad a,b = 1,2, \\
[C_D]_{ab} &= -\frac{c_D}{2}\left[\delta_{ab}
 + \sum_{\bm{X}=\bm{V},\tilde{\bm{V}},\bm{U}} \frac{(\mu^d_{\bm{X}})_a(\mu^d_{\bm{X}})_b^*}{M_{(\bm{X})}^2}\left(\frac{2M_{(\bm{X})}}{M_{\bm D}} + 1\right)\right],
\qquad a,b = 1,2, \\
[C_D]_{33} &= -\frac{c_{B}}{2}\left(1 - 2\,\frac{|\mu^d_{z D_3 d_3}|^2}{M_{B}^2}\right),
\end{align}
where $M_{B}$ is the mass of the singlet $D_3$ to which $q_3$ couples through $y_{q3}^d$. $M_{U_{11}}$ is the mass of the singlet $U_1^1$ into which $\bm{U}$ mixes through $\mu^u_{\bm V}$. Similarly, $M_{(\bm X)}$ denotes the mass of the singlet into which $\bm{D}$ mixes through $\mu^d_{\bm X}$: $M_{(\bm V)}\equiv M_{D_{11}}$ for $D_1^1$, $M_{(\tilde{\bm V})}\equiv M_{D_{21}}$ for $D_2^1$ and $M_{(\bm U)}\equiv M_{D_{31}}$ for $D_3^1$. These are the general expressions of \cref{eq:WC_off-diagonals}. We see that for universal masses within a subblock, the prefactors of the spurion suppressed terms simplify. If we also set $\bm\mu^u_{\bm V} = \bm\mu^d_{\bm V} \equiv \bm\mu_{\bm V}$ and relate $\bm\mu^d_{\bm{\tilde V}} = i \sigma_2 \bm\mu_{\bm V}^*$, we can define the spurion vectors universally. The only spurion that is left non-universal on purpose is the $z$ in order to leave enough freedom to reproduce the correct quark mass ratios with the benchmark in \cref{eq:app_spurion_benchmark}. If we choose to define $z = \mu_{z D_3 d_3}/M_B$ by the coupling of $D_3$ with $d_3$, we obtain the simple case $[C_D]_{33} = -c_B/2 \times [1 - 2 z^2]$ as discussed in the main text in \cref{sec:flavour_direct}. The simplified form shows that every heavy--heavy chain with mixings adds a term $\mp 3 X_a X_b^*$ to the 1,2-block where $X$ is the corresponding spurion and every heavy--light chain adds $\pm 2 X_i X_j^*$ in $[C_U]_{ab}$ and $[C_D]_{33}$.

These coefficients must be rotated to the mass basis in order to match them to the low-energy flavour-changing operators. Applying the rotation matrices in \cref{eq:app_rotation_matrices} adds the rotated diagonal contributions to the off-diagonals. In the down basis they always appear proportional to $(c_D-c_B)$ or $c_U$ for $C_D$ and $C_U$ respectively. This shows that for the down quarks, approximately equal masses $M_D$ and $M_B$ can decrease the contributions to FCNCs if they dominate over the non-rotation generated terms. However, for the calculation of the bounds in \cref{fig:flavour_bounds_projections}, we turn on one off-diagonal coefficient in the down basis at a time and extract the bound on each chain by setting the other parameters to zero. This method disregards the possible cancellations mentioned above and should therefore be interpreted as an estimate only.

\paragraph*{Current bounds.}
For the current bounds in \cref{fig:flavour_bounds_projections} we use a regular chi-squared function comparing the NP value of each observable to its measured value and combining the SM uncertainty with the experimental one in quadrature to obtain the total error. The SM errors are estimated by \texttt{flavio} \cite{Straub:2018kue} with a Monte Carlo procedure or taken from the literature. We use the NP predictions provided by \texttt{flavio}. The experimental and SM values used for the bounds are listed in \cref{tab:flavour_refs_current}. Some errors were symmetrised for convenience and are labelled as such. We analyse a selected set of observables including $\Delta F = 1$ and $\Delta F = 2$ processes as expanded on below.

The current bounds mainly use the experimental values from PDG 2026 \cite{ParticleDataGroup:2026mpi} except for the branching ratios $\mathrm{BR}(B_s\to\mu^+\mu^-)$ and $\mathrm{BR}(B^0\to\mu^+\mu^-)$ for which we use the LHCb combined fit results for run 1+2+3 \cite{LHCb:2026oio}, which are compatible with the SM.

For $\mathrm{BR}(K^+\to\pi^+\nu\bar{\nu})$ the updated measurement from the NA62 experiment is included~\cite{NA62:2026rwr}. One caveat is that for $K_L\to\mu^+\mu^-$ we use the SM prediction and uncertainty of Ref.~\cite{Hoferichter:2023wiy}, while the NP shift is computed with \texttt{flavio}, which sets the dispersive long-distance contribution to zero at its central value (with the two sign solutions of Ref.~\cite{Chobanova:2017rkj} treated as an uncertainty).

\paragraph*{Future projections.}
The projected bounds assume SM-centred measurements and apply the prospective values in \cref{tab:flavour_refs_projections} which either ameliorate the experimental or the theoretical uncertainties, depending on the expected improvements in the corresponding observable.

Observables that are expected to benefit from improved experimental errors include $\mathrm{BR}(B_s\to\mu^+\mu^-)$ and $\mathrm{BR}(B^0\to\mu^+\mu^-)$. They are foreseen to be most precisely measured by LHCb \cite{ATLAS:2025lrr}, which we use as the projection.
Another experimental improvement is expected for $\mathrm{BR}(K^+\to\pi^+\nu\bar{\nu})$ with the evaluation of the full dataset of the NA62 experiment, resulting in a relative uncertainty of 15\% \cite{Anzivino:2023bhp, Aebischer:2025mwl}. We apply this to the current measurement central value $9.60\times 10^{-11}$ \cite{NA62:2026rwr} to obtain the future uncertainty.

For all other observables we keep the current experimental uncertainties and focus on the improvements from the theory side.
An example is the SM value of the kaon decay observable $\epsilon^\prime/\epsilon$ which is calculated from lattice QCD \cite{RBC:2020kdj} and expected to improve due to a reduced theory error in $\epsilon^\prime$ of 10\% in the future \cite{Blum:2022wsz}\footnote{We neglect the error in $\epsilon$ and thus treat the relative error in $\epsilon^\prime$ as the error of the ratio directly, multiplying it with the current SM value.}.
The meson mixing observable bounds for $\epsilon_K$, $\Delta M_d$ and $\Delta M_s$ are determined from the current ones by applying the same improvement factors of around $3$ given by Fig.~5.15 in the Physics Briefing Book \cite{deBlas:2025gyz}. They are mainly sourced by more precise CKM determinations and lattice QCD.
For the branching ratio $\mathrm{BR}(K_L\to\mu^+\mu^-)$, we estimate the theoretical uncertainty by linearly propagating the parametric uncertainties of the CKM parameters $\sin 2\beta$ and $|V_{cb}|$ from Belle II and LHCb \cite{ATLAS:2025lrr} and assume that the error in the real part of the long-distance amplitude $\mathrm{Re}(\mathcal{A}_{\rm LD})$ will be able to match experimental uncertainties in the future \cite{Hoferichter:2023wiy}.

\paragraph*{Additional contributions.} In the calculation of the current bounds for the $\Delta F = 2$ observables, only the RG running induced by the tree-level $C^{(1,3)}_{Hq}$ operators is included in \cref{fig:flavour_bounds_projections}. However, there are also contributions from the one-loop matching of the UV model to the SMEFT and from the dimension-eight operator including two flavour-changing $Z$-insertions to the $\Delta F = 2$ observables as well as SMEFT to LEFT one-loop matching, which are not included.
For the $D$ and $B$ chains, the tree-level coefficients $C_{Hq}^{(1)}=C_{Hq}^{(3)}$ do not contribute to $\Delta F=2$ processes at linear order, since $C_D$ only mixes into $C^{(1)}_{qq}-C^{(3)}_{qq}$. The leading contribution then arises from the one-loop matching of the heavy-quark boxes onto $C^{(1,3)}_{qq}$~\cite{Ishiwata:2015cga}.
The obtained bounds from $\epsilon_K$ and $\Delta M_{d,s}$ are far below the bounds from $\Delta F=1$ observables such as $B_s\to\mu^+\mu^-$.
For the $U$ chain, $\Delta F=2$ operators are generated through the $y_t^2$ running of $C_U$ into $C^{(1)}_{qq}+C^{(3)}_{qq}$~\cite{Bobeth:2017xry}. The finite one-loop matching contributions, from the heavy-$U$ box and the non-logarithmic part of the heavy-$U$–top box, are of comparable size but of opposite sign. Effectively they replace $\ln(M_U^2/\mu^2)\to\ln(M_U^2/\mu^2)-\tfrac32-(y_q^u)^2/(2y_t^2)$, where $\mu$ is the renormalisation scale, and thereby reduce the effect of the running. The meson-mixing bounds on the $U$ chain obtained from the running alone are therefore optimistically tight estimates.

At second order in $C_{Hq}$, which is formally of dimension eight, tree-level $Z$ exchange with two flavour-changing $Z$ couplings generates $\Delta F=2$ operators at the electroweak scale. This affects only the $D$ and $B$ chains, since the $U$ chain does not modify the $Z\bar dd$ couplings. It yields $M/|y|\simeq1$–$1.4$ TeV from $\epsilon_K$ and $\Delta M_{d,s}$, which is still not competitive with the bounds from $B_s\to\mu^+\mu^-$ in the multi-TeV range.

In addition, one can investigate the effect of one-loop UV to SMEFT matching for the dipole operators $C_{fV}$ with $f \in \{e,u,d\}$ and $V \in \{B, G, W\}$. As discussed in the main text, there are no LFV transitions in the lepton sector and no CP-violating phases due to the disconnected, open chains for the three generations. For the quarks, the generated one-loop diagrams contributing to the dipole operators always require a Yukawa insertion to flip the chirality in the chain, which leads to a spurion suppression. We did not obtain any significant bounds from the dipole observables. The strongest constraint comes from $b \to s \gamma$ transitions for the $\bm U$ chain but lies below the bounds from the meson mixing observables.

\begin{table}[tb]
    \centering\small\setlength{\tabcolsep}{4pt}
    \begin{tabular}{llll}
\hline
$[C_{U,D}]_{ij}$ & Observable & Exp. value & SM value \\
\hline
$[C_D]_{12}$ & $\varepsilon'/\varepsilon$ & $(16.7 \pm 2.3)\times 10^{-4}$ \cite{ParticleDataGroup:2026mpi} & $(21.7 \pm 8.4)\times 10^{-4}$ \cite{RBC:2020kdj} \\
$[C_D]_{12}$ & $\mathcal{B}(K_L\to\mu^+\mu^-)$ & $(6.84 \pm 0.11)\times 10^{-9}$ \cite{ParticleDataGroup:2026mpi} & $(7.44 \pm 0.38)\times 10^{-9}$ (sym.) \cite{Hoferichter:2023wiy} \\
$[C_D]_{12}$ & $\mathcal{B}(K^+\to\pi^+\nu\bar\nu)$ & $(9.60 \pm 1.85)\times 10^{-11}$ (sym.) \cite{NA62:2026rwr} & $(8.65 \pm 0.42)\times 10^{-11}$ \cite{Buras:2026vbp} \\
$[C_D]_{13}$ & $\mathcal{B}(B^0\to\mu^+\mu^-)$ & $(1.14 \pm 0.41)\times 10^{-10}$ (sym.) \cite{LHCb:2026oio} & $(1.03 \pm 0.05)\times 10^{-10}$ (sym.) \cite{Beneke:2019slt} \\
$[C_D]_{23}$ & $\mathcal{B}(B_s\to\mu^+\mu^-)$ & $(3.25 \pm 0.28)\times 10^{-9}$ (sym.) \cite{LHCb:2026oio} & $(3.66 \pm 0.14)\times 10^{-9}$ (sym.) \cite{Beneke:2019slt} \\
$[C_U]_{12}$ & $\varepsilon_K$ & $(2.229 \pm 0.012)\times 10^{-3}$ \cite{ParticleDataGroup:2026mpi} & $(2.18 \pm 0.15)\times 10^{-3}$
\cite{Buras:2024mnq} \\
$[C_U]_{13}$ & $\Delta M_d$ [ps$^{-1}$] & $(0.5068 \pm 0.0019)$ \cite{ParticleDataGroup:2026mpi} & $(0.533 \pm 0.029)$ (sym.) \cite{DiLuzio:2019jyq} \\
$[C_U]_{23}$ & $\Delta M_s$ [ps$^{-1}$] & $(17.766 \pm 0.006)$ \cite{ParticleDataGroup:2026mpi} & $(18.40 \pm 0.95)$ (sym.) \cite{DiLuzio:2019jyq} \\
\hline
\end{tabular}
    \caption{Numbers used for the flavour bounds in \cref{fig:flavour_bounds_projections} for each off-diagonal turned on one at a time with selected observables. All coefficients are probed in the down basis.}
    \label{tab:flavour_refs_current}
\end{table}

\begin{table}[tb]
    \centering\small\setlength{\tabcolsep}{4pt}
    \begin{tabular}{lllll}
\hline
$[C_{U,D}]_{ij}$ & Observable & SM-centred & $\sigma_{\rm exp}$ & $\sigma_{\rm th}$ \\
\hline
$[C_D]_{12}$ & $\varepsilon'/\varepsilon$ & $21.7\times 10^{-4}$ \cite{RBC:2020kdj} & $2.3\times 10^{-4}$ \cite{ParticleDataGroup:2026mpi} &
\cellcolor[HTML]{DCF0DC} 10\% \cite{Blum:2022wsz} $\times (21.7 \times 10^{-4})$ \cite{RBC:2020kdj} \\
$[C_D]_{12}$ & $\mathcal{B}(K_L\to\mu^+\mu^-)$ & $7.44\times 10^{-9}$ \cite{Hoferichter:2023wiy} & $0.11\times 10^{-9}$ \cite{ParticleDataGroup:2026mpi} & \cellcolor[HTML]{DCF0DC} $0.23\times 10^{-9}$ (see text) \\
$[C_D]_{12}$ & $\mathcal{B}(K^+\to\pi^+\nu\bar\nu)$ & $8.65\times 10^{-11}$ \cite{Buras:2026vbp} &
\cellcolor[HTML]{DCF0DC} 15\% \cite{Anzivino:2023bhp, Aebischer:2025mwl} $\times (9.60\times 10^{-11})$ \cite{NA62:2026rwr}  & $0.42\times 10^{-11}$ \cite{Buras:2026vbp} \\
$[C_D]_{13}$ & $\mathcal{B}(B^0\to\mu^+\mu^-)$ & $1.03\times 10^{-10}$ \cite{Beneke:2019slt} & \cellcolor[HTML]{DCF0DC} $0.12\times 10^{-10}$ \cite{ATLAS:2025lrr} & $0.05\times 10^{-10}$ \cite{Beneke:2019slt} \\
$[C_D]_{23}$ & $\mathcal{B}(B_s\to\mu^+\mu^-)$ & $3.66\times 10^{-9}$ \cite{Beneke:2019slt} & \cellcolor[HTML]{DCF0DC} $0.16\times 10^{-9}$ \cite{ATLAS:2025lrr} & $0.14\times 10^{-9}$ \cite{Beneke:2019slt} \\
\hline
$[C_U]_{12}$ & $\varepsilon_K$ &  \multicolumn{3}{c}{\cellcolor[HTML]{DCF0DC} current bound on $M$ $\times\,3.29$ \cite{deBlas:2025gyz}} \\
$[C_U]_{13}$ & $\Delta M_d$ &  \multicolumn{3}{c}{\cellcolor[HTML]{DCF0DC} current bound on $M$ $\times\,3.35$ \cite{deBlas:2025gyz}} \\
$[C_U]_{23}$ & $\Delta M_s$ &  \multicolumn{3}{c}{\cellcolor[HTML]{DCF0DC} current bound on $M$ $\times\,3.26$ \cite{deBlas:2025gyz}} \\
\hline
\end{tabular}
    \caption{Numbers used for the projected flavour bounds in \cref{fig:flavour_bounds_projections} where prospective errors are marked in green. We assume either experimental improvement (see $\mathrm{BR}(K^+\to\pi^+\nu\bar{\nu})$, $\mathrm{BR}(B^0 \to\mu^+\mu^-)$ and $\mathrm{BR}(B_s \to\mu^+\mu^-)$) or theoretical improvement (as for $\mathrm{BR}(K_L\to\mu^+\mu^-)$ and $\epsilon^\prime/\epsilon$). The current bounds from the meson mixing observables ($\epsilon_K$, $\Delta M_d$ and $\Delta M_s$) are multiplied by the improvement factor anticipated in Fig.~5.15 of Ref.~\cite{deBlas:2025gyz}.%
    }
    \label{tab:flavour_refs_projections}
\end{table}

\section{Details on the EWPO analysis}
\label{app:EWPO_analysis}
In this section we provide the details on the FCC-ee EWPO fit introduced in \cref{sec:tera-Z_projections}. The observables included in the fit are
\begin{equation}
    \left\{ \Gamma_Z,\, \sigma_{\mathrm{had}}^0,\, R_e^0,\, R_\mu^0,\, R_\tau^0,\, R_b^0,\, R_c^0,\, R_s^0,\, A_{\mathrm{FB}}^{0,e},\, A_{\mathrm{FB}}^{0,\mu},\, A_{\mathrm{FB}}^{0,\tau},\, A_{\mathrm{FB}}^{0,b},\, A_{\mathrm{FB}}^{0,c},\, A_{\mathrm{FB}}^{0,s},\, \mathcal{A}^{\tau\rm{pol}}_e, \mathcal{A}^{\tau\rm{pol}}_\tau,\, m_W,\, \Gamma_W \right\}.
    \label{eq:app_EWPO_list}
\end{equation}
These measurements correspond to the $Z$ line shape and leptonic forward--backward asymmetries, the leptonic asymmetry measurements, the heavy-flavour observables and the $W$ mass and width. The leptonic asymmetries $\mathcal A^{\tau\rm{pol}}_\tau$ and $\mathcal{A}^{\tau\rm{pol}}_e$ are measured through the $\tau$ polarisation. All other asymmetries such as $\mathcal{A}_{\mu,c,b,s}$ are not included because, without a polarised beam, they must be extracted from the measurements of the corresponding forward--backward asymmetries. Hence they are fully correlated with the latter observables and do not add any new information to the fit.
As shown in \cref{app:integrating_out}, the Wilson coefficients $C_{H q}^{(1,3)}$ and $C_{H\ell}^{(1,3)}$ are given at leading order by
\begin{align}
    &[C_{H\ell}^{(1,3)}]_{11} = - \frac{1}{4} c_e, \quad &[C_{H\ell}^{(1,3)}]_{22} &= - \frac{1}{4} c_\mu, \quad &[C_{H\ell}^{(1,3)}]_{33} = - \frac{1}{4} c_\tau, \nonumber\\
    &[C_{Hq}^{(1)}]_{11, 22} = - \frac{1}{4} (c_{D} - c_{U}), \quad &[C_{Hq}^{(1)}]_{33} &= - \frac{1}{4} c_{B}, \nonumber\\
    &[C_{Hq}^{(3)}]_{11, 22} = - \frac{1}{4} (c_{D} + c_{U}), \quad &[C_{Hq}^{(3)}]_{33} &= - \frac{1}{4} c_{B} \, , \label{eq:app_EWPO_fit_params}
\end{align}
with $c_{e,\mu,\tau}=|y_{\ell_{1,2,3}}|^2/M_{e,\mu,\tau}^2$, $c_U=|y_q^u|^2/M_U^2$, $c_D=|y_q^d|^2/M_D^2$ and $c_{B}=|y_{q_3}^d|^2/M_B^2$, all positive by construction.
The fit parameters are $\{ c_{e}, c_{\mu}, c_{\tau}, c_{U}, c_{D}, c_{B} \}$, where $c_{U}$ and $c_{D}$ are common to the first two generations due to the $\mathrm{SU}(2)_q$ flavour symmetry.
Note that we would technically also have the spurion-suppressed or rotation-suppressed diagonal and off-diagonal components. However, these are subleading to the diagonal pieces above, which is why we omit any off-diagonal entries and diagonal corrections in the EWPO fit. Instead, off-diagonal contributions are probed by the flavour observables as detailed in \cref{app:flavour_obs}.
The Wilson coefficients are defined at the matching scale $\mu_0=50\,\mathrm{TeV}$ and evolved to the $Z$ pole with the one-loop RGE implemented in \texttt{wilson}~\cite{Aebischer:2018bkb}.\footnote{Since the \texttt{Warsaw} basis of the WCxf convention~\cite{Aebischer:2017ugx} used by \texttt{wilson} is defined to have diagonal down-quark and lepton Yukawa matrices, we need to rotate the Wilson coefficients to the down basis.}

\subsection{Tera-\texorpdfstring{$Z$}{Z} projection methodology}
\label{app:tera-Z_details}

To perform the analysis for a Tera-$Z$ collider, we build a $\chi^{2}$ function using the SMEFT implementations of the observables from \texttt{flavio}~\cite{Straub:2018kue}.
We tested two different approaches, juxtaposing the full $\chi^{2}$ and a quadratic approximation in the Wilson coefficients using the Fisher matrix. For an observable $X$, we write its prediction as the sum of the SM value $\hat{X}$ and the SMEFT correction $\delta X$,
\begin{equation}
    \bar{X}(C) = \hat{X} + \delta X(C)
\end{equation}
where $C$ are the parameters entering the NP shifts of the observable such that for $C=0$ we obtain the SM value $\bar{X}(C=0) = \hat{X}$. We assume the observables to be Gaussian with total uncertainty $\sigma_X = \sqrt{\sigma_\mathrm{exp}^2 + \sigma_{\mathrm{th}}^2}$. We further assume, unlike for LEP, that the observables are uncorrelated. In these two approaches the global $\chi^{2}$ functions are then given by
\begin{align}
\chi^{2}_{\mathrm{exact}}(C) = \sum_{X\in \mathcal{O}_\mathrm{EW}^{\rm FCC}} \left( \frac{\delta X(C) }{\sigma_{X}} \right)^{2}, \quad
\chi^{2}_{\mathrm{quad}}(\bm C) = \bm C^T \bm H \bm C\, ,
\end{align}
where $\bm C$ stands for the vector of the parameters and $\bm H$ is the Fisher matrix given by
\begin{equation}
    H_{ij} = \sum_{X\in \mathcal{O}_\mathrm{EW}^{\rm FCC}} \frac{1}{\sigma_X^2}\frac{ \partial (\delta X )}{ \partial c_{i} } \frac{ \partial (\delta X) }{ \partial c_{j} }.
\end{equation}
For our analysis the expressions for the partial derivatives are approximated by a numerical derivative. Since the EWPO are approximately linear in the Wilson coefficients at the order considered, one can verify that the two forms of the above $\chi^2$ agree in the proximity of $C = 0$, which is the region we analyse. This justifies the use of the computationally more efficient quadratic approximation for calculating the results shown in \cref{fig:FCCee_fit}.

\begin{figure}[tb]
    \centering
    \includegraphics[width=\linewidth]{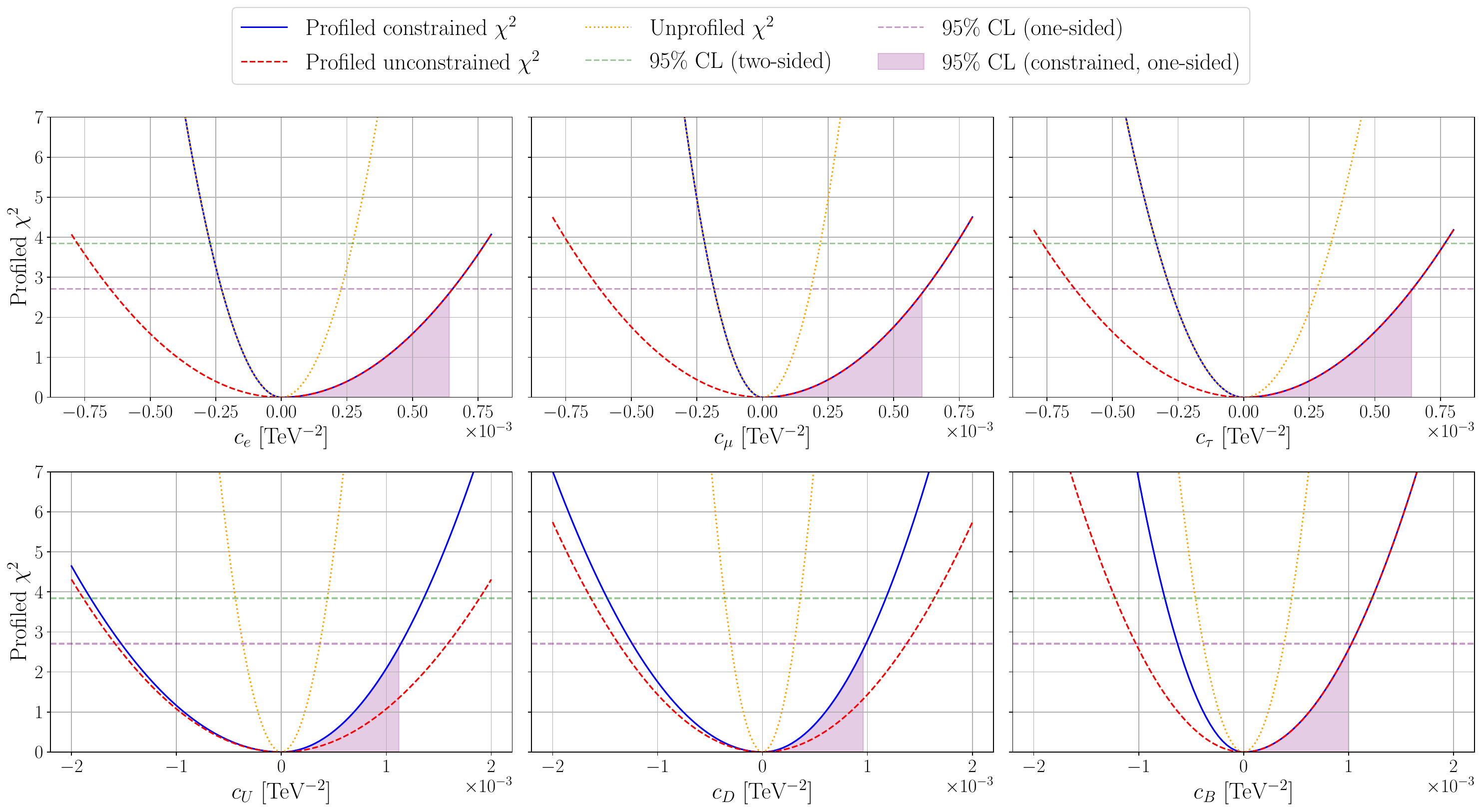}
    \caption{Profiles of $\Delta\chi^2$ as a function of each of the six fit parameters $c_F$ for the aggressive uncertainty scenario S2. Blue solid lines: profiled over the remaining parameters subject to the positivity constraints $c_F \geq 0$. Red dashed lines: profiled without positivity constraints. Orange dotted lines: unprofiled, with the remaining parameters set to zero. Purple (green) dashed lines mark $\Delta \chi^2 = 2.71$ (3.84). Purple shaded regions indicate the $95\%$~CL intervals of the profiled, constrained curve.}
    \label{fig:1d-scan}
\end{figure}

\paragraph*{1d-scans.} In order to obtain the bounds on the VLF masses in \cref{fig:FCCee_fit}, we perform one-dimensional scans over each of the parameters, comparing three different cases. The
\textit{unprofiled} case looks at the allowed 95\% CL for one parameter while setting the other parameters to zero. This corresponds to ignoring all possible cancellation or correlation effects from the residual parameters. In contrast to this stands the \textit{profiled} method which allows the other parameters to float for each scanned point. Thus it is possible to find a more optimal solution (here we use the \texttt{iminuit} package~\cite{iminuit}), for instance by leveraging cancellations between parameters that reduce the bound. This analysis is conducted once with and once without imposing the positivity constraint.
\Cref{fig:1d-scan} illustrates this behaviour for the aggressive scenario S2. As seen well for the $c_U$ parameter, the bounds from the profiled curves (\textit{blue/red}) are relaxed with respect to the unprofiled case (\textit{orange}). It also displays that the physical positivity constraint effectively tightens the bound compared to leaving the parameters unconstrained. This can be attributed to the fact that fixing the signs rules out cancellations from those combinations that require the parameters to have opposite signs.
In addition, the $\Delta \chi^2$ threshold changes when restricting the domain of the parameters to the positive axis. Instead of the two-sided $\Delta \chi^2 \approx 3.84$ for one degree of freedom, we need to apply the one-sided limit $\Delta \chi^2\approx 2.71$, leading to an additional tightening in the profiled bounds.
The bounds quoted in the main text are the $95\%$~CL upper limits on each $c_F$ obtained from the constrained, profiled (hatched bars in \cref{fig:FCCee_fit}) and unprofiled (unhatched bars) curves, translated into lower limits on $M_F/|y_F|=c_F^{-1/2}$. Although here we have only presented the detailed analysis for S2, we performed the same analysis for the three different scenarios.

\begin{figure}[tb]
    \centering
    \includegraphics[width=\linewidth]{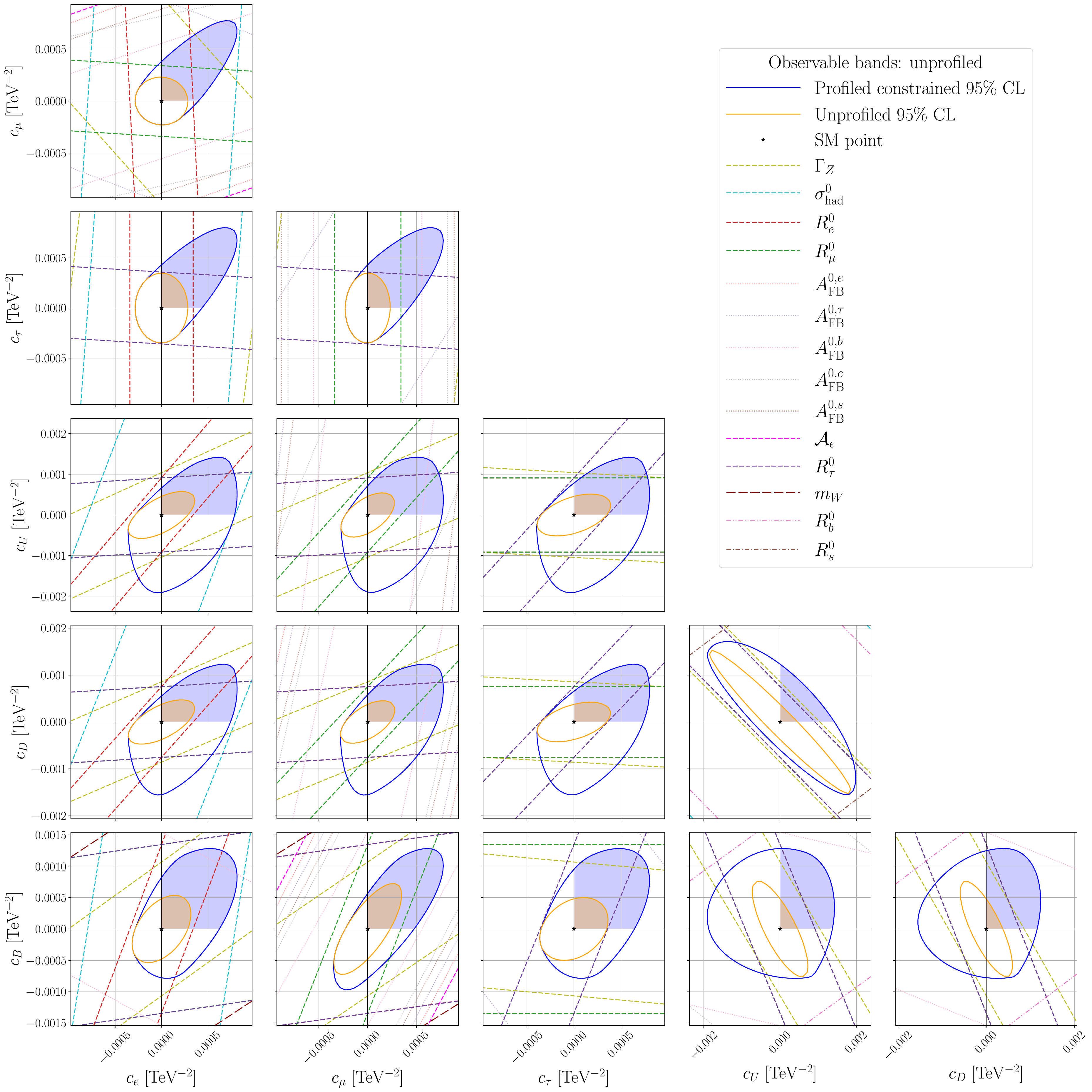}
    \caption{Triangle plot with 2-d scans of each parameter pairing with profiled, constrained (\textit{blue}) and unprofiled (\textit{orange}) 95\% CL ellipses. Unprofiled observable bands for the EWPO observables falling in the scan range are shown as dashed and dotted lines. The shaded regions in the positive quadrant mark the allowed region for which $c_F \geq 0$.}
    \label{fig:triangle-plot_S2}
\end{figure}

\paragraph*{2d-scans.} Analogous to the one-dimensional scans in \cref{fig:1d-scan}, we provide two-dimensional scans of each parameter pair for scenario S2 in \cref{fig:triangle-plot_S2}.
From these plots we can read off the correlations between the parameter pairs and see which observable constrains which direction in the corresponding two-dimensional space. The orange ellipses correspond to the unprofiled 95\% CLs while the blue ellipse shows the profiled, constrained case. The observable bands are determined from each unprofiled single-observable chi-squared which would become kinked and non-linear for the profiled case. Hence the observable bands should be interpreted as an estimate of the most constraining observable, not as the exact 95\% CL band for the profiled parameters. The shaded regions in the positive quadrant denote the physical region where $c_F \geq 0$ for both parameters.
We see that here the strongest observables are given by the ratios of widths $R_f$ together with $\Gamma_Z$.

\begin{figure}[tb]
    \centering
    \includegraphics[width=\linewidth]{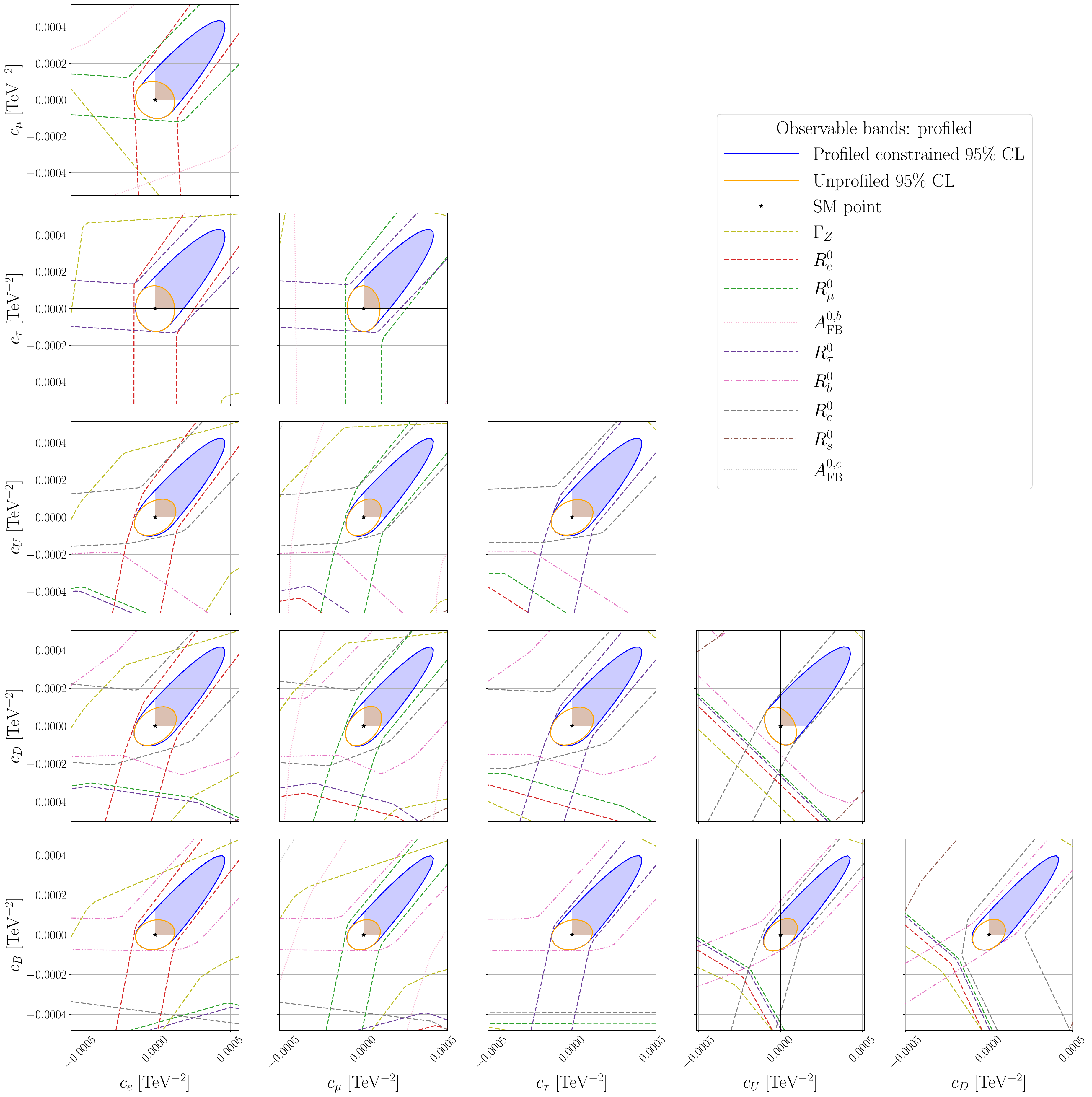}
    \caption{Triangle plot with 2-d scans of each parameter pairing with profiled, constrained (\textit{blue}) and unprofiled (\textit{orange}) 95\% CL ellipses. Profiled observable bands for the EWPO observables falling in the scan range are shown as dashed and dotted lines. They are calculated from the single observable $\chi^2$ function evaluated at the profiled parameter points minimising the global $\chi^2$ and hence show the individual observable contribution to the profiled global $\chi^2$. The shaded regions in the positive quadrant mark the allowed region for which $c_F \geq 0$.}
    \label{fig:triangle-plot_S3}
\end{figure}

The corresponding plot for scenario S3 in \cref{fig:triangle-plot_S3} shows a pronounced preferred direction arising from an approximately flat direction in parameter space, which is discussed below. Here, we also show the profiled bands for each observable, i.e. the observable evaluated at the profiled solution of the global one, whereas those in \cref{fig:triangle-plot_S2} are unprofiled.

\paragraph*{Higgs tuning.} Connected to the bounds on the VLF masses we scrutinise the Higgs tuning in \cref{eq:BG_measure} that they impose, shown on the right y-axis of \cref{fig:LEP_fit,fig:FCCee_fit}:
\begin{equation}
\label{eq:app_tuning_measure}
    \frac{1}{\Delta_{H,F}}=\frac{2\pi^2m_h^2}{N_cN_F|y_F|^2M_F^2}
    =\frac{2\pi^2m_h^2}{N_cN_F|y_F|^4}\left(\frac{M_F}{|y_F|}\right)^{-2}.
\end{equation}
Here, the combination $N_cN_F|y_F|^4/\Delta_{H,F}$ is a function of $M_F/|y_F|$ only, and the tuning for a given multiplicity and Higgs coupling is obtained by dividing by $N_cN_F|y_F|^4$. For the individual bounds only the contribution of the corresponding VLF class is included. The total tuning of the model follows from combining the six classes in quadrature according to \cref{eq:tuning_total}, which at the current marginalised bounds and for $|y_F|=1$ gives $1/\Delta_{\rm total}\simeq0.3\%$, as quoted in \cref{sec:current_constraints}. Note that the six classes need not all sit simultaneously at their profiled bounds within the $95\%$~CL region of the six-dimensional fit, so that this estimate should be regarded as indicative of the order of magnitude of the total tuning rather than as a strict bound.

We can, nonetheless, determine the maximal tuning compatible with the FCC-ee projections by minimising
$\Delta_{\rm total} = \big(\sum_F \Delta_{H,F}^2\big)^{1/2}$ over the 95\% CL region of the
linearised fit, $\chi^2 = C^T H\, C \leq \bar\chi^2_{95}$ with $c_F \geq 0$, where
$\bar\chi^2_{95} \simeq 8.41$ is the one-sided threshold for six parameters.\footnote{The
95\% quantile of the $\bar\chi^2$ distribution
$\sum_{k} \binom{6}{k}\, 2^{-6}\, \chi^2_k$ for six parameters bounded from below,
neglecting correlations in the mixture weights.}
We solve the problem by minimising on a
positive parametrisation of this surface and cross-check the result by randomly sampling
the surface. We find maximal tunings of
$|y|^4/\Delta_{\rm total} \simeq 1.3\times 10^{-2}\,\%$,
$3.3\times 10^{-3}\,\%$ and $2.4\times 10^{-3}\,\%$ for the theory-conservative,
theory-aggressive and experimental-only scenarios. At the optimum found by the minimiser, all states lie at
$M_F/|y_F| \simeq 16$--$17$, $31$--$32$ and $37$--$38\,\text{TeV}$,
respectively. Rescaling the mass bounds from the plot by the threshold ratio $\tilde{M}/|y| = (\Delta\chi^2_{1}/\Delta \chi^2_6)^{1/4} \times M/|y|$ then coincides with the masses obtained from the optimisation. When estimating the total tuning from the individual contributions to $\Delta_{H,F}$ summed in quadrature, we get upper limits on $|y|^4/\Delta_{\rm total}$ of $2.4\times10^{-3}\,\%$, $6.0\times10^{-3}\,\%$ and $2.5\times10^{-2}\,\%$ for the experimental-only, theory-aggressive and theory-conservative scenarios. Again, this is only a naive estimate of the realistically achievable tuning since all the 95\% CL bounds do not need to be saturated at the same time for all VLFs. This estimate shows that for the experimental-only scenario, the optimal tuning is actually achieved by the bounds shown in the plot, i.e. the masses already sit at the collective optimum. This together with the fact that for S3 all masses tend to the same bound indicates that we hit a nearly flat direction in the six-dimensional parameter space. We can see this behaviour in the 2d plot \cref{fig:triangle-plot_S3} where each two-dimensional slice of the six-dimensional space lies along the direction of equal $c_F$.

\paragraph*{The apparent flat direction.} Physically, the almost flat direction can be seen from the NP shifts of the $R_f$ ratios whose uncertainties decrease by roughly an order of magnitude from S2 to S3, confining the Wilson coefficients to a narrow region along the direction left unconstrained by these observables. The total width $\Gamma_Z$ limits the extent of this region, but its precision is insufficient to remove it. This can be seen from the formulas for the observables in terms of the NP parameters
\begin{align}
    R_{f} = \begin{cases}
    \dfrac{\Gamma_f}{\Gamma_{\rm had}} & f = c,s,b \\[10pt]
    \dfrac{\Gamma_{\rm had}}{\Gamma_{f}} & f = \ell
    \end{cases} \quad \implies \frac{\delta R_f}{R_f} = \pm (\frac{\delta\Gamma_f}{\Gamma_f} - \frac{\delta\Gamma_{\rm had}}{\Gamma_{\rm had}}).
\end{align}
From here we see that $R_{f}$ stays the same if all individual widths shift by the same relative amount $\delta = \delta\Gamma_{f}/\Gamma^{\mathrm{SM}}_{f}$ so that
\begin{equation}
   \frac{\delta \Gamma_{\mathrm{had}}}{\Gamma^{\mathrm{SM}}_{\mathrm{had}}} = \frac{1}{\Gamma_{\mathrm{had}}^{\mathrm{SM}}} \sum_{q} \delta \Gamma_{q}=\delta \quad  \implies \delta R_{f} = 0.
\end{equation}
The shifts of the widths come from the modified $Z$-couplings
\begin{equation}
    \delta\Gamma_{ff} \simeq \frac{m_Z^3\, N_c^f}{12\pi}\Big[-2\,T_3^f\, g_L^f\, C_f + A_f\,\big(c_{e}+c_{\mu}\big)\Big]\,,
\end{equation}
where the universal second term stems from the input shift $\delta G_{F}$ and we defined
\begin{equation}
  \kappa \equiv \frac{s_{2\theta}^2}{4\, c_{2\theta}} = \frac{s_\theta^2\,(1 - s_\theta^2)}{1 - 2 s_\theta^2},\qquad A_f \equiv (g_V^f)^2 + (g_A^f)^2 + 2\kappa\, Q_f\, g_V^f\,, \qquad
A_{\rm had} \equiv 2A_u + 2A_d + A_b\,.
\end{equation}
All shifts being equal imposes a set of linear equations that can only be approximately solved by all parameters being approximately equal. Indeed, setting all $c_{F} = \hat{c}$, we obtain a cancellation, reducing the $\delta R_{f}/R_{f}^{\mathrm{SM}}$ shifts by about an order of magnitude with respect to the single parameter shift. Consequently, the simultaneous fit allows sizeable correlated contributions along this direction, explaining why the profiled bounds in \cref{fig:FCCee_fit} approach similar values and remain substantially weaker than the individual bounds. This nearly flat direction is lifted by observables that do not scale like a ratio of the widths above, i.e. that experience a non-zero shift along the $\hat{c}$ direction. One example is $\Gamma_{Z}$ since it includes neutrino widths which do not feel any parameter shift for $C_{H\ell}^{(1)}=C_{H\ell}^{(3)}$ except for the universal contribution from $\delta G_{F}$. This means that in order to tighten the bounds for the experimental-only scenario, one must improve the measurements of observables such as $\Gamma_{Z}$ orthogonal to the flat direction.

\subsection{LEP electroweak methodology}
\label{app:ewpo_codes}

In this subsection, we describe the electroweak fits used to obtain the current bounds in \cref{sec:current_constraints}. At leading order, the contributions of the minimal model are controlled by six combinations of masses and couplings, which determine the Wilson coefficients through the matching relations in \cref{eq:app_EWPO_fit_params}. We therefore fit these six parameters, preserving the relations among the coefficients predicted by the model. In particular, the equality of the first two quark entries follows from the approximate $\mathrm{SU}(2)_q$ flavour symmetry. To implement this matching, we use the Warsaw basis with the WCxf conventions~\cite{Aebischer:2017ugx}, choosing diagonal down-quark and charged-lepton Yukawa matrices. Rotating to this basis generates additional spurion-suppressed off-diagonal terms, which we neglect in the EWPO analysis, as discussed in \cref{app:flavour_obs}. For the baseline comparison, we apply the matching relations directly at $\mu=m_Z$ and retain only contributions linear in the dimension-six coefficients.

With this setup, we compute the observables using \texttt{flavio}, following the conventions adopted in \texttt{smelli}~\cite{Straub:2018kue,Aebischer:2018iyb}, and use \texttt{wilson}~\cite{Aebischer:2018bkb} to handle the Wilson coefficients. The SM predictions include the higher-order corrections available in each implementation, whereas the NP contributions are evaluated at tree level. Although NLO electroweak and QCD corrections to SMEFT EWPO are available~\cite{Hartmann:2016pil,Dawson:2019clf,Biekotter:2025nln,Bellafronte:2026jic}, we do not include them in this comparison.

We then compare the resulting bounds with those obtained using \texttt{SMEFiT}~\cite{Giani:2023gfq,Celada:2024mcf} and \texttt{HEPfit}~\cite{DeBlas:2019ehy}. Since these determinations differ in both their electroweak inputs and statistical methods, their comparison indicates the combined effect of both on the bounds.

\paragraph*{Electroweak likelihoods.}
The fit performed with \texttt{flavio} contains $24$ electroweak observables, following the notation of \cref{eq:app_EWPO_list}. Alongside the forward--backward asymmetries, we include separate measurements of the asymmetry parameters $\mathcal A_f$, enabled by the polarised electron beam at SLD. For the $W$-boson mass, we use $m_W=80.3795\pm0.0121\,\mathrm{GeV}$, excluding the CDF~II measurement~\cite{CDF:2022hxs}.\footnote{The current PDG average, which does not use the CDF~II result either, $m_W=80.3625\pm0.0077\,$GeV~\cite{ParticleDataGroup:2026mpi}, is closer to the SM prediction. Since $[C_{H\ell}^{(3)}]_{ee,\mu\mu}=-c_{e,\mu}/4<0$, these coefficients lower the Fermi constant extracted from muon decay and thus raise the predicted $m_W$. This input, which applies also to the \texttt{HEPfit} value below, would tighten the bounds on $c_e$ and $c_\mu$; we have not repeated the fits with it.} Unlike for FCC-ee, where we do not know the possible experimental correlations, to account for the correlations among these measurements, we construct the experimental covariance from the LEP/SLD combinations~\cite{ALEPH:2005ab,ALEPH:2013dgf}. We then add the SM uncertainties as a diagonal theory covariance, obtaining the total covariance $V=V_{\rm exp}+V_{\rm th}$.

With these inputs, we write the linearised predictions as $\bm{\mathcal O}_{\rm th}=\bm{\mathcal O}_{\rm SM}+J\bm c$, where $J$ describes the response of the observables to the six fit parameters. The Gaussian likelihood is therefore characterised by
\begin{equation}
 \chi^2(\bm c)
 =\bm c^TH\bm c-2\bm b^T\bm c+\chi^2_{\rm SM},\quad \textrm{with }
 H=J^TV^{-1}J, \quad \textrm{and }
 \bm b=J^TV^{-1}
 (\bm{\mathcal O}_{\rm exp}-\bm{\mathcal O}_{\rm SM}).
\label{eq:app_flavio_chi2}
\end{equation}
The quadratic dependence on $\bm c$ follows from the linearised predictions and the use of a fixed covariance.

To derive bounds on the model parameters, we restrict the likelihood to the physical domain $c_F\geq0$ and consider two fitting strategies. In the individual fit, we vary one parameter while setting the other five to zero. To allow their contributions to vary simultaneously, we also perform a profiled fit: at each fixed value of the parameter of interest, we minimise the $\chi^2$ over the remaining five parameters using \texttt{iminuit}~\cite{iminuit}, retaining their non-negativity restrictions. For each scan, we then obtain the one-sided $95\%$ confidence limit from the crossing at $\Delta\chi^2=2.706$, measured relative to the minimum of the corresponding fit within the physical domain.

For comparison, the Bayesian fits described below explore the same physical domain with flat priors in the coefficients $c_F$. Here, the simultaneous bounds are obtained by marginalising over the remaining parameters rather than profiling over them. We define the one-sided upper limit as the $95$th percentile of the resulting posterior, applying the same prescription to the single-parameter fits.\footnote{This differs from the upper endpoint of a central $95\%$ credible interval, which corresponds to the $97.5$th percentile.} Since profiling selects the best-fitting values of the other parameters, whereas marginalisation integrates over their posterior distribution, the two approaches need not yield identical bounds, particularly near the physical boundary. In both approaches, we finally convert the upper limits on $c_F$ into lower bounds on $M_F/|y_F|$.

For the independent \texttt{SMEFiT} determination, we use the published \texttt{smefit\_database} tables~\cite{Celada:2024mcf}, which provide both the measurements and their theoretical predictions. The dataset contains $84$ measurements, comprising $20$ $Z$-pole and electroweak inputs, including $\alpha_{\rm EW}(m_Z)$, $21$ Bhabha-scattering measurements, three leptonic $W$ branching fractions, and $40$ $W$-pair-production measurements at centre-of-mass energies between $182$ and $206\,\mathrm{GeV}$. Its experimental content therefore differs from that of the \texttt{flavio} fit: it includes additional LEP-II measurements but omits $m_W$, $\Gamma_W$, $\mathcal A_s$ and $R_{uc}^0$, while retaining separate SLD and $\tau$-polarisation inputs.

To fit the same model to this dataset, we translate the six parameters of \cref{eq:app_EWPO_fit_params} into the \texttt{SMEFiT} operator basis. We enforce the resulting relations, so that only the six model parameters are varied. As in the \texttt{flavio} analysis, we retain contributions linear in the dimension-six coefficients, together with the higher-order SM predictions supplied in the database. With these predictions, we sample the posterior using \texttt{ultranest} and flat priors $c_F\in[0,1]\,\mathrm{TeV}^{-2}$. Following the prescription described above, we obtain individual limits from single-parameter fits and simultaneous limits from the one-dimensional marginal posteriors of the six-parameter fit. We then estimate the sampling uncertainties on the quoted quantiles by bootstrap resampling.

We complement these determinations with a custom \texttt{HEPfit} model implementing the same six parameters. This provides a further comparison using an independent calculation of the electroweak observables and their higher-order SM corrections. We adopt the measured central values, uncertainties and correlations of Ref.~\cite{DeBlas:2019ehy}, including $m_W=80.379\pm0.012\,\mathrm{GeV}$, and vary the SM inputs as nuisance parameters subject to their corresponding constraints.

\begin{figure}[tb]
 \centering
 \includegraphics[width=0.9\linewidth]{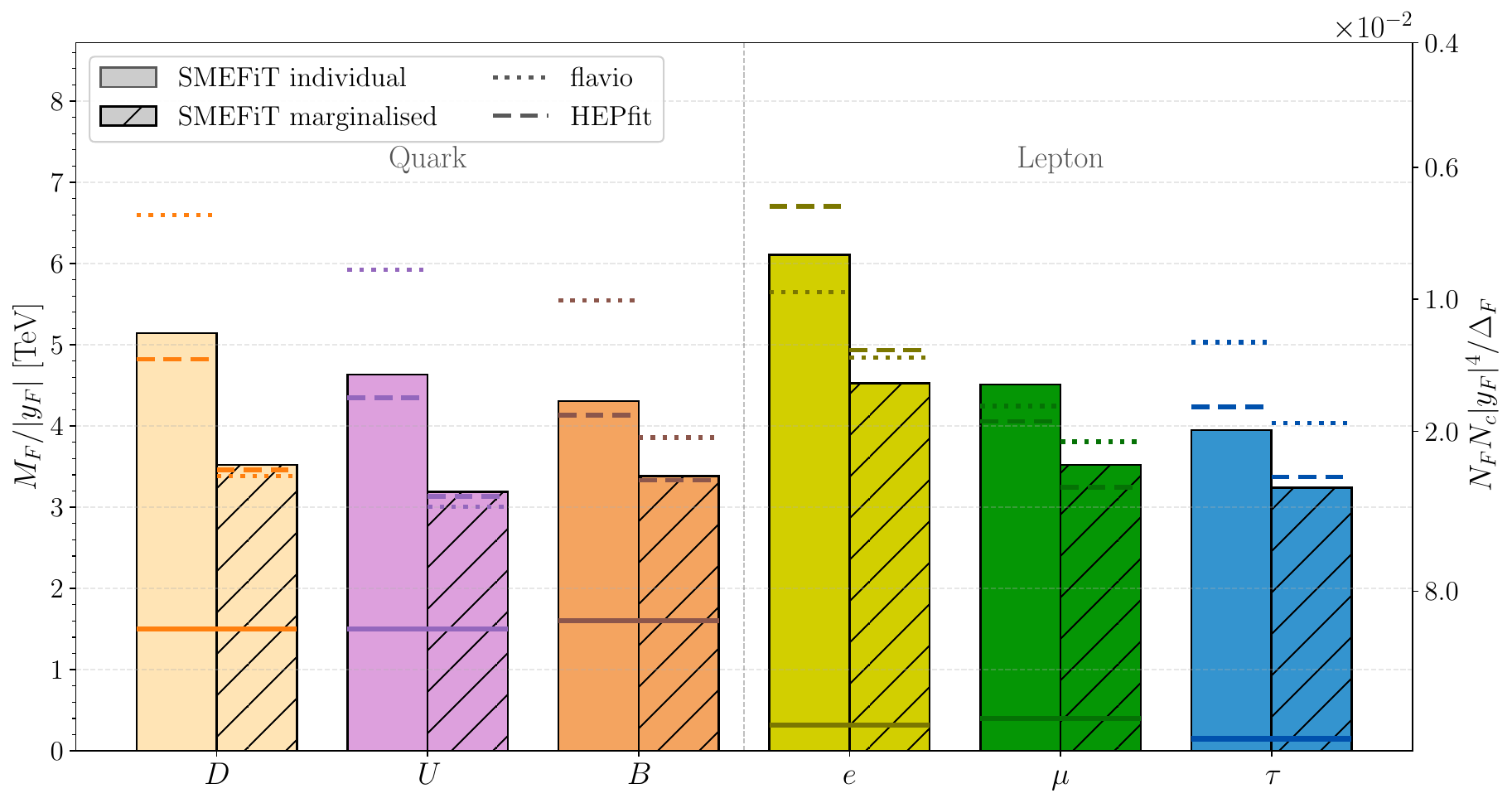}
 \caption{Current $95\%$ CL lower limits on $M_F/|y_F|$ for the quark (left) and lepton (right) VLF classes. Filled and hatched bars denote individual and marginalised \texttt{SMEFiT} limits. Dotted and dashed lines show the corresponding \texttt{flavio} and \texttt{HEPfit} results, with profiling used for the simultaneous \texttt{flavio} fit. The interval conventions are specified in the text. Solid horizontal lines show direct-search limits on $M_F$, allowing a comparison for $|y_F|=1$. The right axis gives $N_cN_F|y_F|^4/\Delta_{H,F}$.}
 \label{fig:LEP_fit}
\end{figure}

Before constructing the \texttt{HEPfit} posterior, we ensure that the SM inputs are interpreted consistently with the other implementations. In particular, we treat the top mass as a pole mass by setting \texttt{FlagMtPole=true}. This choice has an appreciable effect on the predicted $W$ mass, since omitting the flag gives $m_W=80.419\,\mathrm{GeV}$, approximately $3.3$ experimental standard deviations above the measured value. With the pole-mass convention imposed, the SM predictions for $m_W$ from the three implementations agree within $0.4$ experimental standard deviations. Having fixed these input conventions, we sample the posterior with the Markov-chain Monte Carlo implementation, again adopting flat priors on $c_F\in[0,c_F^{\rm max}]$. We require a Gelman--Rubin statistic $R<1.01$ and $10^5$ iterations per chain, and extract the limits only from converged chains. To check the posterior extraction, we also repeat the fits with SM-centred pseudo-data. In this case, a single-parameter Gaussian likelihood with a flat non-negative prior gives a half-normal posterior, provided the upper prior boundary lies sufficiently far from the allowed region. We extract the upper limit by finding the value of $c_F$ below which $95\%$ of the posterior probability lies.

\paragraph*{Comparison of the bounds.}
The resulting lower limits on $M_F/|y_F|$ are shown in \cref{fig:LEP_fit}. Filled and hatched bars denote the individual and marginalised \texttt{SMEFiT} limits, while dotted and dashed lines show the corresponding \texttt{flavio} and \texttt{HEPfit} determinations. For the simultaneous \texttt{flavio} fit, the limits are obtained by profiling rather than marginalisation. To interpret the differences between these determinations, it is important to recall that they use different datasets as well as different statistical prescriptions, as summarised in \cref{tab:app_code_comparison}. In particular, $m_W$ enters the \texttt{flavio} and \texttt{HEPfit} fits but is absent from the \texttt{SMEFiT} dataset used here. The comparison in the figure therefore reflects both effects, which we do not disentangle here. The individual limits of the three implementations differ by up to about $40\%$ in $M_F/|y_F|$, with \texttt{flavio} giving the strongest bounds on the quark classes.

Despite these differences, all three determinations show that EWPO set stringent bounds on the scale of the VLF chains. The electron-like VLF is the most constrained, with LEP/SLD data probing $M_e/|y_e|\gtrsim6\,\mathrm{TeV}$. Most individual bounds lie in the $4$--$5\,\mathrm{TeV}$ range, while the profiled or marginalised bounds typically lie between $3$ and $4\,\mathrm{TeV}$. Although some flavour bounds reach similar scales, see \cref{sec:flavour_direct,app:flavour_obs}, EWPO probe all the VLF chains of the model, whereas flavour observables constrain only a few VLF classes and combinations of parameters. They therefore provide a broader test of these chain-based flavour models, whose implications are discussed in \cref{sec:current_constraints}.

To place these indirect bounds in context, we also show the direct-search benchmarks of \cref{sec:flavour_direct}, namely $1.5\,\mathrm{TeV}$ for $U$ and $D$, $1.6\,\mathrm{TeV}$ for $B$, and $320$, $400$ and $150\,\mathrm{GeV}$ for the electron, muon and tau partners, respectively. The lepton limits correspond to weak singlets, as required by the minimal model. Since direct searches constrain $M_F$, their comparison with the electroweak bounds assumes $|y_F|=1$. Finally, the right axis translates the mass bounds into the normalised inverse sensitivity $N_cN_F|y_F|^4/\Delta_{H,F}$. The tuning is defined in \cref{eq:BG_measure}, and its conversion for each VLF class is given in \cref{eq:app_tuning_measure}.

\begin{table*}[tb]
 \centering
 \renewcommand{\arraystretch}{1.2}
 \begin{tabular}{lccc}
  \hline\hline
  & \texttt{flavio} & \texttt{SMEFiT} & \texttt{HEPfit} \\
  \hline
  Data & $24$ observables & $84$ measurements & Electroweak dataset \\
    $m_W$ included & Yes & No & Yes \\
  Simultaneous limits & Profile likelihood & Marginal posterior & Marginal posterior \\
  Numerical method & Constrained minimisation & Nested sampling & MCMC \\
  Upper-limit prescription & $\Delta\chi^2=2.706$ & $95$th percentile & $95$th percentile \\
  Physical domain & $c_F\geq0$ & $c_F\geq0$ & $c_F\geq0$ \\
  \hline\hline
 \end{tabular}
 \caption{Electroweak fits entering \cref{fig:LEP_fit}. The \texttt{SMEFiT} column refers to the independent database fit used.}
 \label{tab:app_code_comparison}
\end{table*}
\twocolumngrid
\bibliographystyle{JHEP}
\bibliography{refs}

\end{document}